\documentclass{ws-ijmpa}

\def\be{\begin{equation}}
\def\ee{\end{equation}}
\def\ba{\begin{array}}
\def\ea{\end{array}}
\def\beqn{\begin{eqnarray}}
\def\eeqn{\end{eqnarray}}
\def\nonum{\nonumber}
\def\bt{\begin{tabular}}
\def\et{\end{tabular}}
\def\bc{\begin{center}}
\def\ec{\end{center}}

\begin{document}

\markboth{Manmohan Gupta, Gulsheen Ahuja} {Flavor mixings and
textures of the fermion mass matrices}

%
\catchline{}{}{}{}{}
%

\title{FLAVOR MIXINGS AND TEXTURES OF THE FERMION MASS MATRICES}

\author{MANMOHAN GUPTA}

\address{Department of Physics, Centre of Advanced Study, P.U., Chandigarh,
India.\\ mmgupta@pu.ac.in}

\author{GULSHEEN AHUJA}

\address{Department of Physics, Centre of Advanced Study, P.U., Chandigarh,
India.}

\maketitle

\begin{history}
\received{Day Month Year}
\revised{Day Month Year}
\end{history}

\begin{abstract} A comprehensive review of several aspects of
fermion mixing phenomenon  and texture specific mass matrices have
been presented. Regarding fermion mixings, implications of
unitarity and certain new developments for the CKM paradigm have
been discussed. In the leptonic sector, the question of
possibility of CP violation has been discussed in detail from the
unitarity triangle perspective. In the case of texture specific
mass matrices, the issues of viability of Fritzsch-like as well as
non Fritzsch-like mass matrices have been detailed for both the
quark and leptonic sectors. The relationship of textures,
naturalness and weak basis rotations has also been looked into.
The issue of the compatibility of texture specific mass matrices
with the SO(10) based GUT mass matrices has also been discussed.

\keywords{Fermion mixings, unitarity and CP violation, fermion
mass matrices, SO(10) based mass matrices}
\end{abstract}

\ccode{PACS numbers: 12.15.Hh, 12.15Ff, 12.10.-g, 14.60Pq}

\section{Introduction}
Understanding fermion masses and mixings, an important aspect of
flavor physics, is one of the outstanding problems of present day
High Energy Physics. Fermion mixing phenomena in the context of
quarks was initiated by Cabibbo in 1963\cite{cabibbo},
subsequently generalized to two generations by Glashow,
Illiopoulos, Maiani\cite{gim} and finally to three generations by
Kobayashi and Maskawa\cite{kobayashi}. This is very well
incorporated in the Standard Model (SM)\cite{gla}\cdash
\cite{smrev} and has been tested to a great accuracy. Recently,
flavor mixing has also been observed in the case of neutrinos
implying the existence of non zero and non degenerate neutrino
masses.

The complexity of the problem can be understood by considering the
fermion masses which span over many orders of magnitudes. For
example, considering the smallest neutrino mass to be of the order
of a fraction of an eV and the top quark mass being of the order
of 10$^{11}$ eV, the range of fermion masses looks to span over 13
orders of magnitudes. Similarly, fermion mixing angles again span
several orders of magnitudes, these being quite small in the quark
sector whereas in the leptonic sector these are somewhat `large'.
The recent T2K\cite{t2k}, MINOS\cite{minos}, DAYA
BAY\cite{dayabay} and RENO\cite{reno} measurements regarding the
mixing angle $s_{13}$ also suggest a not so `small' value. In
fact, understanding the different patterns of mixing angles in the
quark and lepton sector is a problem in itself. One may also
mention that the non zero value of angle $s_{13}$ has given a big
impetus to the sharpening of the implications of the neutrino
oscillations and has added another dimension to neutrino physics
by implying the possibility of existence of CP violation in the
leptonic sector.

At present, it seems that fermion masses and mixings provide a
fertile ground to hunt for physics beyond the SM as well as pose a
big challenge to understand these from more fundamental
considerations. It may be noted that mixing angles and CP
violating phases are very much related to the corresponding mass
matrices, therefore in view of this relationship one has to
essentially formulate the fermion mass matrices to unravel some of
the deeper aspects of flavor physics. This becomes more
challenging and interesting in case one attempts to understand the
fermion masses and mixings in a unified framework. To this end,
several ideas regarding the possible connections between quarks
and leptons\cite{qlepuni} are under investigation. One of the most
seriously pursued idea is the quark-lepton symmetry wherein quarks
and leptons may be two different manifestations of the same form
of matter. This principle is enshrined in the Grand Unified
Theories (GUTs) wherein both quarks and leptons form multiplets of
the extended group. The most appealing such a group is SO(10)
where all known components of quarks and leptons, including the
right handed neutrinos, fit into the unique 16-plet spinor
multiplet.

It may be noted that while on the one hand, GUTs have provided
vital clues for understanding the relationship of fermion mass
matrices between quarks and leptons, on the other hand, horizontal
symmetries\cite{horizontal1}\cdash \cite{horizontal5} have given
clues for the relationship between different generation of
fermions. Ideas such as extra dimensions\cite{extradim1,extradim2}
have also been invoked to understand fermion masses and mixings.
Unfortunately, at present it seems that we do not have any
theoretical framework which provides a viable and satisfactory
description of flavor physics. The lack of a convincing fermion
flavor theory from the `top down' perspective necessitates a
re-look at the issue from a `bottom up' approach. The essential
idea behind this complimentary approach is that one tries to find
the phenomenological fermion mass matrices which are in tune with
the low energy data and can serve as guiding stone for developing
more ambitious theories. One may note that at present a `bottom
up' approach perhaps may be more useful, wherein apart from
understanding the subtleties of mixing matrices of quarks and
leptons, one has to also understand the texture structure,
hierarchy of elements, phase structure, etc. of the corresponding
mass matrices\cite{9912358,0307359}. In particular, a viable
strategy would be to examine general features of texture specific
mass matrices of quarks and leptons which are in tune with the
mixing parameters as well as are in accordance with the GUTs.

One may note that vital constraints on mass matrices can be
obtained through continuous refinements of the mixing parameters.
In particular, in the last few years, the recent success of B
factories\cite{bfactories1,bfactories2}, analyzing over $10{^9}$
decays of B-hadrons, has provided a wealth of data leading to a
good deal of progress in our understanding of flavor physics.
Precise measurements of CP asymmetry parameter sin$\,2\beta$,
characterizing CP asymmetry $a_{\psi K_s}$ in the $B^o_d
\rightarrow \psi K_s$ decay, also referred to as the Golden Mode,
the Cabibbo-Kobayashi-Maskawa (CKM)\cite{cabibbo,kobayashi} matrix
elements $V_{us}, V_{cb}, V_{ud}$ as well as of several other
parameters have been carried out. Based on these, several
phenomenological analyses\cite{pdg10}\cdash\cite{hfag} have
allowed us to conclude that the single CP violating phase, encoded
in the CKM matrix, appears to be the dominant source of CP
violation in the meson decays, at least to the leading order.
Further, these refinements have also brought up several issues
having potentiality of New Physics (NP). Therefore, one may
conclude that continuous refinements of the mixing parameters,
coupled with certain theoretical improvements in the lattice QCD
calculations, not only sharpen the constraints on the mass
matrices but also provide clues for finding NP.

The detailed plan of the article is as follows. Essentials
regarding quark mixing phenomenology, unitarity and CP violation
in the quark sector have been presented in Section (\ref{qmph}).
Section (\ref{uninew}) presents the implications of unitarity,
precision measurements and certain new developments for the CKM
paradigm. Similarly, in Section (\ref{unilep}) unitarity and CP
violation in the leptonic sector have been discussed. Further,
details pertaining to texture 6, 5, 4 zero quark and lepton mass
matrices have been presented in Section (\ref{tsmm}). The concepts
of natural mass matrices and weak basis (WB) transformations have
been discussed in Section (\ref{nmmwb}). Section (\ref{texso10})
discusses the issue of compatibility of texture specific mass
matrices with SO(10) inspired matrices. Finally, in Section
(\ref{summ}) we summarize and conclude.

\section{Essentials of quark mixing phenomenology \label{qmph}}
The idea of quark mixing was introduced by Cabibbo\cite{cabibbo}
in 1963 in order to explain the suppression of the
strangeness-changing ($\bigtriangleup S= 1$) weak interactions in
comparison to the strangeness-conserving ($\bigtriangleup S= 0$)
weak interactions involving hadrons. To explain this anomalous
behaviour, Cabibbo suggested that the electroweak eigenstates are
a mixture of the flavor eigenstates ($ u, d, s$), e.g., \be \left(
\ba{c}u
\\d
 \ea \right) \longrightarrow  \left( \ba {c}
u \\ d' \\
 \ea \right)\longrightarrow
\left( \ba {c} u\\ d cos \theta_ {c}+ s  sin  \theta _{c} \ea
\right),\ee where $\theta_{c} $ is the Cabibbo angle. The Cabibbo
hypothesis was generalized by Glashow, Illiopoulos and Maiani
(GIM)\cite{gim} to explain the absence of the flavor changing
neutral currents by introducing a new quark, later on called the
charm quark. In accordance with the GIM mechanism, the doublet
consisting of c and s quarks takes the form \be \left( \ba{c}c
\\ s' \ea \right)\longrightarrow \left( \ba {c} c \\ s cos \theta_{c}- d
sin  \theta _{c} \ea \right),\ee leading to \be \left( \ba{c}d'
\\ s'
\ea \right)= V \left( \ba{c}d
\\ s
\ea \right)= \left( \ba {cc} cos \theta_{c} & sin \theta_{c} \\ -
sin \theta_{c}& cos \theta_{c}\ea \right)  \left( \ba {c}d
\\s \ea \right),\ee where $V$ is the $2 \times 2$ quark mixing
matrix.

In 1974, Kobayashi and Maskawa\cite{kobayashi} generalized the
above mixing matrix to the case of three generations by defining
the weak interaction eigenstates ($ d', s', b'$) in terms of the
flavor eigenstates ($ d, s, b$), e.g.,
  \be
   \left( \ba{c}d'
\\ s'
\\ b'
 \ea \right)= V
\left( \ba{c}d
\\ s
\\ b
\ea \right)= \left( \ba {lll} V_{ud} & V_{us} & V_{ub} \\ V_{cd} &
V_{cs} & V_{cb} \\ V_{td} & V_{ts} & V_{tb} \\ \ea \right) \left(
\ba {c} d\\ s \\ b \ea \right). \ee This quark mixing matrix is a
unitary matrix which describes the transition from one quark to
another quark, mediated by the charged weak gauge currents. It may
be added that a general $n \times n$ unitary matrix has $n^2$
parameters, $n(n-1)/2$ of these are the Eulers angles and the
remaining parameters are the phases. However, in the present case,
some of the phases can be rotated away, leaving only
$(n-1)(n-2)/2$ measurable physical phases. Thus, for the case of
two generations, it may be noted that one is left with no CP
violation as the phase gets rotated away. However, for the case of
three families, the mixing matrix is expressed in terms of three
mixing angles and one phase, the latter being responsible for CP
violation.

Originally, Kobayashi and Maskawa considered the following
parameterization\cite{kobayashi}, obtained by taking the product
of three rotations, namely \beqn
V_{\rm{KM}}&=&R_{23}(\theta_3,\phi)\, R_{12}(\theta_1,0)\,
R_{23}(\theta_2,0) \nonumber \\ & = & \left( \ba{ccc} c_1 & -s_1
c_3 & -s_1s_3 \\ s_1c_2 & c_1c_2c_3-s_2s_3\,e^{i\phi} & c_1c_2s_3
+ s_2c_3\,e^{i\phi}
\\
s_1s_2 & c_1s_2c_3+c_2s_3\,e^{i\phi} & c_1s_2s_3 -
c_2c_3\,e^{i\phi} \ea \right)\,, \eeqn where $s_i$=${\rm
sin}\,\theta_i$ and $ c_i$=${\rm cos}\,\theta_i$ for $i$=1,2,3. It
may be noted that apart from the above mentioned form, several
other parameterizations of the quark mixing matrix have been
proposed in the literature. In particular, altogether there are 36
different possible parameterizations which are all equivalent.
These parameterizations can be arrived at easily if one notes that
rotations in different planes do not commute, therefore for a
given central rotation there are four possibilities related to
left and right rotations. One may also mention that there are
three possibilities for the central rotation, leading to 12
possible rotations. Further, corresponding to each of these there
are three ways in which the phase can be introduced leading to 36
parameterizations in all. For details, the readers are referred to
Ref.~\refcite{jarl}. In the following, we discuss some of the
commonly used parameterizations of the quark mixing matrix.

One of the popular parameterization based on hierarchical
expansion was proposed by Wolfenstein\cite{wolf}. In this
representation, each element is expanded as a power series in the
small parameter $\lambda $, leading to the quark mixing matrix
$V_{\rm {CKM}}$ being
 \be V_{{\rm CKM}} =
 \left( \ba {ccc} 1-\frac{1}{2} \lambda ^2 & \lambda  &
  A \lambda ^3 (\rho- i\eta )\\
  -\lambda &
 1-\frac{1}{2} \lambda ^2
  & A \lambda ^2 \\
  A \lambda ^3 (1- \rho- i\eta )&
  -  A \lambda ^2 &
 1 \ea \right),  \label{wolfeq}  \ee
where $\lambda \sim 0.22, A \sim 0.83$ and the magnitudes of
$\rho$ and $\eta$ are smaller than one. It may be added that in
the context of flavor physics, this parameterization has been
extensively used, however, one may note that in this case the
constraints of unitarity have to be satisfied order by order.

It needs to be mentioned that because of the smallness of
$\lambda$ and the fact that for each element the expansion
parameter is actually $\lambda^2$, the Wolfenstein
parameterization is a rapidly converging expansion. In case one
requires sufficient level of accuracy, the terms of
${\cal{O}}(\lambda^4)$ and ${\cal{O}}(\lambda^5)$ have to be
included in phenomenological applications. This has been carried
out by generalizing the above mentioned Wolfenstein
parameterization to the Wolfenstein-Buras
parameterization\cite{wolfbur1,wolfbur2} by including
${\cal{O}}(\lambda^4)$ and ${\cal{O}}(\lambda^5)$ terms, e.g.,
\begin{equation}\label{2.775}
V_{{\rm CKM}}= \left(\begin{array}{ccc}
1-\frac{1}{2}\lambda^2-\frac{1}{8}\lambda^4               &
\lambda+{\cal{O}}(\lambda^7)                                   & A
\lambda^3 (\rho-i \eta)                              \\
-\lambda+\frac{1}{2} A^2\lambda^5 [1-2 (\rho+i \eta)]  &
1-\frac{1}{2}\lambda^2-\frac{1}{8}\lambda^4(1+4 A^2)     &
A\lambda^2+{\cal{O}}(\lambda^8)                                \\
A\lambda^3(1-\overline\rho-i\overline\eta) &
-A\lambda^2+\frac{1}{2}A\lambda^4[1-2 (\rho+i\eta)]   &
1-\frac{1}{2} A^2\lambda^4
\end{array}\right),
\end{equation}
where
\begin{equation}\label{2.88d}
\overline\rho\simeq \rho (1-\frac{\lambda^2}{2})+{\cal
O}(\lambda^4), \qquad \overline\eta=\eta
(1-\frac{\lambda^2}{2})+{\cal O}(\lambda^4).
\end{equation}
By definition the expression for $V_{ub}$ remains unchanged
relative to the original Wolfenstein parameterization, given in
Eq.~(\ref{wolfeq}), and the corrections to $V_{us}$ and $V_{cb}$
appear only at ${\cal{O}}(\lambda^7)$ and ${\cal{O}}(\lambda^8)$
respectively. The advantage of this generalization of the
Wolfenstein parameterization is the absence of relevant
corrections to $V_{us}$, $V_{cd}$, $V_{ub}$ and $V_{cb}$ and an
elegant change in $V_{td}$, however it may be noted that the
constraints of unitarity are not explicit.

Another popular parameterization, the `standard parameterization'
advocated by Chau, Keung\cite{chau}, adopted by Particle Data
Group (PDG)\cite{pdg10}, is given by \beqn V_{{\rm
CKM}}=R_{23}(\theta_{23},0)\, R_{13}(\theta_{13},-\delta)\,
R_{12}(\theta_{12},0)~~~~~~~~~~~~~~~~~~~~~~~~~ \nonumber \\
~~~~~~~~~~~~=
 \left( \ba {ccc} c_{12} c_{13} & s_{12} c_{13} &
  s_{13}e^{-i{ \delta}} \\
  -s_{12} c_{23} - c_{12} s_{23} s_{13}e^{i{\delta}} &
 c_{12} c_{23} - s_{12} s_{23}s_{13}e^{i{\delta}}
  & s_{23} c_{13} \\
  s_{12} s_{23} - c_{12} c_{23} s_{13}e^{i{\delta}} &
  - c_{12} s_{23} - s_{12}c_{23} s_{13}e^{i{\delta}} &
  c_{23} c_{13} \ea \right),  \label{1ckm}  \eeqn
  with $c_{ij}={\rm cos}\,\theta_{ij}$ and
   $s_{ij}={\rm sin}\,\theta_{ij}$.  The angles $\theta_{12}, \theta_{23}$
and $\theta_{13}$ can be chosen to lie in the first quadrant. The
parameter $ \delta $ is the CP violating phase which may vary in
the range $0 \leq \delta \leq 2\pi$. However, measurements of CP
violation in K decays force it to be in the range $0 < \delta <
\pi$ as the sign of the relevant hadronic parameter is fixed. The
relationship of the mixing angles and the CP violating phase with
the Wolfenstein parameters, $\lambda, A, \rho$ and $\eta$ is given
by
  \be s_{12}=\lambda,~~~ s_{23}=A \lambda^2,
  ~~~\rho=\frac{s_{13}}{s_{12}s_{23}}{\rm cos}\,\delta,
  ~~~\eta=\frac{s_{13}}{s_{12}s_{23}}{\rm sin}\,\delta. \ee
It may be mentioned that in the present era of precision
measurements this parameterization is found to be very useful for
numerical evaluations. Noting that $c_{13}$ being almost close to
unity, one can consider the three mixing angles $\theta_{12},
\theta_{23}$ and $\theta_{13}$ respectively being represented by
the three elements of the mixing matrix, $V_{us}$, $V_{cb}$ and
$V_{ub}$, measurable at tree level. Therefore, whenever one has to
consider issues related to New Physics, the use of this
parameterization is recommended as the CKM matrix can be easily
constructed by making use of the measured values of the above
mentioned mixing matrix elements and the CP violating phase
$\delta$. Also, for phenomenological analyses wherein emphasis is
on unitarity it becomes more convenient to use this representation
wherein the unitarity is built-in.

Another important parameterization discussed in the literature is
the Fritzsch-Xing parameterization\cite{frzans1,frzans2} given by
                \be V_{{\rm CKM}} =
 \left( \ba {ccc} s_{u} s_{d} c + c_{u}c_{d} e^{-i{\phi}} & s_{u} c_{d} c + c_{u}s_{d} e^{-i{\phi}} &
  s_{u}s \\
 c_{u} s_{d} c - s_{u}c_{d} e^{-i{\phi}} &
 c_{u} c_{d} c + s_{u}s_{d} e^{-i{\phi}} & c_{u}s
  \\
 -s_{d} s  &
  - c_{d} s &
  c
  \ea \right),  \ee
where $ s_{u} = {\rm sin}\theta_{u}, $
        $ s_{d} = {\rm cos} \theta_{d}, $
         $ c = {\rm cos}\theta. $
It may be mentioned that in contrast to the other
parameterizations, there is no phase in the third row and the
third column of the mixing matrix, thus the CP violating phase
resides only in the $2 \times 2$ submatrix involving light quarks
$u, d, s$ and $c$. Fritzsch and Xing recommend the use of this
parameterization for the study of flavor mixing and CP violating
phenomena.

\subsection{Unitarity and unitarity triangles in the quark sector
\label{uniquark}} After having looked at some of the
parameterizations of the quark mixing matrix, we discuss its
unitarity which is the only powerful constraint, imposed by the SM
itself, on the quark mixing matrix. It may be noted that unitarity
and unitarity triangles have played a crucial role in
understanding the implications of CKM phenomenology. The unitarity
triangles, in particular, have also played an important role in
establishing the CKM paradigm as well as the fact that a single CP
violating phase $\delta$ is largely responsible for understanding
the CP violation in the K and B sector. In this context, several
well known groups\cite{pdg10}\cdash\cite{hfag} have been
periodically updating their analyses which have played crucial
role in affecting refinements of the CKM paradigm. It is outside
the scope of the present review to delve into the details of their
analyses, however, we would like to look into some of the aspects
of the unitarity triangles which have facilitated understanding of
fermion mixing phenomena and its relation to the fermion mass
matrices.

\begin{figure}[hbt]
\vspace{0.10in} \centerline{\epsfysize=4in\epsffile{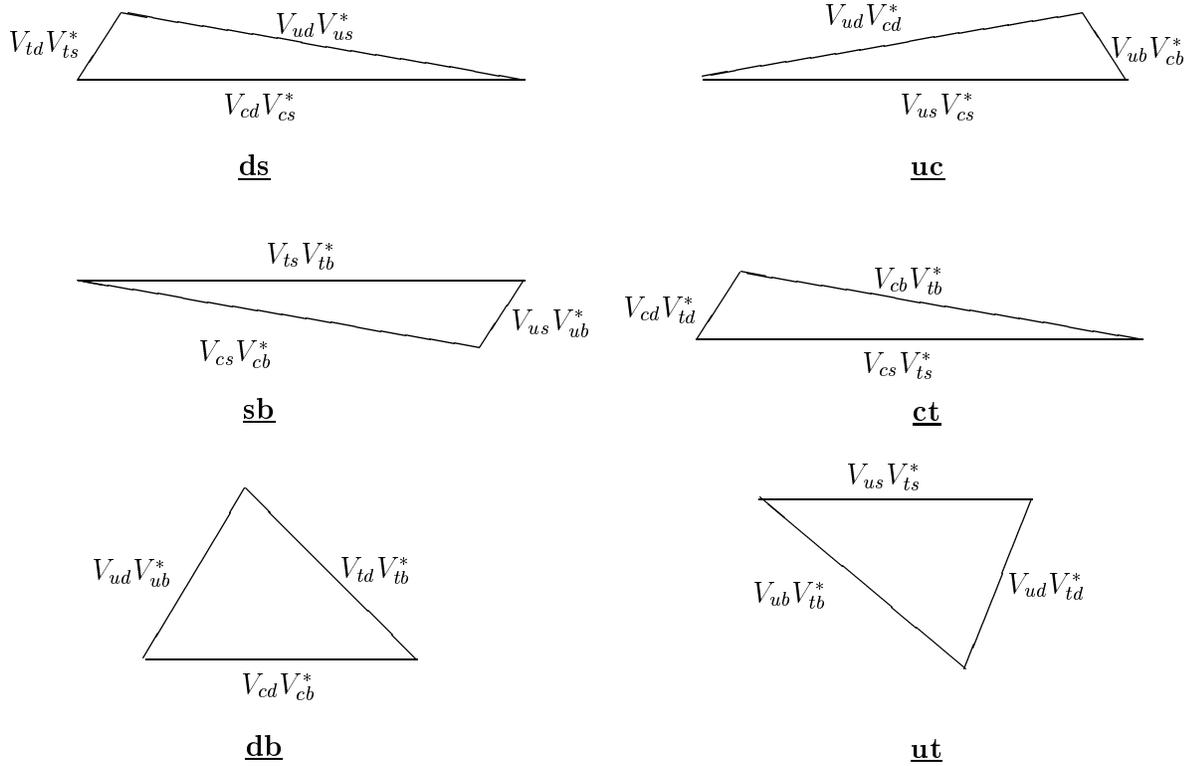}}
\vspace{0.08in} \caption{The six unitarity triangles in the quark
sector. The symbols {\bf ds} etc. indicate
 the pair of rows or columns, whose orthogonality the triangle
 depicts. The magnitudes of the sides are not to scale.}
\label{figsix}
\end{figure}

Unitarity of the CKM matrix implies nine relations, three in terms
of normalization conditions also referred to as `weak unitarity
conditions', and the other six are usually expressed through
unitarity triangles in the complex plane. Because of the strong
hierarchical nature of the CKM matrix elements as well as the
limitations imposed by the present level of measurements, it is
difficult to study the implications of normalization relations,
therefore, the six non-diagonal relations are used to study the
implications of unitarity on CKM phenomenology. These six
non-diagonal relations can be expressed through six independent
unitarity triangles in the complex plane and can be defined as
  \be   \sum_{\alpha=d,s,b} V_{i\alpha}V^*_{j\alpha}=\delta_{ij}\,, \ee
 \be  \sum_{i=u,c,t} V_{i\alpha}V^*_{i\beta}=\delta_{\alpha\beta}\,,
    \label{unit2} \ee
where Latin indices run over the up type quarks $(u,c,t)$ and
Greek ones run over the down type quarks $(d,s,b)$. The six
unitarity triangles in the complex plane can be expressed as \beqn
  uc \qquad  V_{ud}V_{cd}^{*} + V_{us}V_{cs}^{*} + V_{ub}V_{cb}^{*}
& = & 0 \,, \label{uc2} \\ db \qquad V_{ud}V_{ub}^{*} +
V_{cd}V_{cb}^{*} + V_{td}V_{tb}^{*}
 & = & 0\,,
\label{db2} \\
  ds \qquad V_{ud}V_{us}^{*} + V_{cd}V_{cs}^{*} + V_{td}V_{ts}^{*}
& =  & 0\,, \label{ds2}  \\
  sb \qquad V_{us}V_{ub}^{*} + V_{cs}V_{cb}^{*} + V_{ts}V_{tb}^{*}
 & = & 0\,,
\label{sb2}  \\
  ut \qquad V_{ud}V_{td}^{*} + V_{us}V_{ts}^{*} + V_{ub}V_{tb}^{*}
& = & 0\,, \label{ut2}  \\
  ct \qquad V_{cd}V_{td}^{*} + V_{cs}V_{ts}^{*} + V_{cb}V_{tb}^{*}
& = & 0\,,
 \label{ct2}
\eeqn where the letters $uc$ etc. represent the corresponding
unitarity triangle. These triangles are schematically shown in
Fig.~\ref{figsix}.

\subsection{Unitarity and CP violation in the quark sector}
In the context of CP violation, phenomenologically there is an
important parameter known as the Jarlskog's rephasing invariant
parameter and denoted as $J$\cite{jarl}. The significance of $J$
lies in the fact that all the CP violating effects in the Standard
Model (SM) are proportional to it. Also, $J$ is a universal
quantity in the sense that it does not depend on the specific
parameterization of the CKM matrix, therefore it is rephasing
invariant. Because of the above mentioned reasons, $J$ is of much
interest for the study of CP violation in the CKM phenomenology.
Further, the parameter $J$ is related to area of any of the
unitarity triangle as \be |J| = 2 \times {\rm Area~ of~any~ of~
the~ unitarity~ triangle.}
 \label{area} \ee
In terms of the elements of the CKM matrix, $J$ can be written in
a form that is explicitly parameterization independent, e.g.,
\begin{equation}
{\rm Im} \left (V_{i\alpha} V_{j\beta} V^*_{i\beta} V^*_{j\alpha}
\right ) \; =\; J \sum_{k,\gamma} \left (\varepsilon_{ijk}
\varepsilon_{\alpha\beta\gamma} \right ) \; \label{j3},
\end{equation}
in which each Latin subscript ($i,j,k$) runs over the up-type
quarks $(u,c,t)$ and each Greek subscript ($\alpha, \beta$,
$\gamma$) runs over the down-type quarks $(d,s,b)$. Thus knowing
the elements of the CKM matrix, $J$ can be easily evaluated.
Further, in the standard parameterization $J$ is given as
\be
J=s_{12}s_{23}s_{13}c_{12}c_{23}c_{13}^2 \,{\rm sin}\,\delta,
\label{jdq}\ee with $c_{ij}= {\rm cos}~ \theta_{ij}$ and $s_{ij}=
{\rm sin}~ \theta_{ij}$ for $i,j=1,2,3$. $\theta_{12},
\,\theta_{23}$ and $\theta_{13}$ are the mixing angles and
$\delta$ is the CP violating phase.

\section{Unitarity and CKM phenomenological parameters \label{uninew}}
 In the last few years, many important developments have taken place in the context
of phenomenology of Cabibbo-Kobayashi-Maskawa (CKM) matrix, both
from theoretical as well as experimental point of view. As
mentioned earlier, several detailed and extensive phenomenological
analyses\cite{pdg10}\cdash\cite{hfag} have allowed us to conclude
that the single CKM phase looks to be a viable solution of CP
violation not only in the case of K-decays but also in the context
of B-decays, at least to the leading order. In this context,
unitarity triangles have been widely used to check the CKM
paradigm as well as to test the predictions of the SM. To this
end, the readers are referred to Ref.~\refcite{0907.5386} wherein
the author presents extensive details of several analyses
incorporating global analysis tools to determine the CKM
parameters in the framework of the Standard Model. In particular,
employing Wolfenstein parameterization, global fits comparing the
data to theoretical predictions have been carried out and the
results obtained by the UTfit\cite{utfit} group as well as the
CKMfitter\cite{ckmfitter} group have been presented. These two
groups employ different techniques, Bayesian approach detailed in
Ref.~\refcite{bayesian} has been followed by the UTfit group and
the CKMfitter group employs the recently introduced
RangeFit\cite{rangefit} procedure to combine measurements with
very different statistical errors and extract the best
information. Combining results from these two approaches one can
reach at the conclusion that a good overall consistency between
the various measurements at 95$\%$ C.L. is observed, thus
establishing the CKM mechanism as the dominant source of CP
violation in B-meson decays.

\subsection{Implications of unitarity and sin$\,2\beta$ on $V_{ub}$
and $\delta$} It may be noted that most of the present day
analyses related to CKM phenomenology, including the one mentioned
above\cite{0907.5386}, invoke global inputs, wherein the
implications of unitarity are not obvious. This is further
complicated by the use of the Wolfenstein-Buras parameterization,
wherein unitarity of the CKM matrix has to be satisfied order by
order making it further complicated to examine its implications.
However, recently, an interesting analysis has been carried
out\cite{ouruni} wherein the implications of unitarity on CKM
phenomenology have been examined explicitly, unlike the other
approaches. This analysis employs the PDG representation of the
CKM matrix which is more convenient to use in this context as the
unitarity is built-in. The analysis\cite{ouruni} investigates the
implications of unitarity along with the well measured $V_{us}$,
$V_{cb}$, sin$\,2\beta$ and angle $\alpha$ of the unitarity
triangle on some of the lesser known elements of the CKM matrix
such as $V_{ub}, V_{cs}, V_{ts}$ and $V_{td}$. Using minimal
inputs, the possibility of constructing a `precise' CKM matrix has
also been explored. It is instructive to discuss some of the
essential details of this analysis.

In this context, it may be mentioned that out of the earlier
mentioned six unitarity triangles, the triangles $uc$, $ds$, $sb$
and $ct$ are highly skewed which means that one side of these is
very small as compared to the other two, therefore it is difficult
to study their implications\cite{mon3}\cdash\cite{botella} with
the present level of accuracy of the CKM matrix elements. Out of
the other two, it is usual to consider the $db$ triangle expressed
as
 \be V_{ud} V_{ub}^* + V_{cd} V_{cb}^* + V_{td}
V_{tb}^* =0\,.\label{db} \ee The angles of this triangle in terms
of the CKM matrix elements, mixing angles and CP violating phase
$\delta$ are expressed as
 \begin{eqnarray}\alpha\equiv{\rm arg}\left[-\frac{V_{td} V_{tb}^*}{V_{ud}
 V_{ub}^*}\right]=\tan^{-1}\left[\frac{s_{12}
 s_{23}\, {\rm sin}\, \delta}{c_{12} c_{23} s_{13}-s_{12} s_{23}\, {\rm cos}\,\delta}\right]
  ,\label{anglealpha} \end{eqnarray}
 \begin{eqnarray} \beta\equiv{\rm arg}\left[-\frac{V_{cd} V_{cb}^*}{V_{td}
 V_{tb}^*}\right]=\tan^{-1}\left[\frac{c_{12}
 s_{12} s_{13}\, {\rm sin}\,\delta}{c_{23} s_{23} (s_{12}^2-c_{12}^2 s_{13}^2)-c_{12} s_{12} s_{13}
 (c_{23}^2-s_{23}^2)\, {\rm cos}\,\delta}\right]
\label{anglebeta} \end{eqnarray}
 \begin{eqnarray} \gamma\equiv{\rm arg}\left[-\frac{V_{ud} V_{ub}^*}{V_{cd}
 V_{cb}^*}\right]=\tan^{-1}\left[\frac{s_{12}
 c_{23}\, {\rm sin}\,\delta}{c_{12} s_{23} s_{13}+s_{12} c_{23}\, {\rm cos}\,\delta}\right]
\label{anglegamma}. \end{eqnarray} To obtain information about the
CP violating phase $\delta$ from the experimentally well
determined angle $\beta$ one can express Eq.~(\ref{anglebeta}) as
\be
{\rm tan}\,\frac{\delta}{2} = \frac{A - \sqrt{A^2-(B^2-A^2
C^2){\rm tan}^2 \beta}}{(B+AC){\rm tan}\,\beta}, \label{tand} \ee
where $A=c_{12} s_{12} s_{13}$, $B=c_{23} s_{23}(s_{12}^2-c_{12}^2
s_{13}^2)$ and $C=c_{23}^2-s_{23}^2$. Using $s_{12}^2 \gg c_{12}^2
s_{13}^2$ and $s_{23}^2 \ll
 c_{23}^2$, the above relation can be re-expressed as
\be
\delta~=-\beta+{\rm
sin}^{-1}\left(\frac{s_{12}s_{23}}{c_{12}s_{13}}{\rm
sin}\beta\right), \label{delb}\ee which can also be written as
\be
\frac{{\rm sin}\,(\delta+\beta)}{{\rm
sin}\,\beta}=\frac{s_{12}s_{23}}{c_{12}s_{13}}. \label{dpb}\ee
From Eq.~(\ref{anglegamma}), one can easily show that $\gamma =
\delta$ with an error of around 2$\%$, therefore, using the
closure property of the angles of the triangle,
$\alpha+\beta+\gamma=\pi$, the above
 equation can be written as
\be
s_{13}=\frac{s_{12}s_{23}\,{\rm sin}\,\beta}{c_{12}\,{\rm
sin}\,\alpha}, \label{s13ab} \ee which can also be derived from
Eq.~(\ref{anglealpha}) by using the closure property of the
triangle. Eq.~(\ref{dpb}) can be used to provide a lower bound on
$s_{13}$, e.g.,
\be
s_{13}~\geq~\frac{s_{12}s_{23}}{c_{12}}\,{\rm sin}\,\beta.
\label{lbs13} \ee

\begin{figure}[hbt]
 \vspace{0.10in} \centerline{\psfig{figure=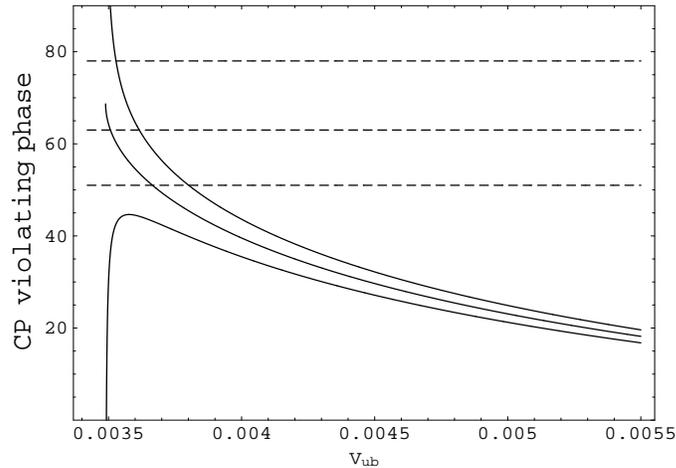,width=3.5in,height=2.5in}}
 \vspace{0.08in} \caption{Plot showing variation of $V_{ub}$ versus CP violating
phase $\delta$, obtained by using Eq.~(\ref{tand}). The central
solid line corresponds to mean value of input parameters, whereas
the other 2 lines correspond to 1$\sigma$ variations.}
 \label{vubdel}
  \end{figure}

On examining unitarity based Eq.~(\ref{tand}), it can be noted
that $\delta$ is dependent on $V_{us}$, $V_{cb}$, angle $\beta$ as
well as it involves $V_{ub}$. Using this equation, the authors
have plotted the CP violating phase $\delta$ versus $V_{ub}$,
presented in Fig.~\ref{vubdel}. Also included in the figure is the
then\cite{pdg06} experimentally measured $\delta=(63.0 + 15.0
-12.0)^{\rm o}$ shown by horizontal dashed lines, inclusive of
results of various global analyses. The solid central line depicts
$\delta$ obtained by using the mean values of $V_{us}$, $V_{cb}$
and  sin$\,2\beta$ whereas the outer lines correspond to the
1$\sigma$ ranges of these inputs.

From the figure one finds that for values of $V_{ub} > 0.00355$,
the central value of $\delta$ shows a smooth decline as well as
the range of $\delta$ gets narrower and narrower with increasing
$V_{ub}$, however for $V_{ub} < 0.00355$ it seems that there is a
sharp broadening of the $\delta$ range, with no restriction on
$\delta$ when $V_{ub} < 0.0035$. It may be noted that as per the
data given by PDG 2006\cite{pdg06}, from the graph one finds that
the 1$\sigma$ range of the inclusive value of $V_{ub}$ restricts
$\delta$ to $23 ^{\rm o}- 39 ^{\rm o}$, whereas the mean value of
the exclusive value does not constrain $\delta$, however the upper
limit of the 1$\sigma$ range of the exclusive value provides only
a lower bound $\delta > 38 ^{\rm o}$. Therefore, it was emphasized
that the precisely known sin$\,2\beta$, for the inclusive value of
$V_{ub}$ implies a narrow range for $\delta$, whereas for the
exclusive value of $V_{ub}$ it implies only a lower bound on
$\delta$. It may be noted that the above mentioned conclusions
remain largely valid even when the recent data pertaining to CKM
matrix elements is included.

The conclusion about $V_{ub}$ can been sharpened further, e.g.,
using Eq.~(\ref{lbs13}) one can easily obtain the following
rigorous lower bound on $V_{ub}$,
\be
V_{ub}\geq 0.0035. \label{alb} \ee It may be noted that this bound
is independent of the value of $\delta$ as well as contamination
of NP in the measurement of $\delta$. Predictions regarding
$V_{ub}$ have been refined further by incorporating angle $\alpha$
of the unitarity triangle, measured from $B \rightarrow \pi \pi$
and $B \rightarrow \rho \rho$ decays. Using its then known
value\cite{gronau}, from Eq.~(\ref{s13ab}) one obtains
\be
 V_{ub}= 0.0035\pm 0.0002. \label{s13r} \ee
Interestingly, this precise value of $V_{ub}$ is a consequence of
unitarity and the precisely measured elements $V_{us}$, $V_{cb}$
and angles $\beta$ and $\alpha$ as well as is in full agreement
with the exclusive $V_{ub}$. It may also be noted that this value
is quite insensitive to a change in the value of angle $\alpha$,
e.g., even if the mean value of $\alpha$ changes by more than
20$\%$, still $V_{ub}$ would register a variation of only a few
percent.

The above discussion also underlines the fact that precisely
measured sin$\,2\beta$ does not lead to any well defined
conclusion regarding $\delta$ because of the persistent difference
between exclusive and inclusive values of $V_{ub}$. Therefore, the
unitarity based $\delta$ can be obtained through the angles
$\alpha$ and $\beta$ by using the closure property of the angles
of the unitarity triangle, e.g.,
 \be \delta=67.8^{\rm o}\pm 7.3^{\rm o}. \label{delta1}\ee
This unitarity based value of $\delta$ is compatible with its
directly measured value in $B^\pm \rightarrow D K^\pm$
decays\cite{glw} as well as with the one obtained\cite{buras1}
from the $B \rightarrow \pi \pi$ and $B \rightarrow \pi K$ decays.
It may also be mentioned that this value is compatible with the
$\delta$ bound given by exclusive $V_{ub}$, as obtained from
Fig.~\ref{vubdel}, however does not agree with the $\delta$ range
obtained for inclusive $V_{ub}$.

After having found $V_{ub}$ and $\delta$ from unitarity, the
authors have constructed the entire CKM matrix which is obtained
at 1$\sigma$ C.L. as follows
 \be \left( \ba{ccc}
  0.9738- 0.9745 &   0.2244 - 0.2272 &  0.0033 - 0.0036 \\
 0.2243 - 0.2270  &   0.9730 - 0.9736    &  0.0409 - 0.0423\\
0.0082 - 0.0091  &  0.0401 - 0.0415 &  0.9990 - 0.9991 \ea
\right). \label{1sm} \ee It may be mentioned that this matrix is
free from contamination by NP to the extent that the measured
values of angles $\alpha$ and $\beta$ are free from NP effects.
The matrix reveals that the ranges of CKM elements obtained here
are quite compatible with those obtained by global
analyses\cite{pdg10}\cdash\cite{hfag}. This perhaps indicates that
unitarity plays a key role even in the case of global analyses.

Making use of the above mentioned CKM matrix, the authors have
also calculated the ratio $\frac{V_{ts}}{V_{td}}$ which comes out
to be $4.69 \pm 0.23$, having an excellent overlap with $4.7 \pm
0.4$, found from precision measurements of $\Delta
M_{B_s}$\cite{klein}. The measured value of the ratio
$\frac{V_{ts}}{V_{td}}$ can be considered as an over constraining
check on the above unitarity based predictions.

The above discussion leads to the conclusion that a further
precision in the measurement of sin$\,2\beta$, needless to say,
would have far reaching implications for CKM phenomenology,
particularly for CP violating phase $\delta$ and $V_{ub}$. In this
context, measurements of several other CKM parameters are also
reaching at the precision level, making it essential to examine
their implications for the CKM paradigm. In the following, we
briefly discuss some of the analyses which have explored these
issues.

\subsection{Implications of precision measurements for CKM
paradigm \label{precision}} As already emphasized, precise
measurements of sin$\,2\beta$ as well as of the
Cabibbo-Kobayashi-Maskawa (CKM) matrix\cite{cabibbo,kobayashi}
elements $V_{us}, V_{cb}, V_{ud}$ and several other parameters
have been carried out. Similarly, a good deal of data has been
collected for a large number of flavor changing neutral current
processes involving $b \rightarrow d$ and $b \rightarrow s$
transitions and several CP violating asymmetries have also been
studied in detail\cite{decays}. Based on these efforts, one may
now conclude that the larger picture of CKM paradigm appears to be
well confirmed.

In line with the precision measurements regarding CKM parameters,
several recent developments have also taken place on the
theoretical front concerning the calculations of hadronic factors.
In particular, improvements have taken place in the lattice QCD
calculations of hadronic factors in the case of $K - \bar{K}$ and
the $B_d - \bar{B_d}$ mixings along with contribution of the long
distance effects in the $K - \bar{K}$ system. To this end, the new
lattice QCD calculations of the hadronic matrix element $B_k$,
relevant for the determination of $\epsilon_K$, constrain it to an
accuracy of $(4 - 6)$\%\cite{bk}. Also, a correction of $(5 -
10)$\%\cite{ke} in the determination of $\epsilon_K$ has now been
advocated by incorporating the long distance contribution related
to the ratio $K \rightarrow \pi \pi$ decay amplitudes in the
$\Delta I =1/2$ channel leading to the introduction of an overall
multiplicative factor $\kappa_{\epsilon}\sim 0.92$, omitted from
the earlier CKM phenomenological analyses.

Triggered by these theoretical improvements, recently Buras and
Guadagnoli\cite{buras} as well as Lunghi and Soni\cite{soni} have
investigated the implications of these for the CKM phenomenology.
In particular, Ref.~\refcite{buras} points out that the CP
violation in $B_d - \bar{B_d}$ mixing evaluated by considering the
measured ratio $\Delta m_d / \Delta m_s$, the recent value of the
non-perturbative parameter $B_K$ and the additional effective
suppression factor $\kappa_{\epsilon}$ may be insufficient to
describe the measured value of $\epsilon_K$ within the Standard
Model (SM). In other words, the above mentioned theoretical
improvements tend to lower the SM prediction for $\epsilon_K$ if
the amount of CP violation in the $B_d$ system, quantified by
sin$\,2\beta$ from $B^o_d \rightarrow \psi K_s$ decay, is used as
an input. From this, they supposedly obtain a hint for a possible
inconsistency between the size of CP violation in the $K -
\bar{K}$ and/or $B_d - \bar{B_d}$ systems. A closer look at their
analysis reveals that in case one considers the absence of an
additional CP violating phase in the ${B_d}$ system, the value of
$\epsilon_K$ then comes out to be almost 20$\%$ below its measured
value, hinting at New Physics (NP) in the $K - \bar{K}$ mixing. On
the other hand, if absence of an additional CP violating phase in
the $K - \bar{K}$ system is considered, then this implies
sin$\,2\beta$ coming out be 10-20$\%$ larger.

The analysis by Ref.~\refcite{soni} also explores the possibility
of the presence of NP in the  $K - \bar{K}$ and $B_d - \bar{B_d}$
systems as well as in the $b \rightarrow s$ penguin transitions.
In particular, they attempt to predict the value of sin$\,2\beta$
keeping in mind the role played by the parameter $V_{ub}$, e.g.,
without the inclusion of $V_{ub}$ the value of sin$\,2\beta$ comes
out to be $0.87 \pm 0.09$, whereas on including $V_{ub}$ the value
of sin$\,2\beta$ becomes $0.75 \pm 0.04$. These predicted values
of sin$\,2\beta$ point towards possible inconsistencies with the
directly measured value through the gold-plated $B^o_d \rightarrow
\psi K_s$ decay and also by the penguin-dominated modes leading
the authors to conclude that one has to give a re-look at the CKM
paradigm. To summarize the findings of the two above mentioned
analyses\cite{buras,soni}, one can conclude that both the analyses
supposedly incorporate the presence of NP to reconcile the value
of the parameter sin$\,2\beta$ with the $K - \bar{K}$ and the $B_d
- \bar{B_d}$ mixings.

However, in a recent analysis by Ahuja {\it et
al.}~\cite{ourepsilon}, attempts have been made to look at the
above mentioned conclusions from a different perspective. In this
context, one finds that the CKM matrix elements measured at tree
level and those which can be constrained by unitarity can be
considered to be essentially free from NP effects. Similarly, the
ratio $\Delta m _d / \Delta m _s$ is also considered to be largely
free from the effects of NP\cite{{pdg10}}. Also, regarding the
angle $\alpha$ of the unitarity triangle, for the case of its
prediction through the $B \rightarrow \rho \rho$ decay the effects
of the penguin diagrams are considered to be relatively
small\cite{pdg10,grossman}\cdash\cite{nir}. The implications of
these NP free CKM parameters have been examined\cite{ourepsilon}
on the parameters sin$\,2\beta$, $V_{td}$, $\epsilon_K$, etc..
Employing the PDG representation of the CKM mixing matrix and
incorporating the constraints of unitarity, for $c_{13}\cong
0.9999$ implying $V_{us} \cong s_{12}$ and $V_{cb} \cong s_{23}$
up to third place of decimal, the authors consider $V_{us} \cong
V_{cd}$, $V_{ts} \cong V_{cb}$ and $V_{tb} \cong 1$.

It may be noted that the CP asymmetry parameter sin$\,2\beta$ is
generally determined from the asymmetry measurement of the $B^o_d
\rightarrow \psi K_s$ decay. However, as already mentioned, the
analyses by Refs.~\refcite{buras} and \refcite{soni} point towards
the possibility of NP due to a new phase in the $B_o - \bar{B_o}$
mixing. It may also be noted here that information about $\beta$
can also be obtained from a measurement of element $V_{td}$ from
$B_o - \bar{B_o}$ mixing, however this cannot be considered free
from NP as its evaluation involves loop processes. Also, in near
future, the possibility of measurement of the third row elements
of the CKM matrix through the tree level decays is not very
promising. Therefore, the analysis by Ref.~\refcite{ourepsilon}
attempts to determine sin$\,2\beta$ from unitarity and quantities
free from NP effects, essential details of this have been
summarized here.

Using the $db$ unitarity triangle, the authors make use of the law
of sines to avoid the involvement of $V_{td}$ in the evaluation of
angle $\beta$ through Eq.~(\ref{anglebeta}), re-expressed as
\be
\beta=   {\rm sin}^{-1} \left (\frac{V_{ud} V_{ub}^* }{V_{cd}
V_{cb}^*~ {\rm sin} \alpha} \right ). \label{beta1} \ee Using the
PDG 2010 values\cite{pdg10} of the angle $\alpha$ and the CKM
matrix elements, the authors obtain
\be
\beta = (23.94 \pm 2.95)^{\rm o}, \label{betavalue} \ee implying
\be
{\rm sin} 2 \beta = 0.742 \pm 0.103. \label{s2b} \ee This value of
sin$\,2\beta$ seems to be free from contamination of NP and is
inclusive of its recent experimental range $0.673\pm
0.023$\cite{pdg10}. Further, making use of the closure property of
the angles of the unitarity triangle and using the angle $\alpha$
and the above mentioned value of $\beta$, the angle $\delta$ has
been obtained as
 \be \delta=67.1^{\rm o}\pm 5.3^{\rm o}. \label{delta0}\ee
Again, one finds that this unitarity based value of $\delta$ is
unaffected by NP in the $B - \bar{B}$ mixing as well as is
compatible with the directly measured value,
$(73^{+22}_{-25})^{\rm o}$, in $B^\pm \rightarrow D K^\pm$
decays\cite{pdg10} and by many other global
analyses\cite{utfit}\cdash\cite{hfag}.

The CKM matrix element $V_{td}$, calculated usually from  $B_o -
\bar{B_o}$ mixing which is a loop dominated process and therefore
not considered to be NP free has also been obtained free from NP
effects. Using the PDG values of angle $\alpha$, the CKM matrix
elements, the above determined value of angle $\delta \sim \gamma$
and considering $V_{tb} \cong 1$, one gets
 \be
V_{td}= (8.69 \pm 0.61) \times 10^{-3}. \label{vtdfromgamma} \ee
It may be noted that this prediction of $V_{td}$ is based on
unitarity and is through CKM parameters essentially free from NP
effects.

After having checked that the parameter sin$\,2\beta$ and CKM
matrix element $V_{td}$ do not require any additional NP inputs,
the authors\cite{ourepsilon} have examined the calculation of
$\epsilon_K$, defining CP violation in the $K - \bar{K}$ system.
To this end, the recent expression of the CP violating parameter
$\epsilon_K$ has been used which modified by incorporating the
factor $\kappa_{\epsilon}$ now becomes \be |\epsilon_K| =
\kappa_{\epsilon} \frac{G_F^2 F_K^2 m_K m_W^2}
        {6 \sqrt{2} \pi^2 \Delta m_k}B_K{\rm Im}\lambda_t
  [{\rm Re} \lambda_c(\eta_1 S_0(x_c)-\eta_3 S_0(x_c,x_t))
 - {\rm Re} \lambda_t \eta_2 S_0(x_t)], \label{eps} \ee
where $\eta_1$, $\eta_2$, $\eta_3$ are the perturbative QCD
corrections, $S_0(x_i)$ are Inami-Lim functions,
$x_i=m_i^2/M_W^2$, and $\lambda_i=V_{id}V^*_{is}$, $i=c,t$. In
terms of the mixing angles and the phase $\delta$, the quantities
Im$\lambda_t$, Re$\lambda_t$ and Re$\lambda_c$ can be expressed as
\beqn {\rm Im}\lambda_t & = &
 s_{23}s_{13}c_{23}{\rm sin}\,\delta,  \\
{\rm Re}\lambda_t & = &
 s_{23}s_{13}c_{23}(c_{12}^2-s_{12}^2){\rm cos}\,\delta -
s_{12}c_{12}(s_{23}^2-c_{23}^2s_{13}^2),\\
{\rm Re}\lambda_c & = &
 s_{23}s_{13}c_{23}(s_{12}^2-c_{12}^2){\rm cos}\,\delta -
s_{12}c_{12}(c_{23}^2-s_{23}^2s_{13}^2). \eeqn

 \begin{figure}[tbp]
\centerline{\epsfysize=2.6in\epsffile{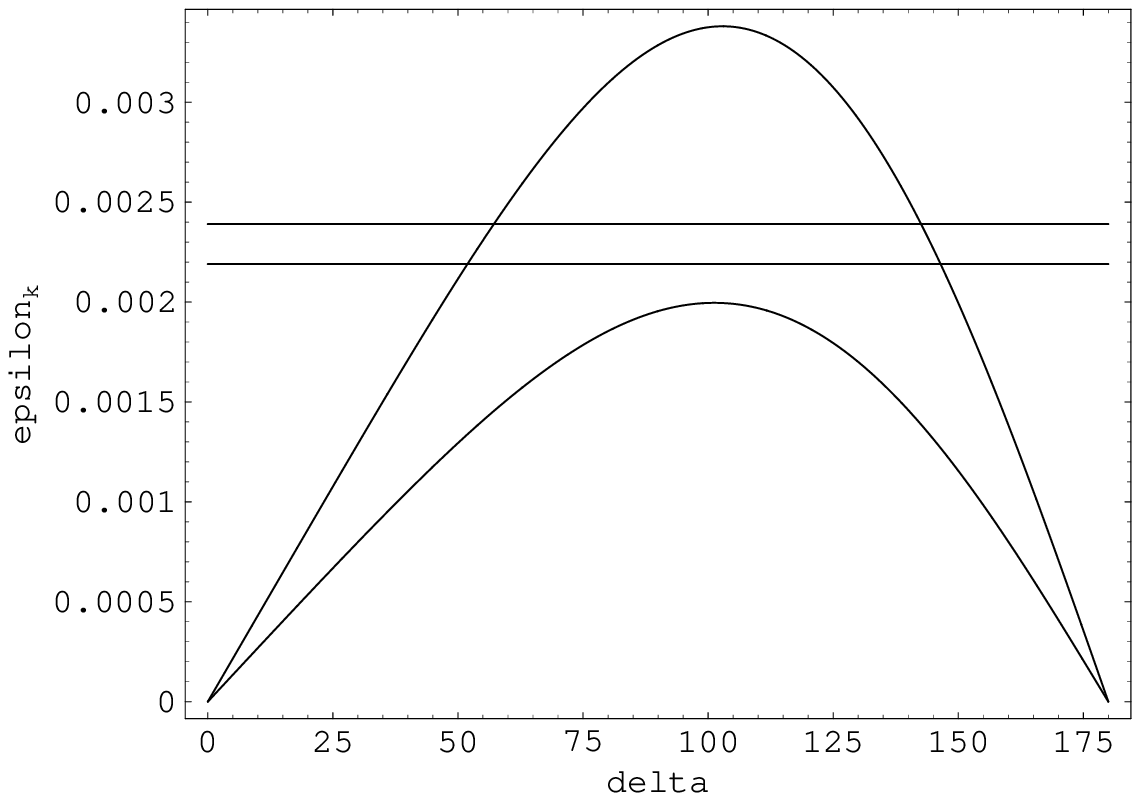}}
\vspace{0.08in}
   \caption{CP violating phase $\delta$ versus $\epsilon_{k}$ using exclusive value of
   $V_{ub}$}
  \label{forexclvub}
  \end{figure}

 \begin{figure}[tbp]
\centerline{\epsfysize=2.6in\epsffile{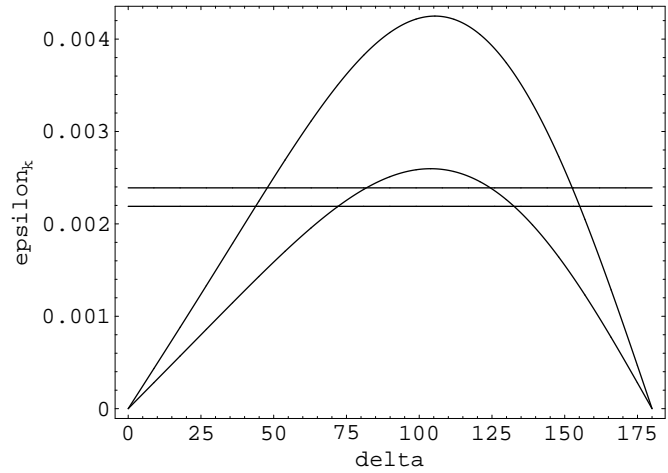}}
\vspace{0.08in}
   \caption{CP violating phase $\delta$ versus $\epsilon_{k}$ using inclusive value of
   $V_{ub}$}
  \label{forinclvub}
  \end{figure}
The implications of $\epsilon_K$ on the CP violating phase
$\delta$ for both the exclusive and inclusive values of  CKM
matrix element $V_{ub}$ have been investigated. In
Figs.~\ref{forexclvub} and \ref{forinclvub}, the plots of phase
$\delta$ versus the parameter $\epsilon_K$ have been presented.
From the graphs, one finds different constraints on phase $\delta$
for the exclusive and inclusive values of $V_{ub}$, e.g., \beqn
{\rm for~exclusive~value~of~}V_{ub}, ~~~~~\delta=(52-147)^{\rm
o},~~~~~~~~~~~~~~~~~~~~~~~~~~\\ \nonumber {\rm
for~inclusive~value~of~}V_{ub},~~~~~\delta=(44-72)^{\rm o},
~~~~(132-155)^{\rm o}.~~~~~~~ \eeqn  It is interesting to note
that the above mentioned ranges of the CP violating phase $\delta$
come out to be compatible with the experimentally determined range
of angle $\delta$. One can therefore conclude that the recent
improvements in the lattice calculations of hadronic parameters,
as well as precision measurements of CKM parameters do not seem to
provide any significant clues regarding the possibility of
existence of NP, as far as compatibility of the CP violating phase
$\delta$ is concerned. If at all effects of NP are present, they
are at only a few percent level.

\section{Unitarity and unitarity triangles in the leptonic sector \label{unilep}}
At present, one of the key issues in the context of neutrino
oscillation phenomenology is to explore the existence of CP
violation in the leptonic sector. After having explored the
implications of Jarlskog's rephasing invariant parameter $J$,
unitarity and unitarity triangles, which facilitate the
understanding of several features of CKM phenomenology, it becomes
interesting to discuss similar attempts in the case of
phenomenology of the Pontecorvo-Maki-Nakagawa-Sakata  (PMNS)
matrix\cite{pmns1}\cdash\cite{pmns4}. To this end, several
attempts\cite{farsm}\cdash\cite{ourlepuni} have been made to
explore the existence of the corresponding unitarity triangle in
the leptonic sector. For example, Farzan and Smirnov\cite{farsm}
have explored the construction of leptonic unitarity triangle for
finding possible clues to the existence of CP violation in the
leptonic sector. In particular, considering the `$e$.$\mu$'
triangle, for $U_{e3}$ ($\equiv s_{13}$) values in the range
$0.09-0.22$ they have examined the detailed implications of
different values of Dirac-like CP violating phase in the leptonic
sector $\delta_l$ on the possible accuracy required in the
measurement of various oscillation probabilities. Further,
recently Bjorken {\it et al.}\cite{bjorken}, by considering a
modified tri-bimaximal scenario, have not only presented a very
useful parameterization of the PMNS matrix but have also proposed
a unitarity triangle, referred to as `$\nu_2.\nu_3$', which could
be leptonic analogue of the much talked about $db$ triangle in the
quark sector. Furthermore, by considering different values of
$U_{e3}$, suggested by various theoretical models, in the
parameterization of the PMNS matrix given by Bjorken {\it et
al.}\cite{bjorken}, Ahuja and Gupta\cite{ourlepuni} have explored
in detail the probability of finding a non zero value of $J_l$,
the Jarlskog's rephasing invariant parameter in the leptonic
sector and the related Dirac-like CP violating phase $\delta_l$.

It may be noted that the analyses, mentioned above, were carried
out before the recent T2K\cite{t2k}, MINOS\cite{minos}, DAYA
BAY\cite{dayabay} and RENO\cite{reno} observations regarding the
mixing angle $s_{13}$, suggesting its not so `small' value. These
observations have given a big impetus to the sharpening of the
implications of the neutrino oscillations, in particular the non
zero value of angle $s_{13}$ implies the possibility of CP
violation in the leptonic sector. Recently, keeping in view these
latest observations, Ahuja\cite{lepcp} has explored the
possibility of existence of CP violation in the leptonic sector
through the `$\nu_1.\nu_3$' leptonic unitarity triangle.

It may be mentioned that while the present manuscript was being
prepared, we came across a very recent review article by Branco
{\it et al.}\cite{cpbranco}, wherein several topics on CP
violation in the leptonic sector have been reviewed. However, in
the present case, we have made an attempt to emphasize those
points which have not been discussed in detail in their
review\cite{cpbranco}. In the sequel, we would like to present
some of the details of the works by Ahuja {\it et al.},
Refs.~\refcite{ourlepuni} and \refcite{lepcp}. In the absence of
information regarding mixing angle $s_{13}$,
Ref.~\refcite{ourlepuni} have made use of the parameterization of
the PMNS matrix in the modified tribimaximal scenario and carried
out the analysis, whereas Ref.~\refcite{lepcp} incorporates the
latest information regarding $s_{13}$ to explore the likelihood of
CP violation in the leptonic sector.

Before presenting the details of these analyses, we first begin
with the neutrino mixing phenomenon, often expressed in terms of a
$3 \times 3$ neutrino mixing Pontecorvo-Maki-Nakagawa-Sakata
(PMNS) matrix\cite{pmns1}\cdash\cite{pmns4} given by \be \left(
\ba{c} \nu_e \\ \nu_{\mu} \\ \nu_{\tau} \ea \right)
  = \left( \ba{ccc} U_{e1} & U_{e2} & U_{e3} \\ U_{\mu 1} & U_{\mu 2} &
  U_{\mu 3} \\ U_{\tau 1} & U_{\tau 2} & U_{\tau 3} \ea \right)
 \left( \ba {c} \nu_1\\ \nu_2 \\ \nu_3 \ea \right),  \label{nm4}  \ee
where $ \nu_{e}$, $ \nu_{\mu}$, $ \nu_{\tau}$ are the flavor
eigenstates and $ \nu_1$, $ \nu_2$, $ \nu_3$ are the mass
eigenstates. Following Particle Data Group (PDG)\cite{pdg10}
representation, wherein the unitarity is built-in, involving three
angles $\theta_{12}$, $\theta_{23}$, $\theta_{13}$ and the
Dirac-like CP violating phase $\delta_l$ as well as the two
Majorana phases $\alpha_1$, $\alpha_2$, the PMNS matrix $U$ can be
written as \beqn U&=&{\left( \ba{ccl} c_{12} c_{13} & s_{12}
c_{13} & s_{13}e^{-i \delta_l} \\ - s_{12} c_{23} - c_{12} s_{23}
s_{13} e^{i \delta_l} & c_{12} c_{23} - s_{12} s_{23} s_{13} e^{i
\delta_l} & s_{23} c_{13}
\\ s_{12} s_{23} - c_{12} c_{23} s_{13} e^{i \delta_l} & - c_{12}
s_{23} - s_{12} c_{23} s_{13} e^{i \delta_l} & c_{23} c_{13} \ea
\right)} \left( \ba{ccc} e^{i \alpha_1/2} & 0 & 0 \\ 0 &e^{i
\alpha_2/2} & 0 \\ 0 & 0  & 1 \ea \right), \label{nmm4} \nonumber
\\&& \eeqn with $c_{ij}= {\rm cos}~ \theta_{ij}$ and $s_{ij}= {\rm
sin}~ \theta_{ij}$ for $i,j=1,2,3$. The Majorana phases $\alpha_1$
and $\alpha_2$ do not play any role in neutrino oscillations and
henceforth would be dropped from the discussion. Further, in this
representation, $|U_{e3}| \equiv s_{13}$, therefore, while
discussing the magnitude of the PMNS matrix elements, $U_{e3}$ and
$s_{13}$ would be used interchangeably.

\begin{figure}[t]
\begin{picture}(400,315)(20,0)
\put(70,300){\line(1,0){120}} \put(120,310){\makebox(0,0){$U_{e
1}U^*_{\mu 1}$}} \put(70,300){\line(1,-4){9.2}}
\put(50,283){\makebox(0,0){$U_{e 3}U^*_{\mu 3}$}}
\put(190,300){\line(-3,-1){111}} \put(157,270){\makebox(0,0){$U_{e
2}U^*_{\mu 2}$}} \put(120,248){\makebox(0,0){$e.\mu$}}
\put(290,300){\line(1,0){100}} \put(340,310){\makebox(0,0){$U_{e
1}U^*_{e 2}$}} \put(290,300){\line(3,-2){50}}
\put(296,275){\makebox(0,0){$U_{\mu 1}U^*_{\mu 2}$}}
\put(390,300){\line(-3,-2){50}}
\put(389,275){\makebox(0,0){$U_{\tau 1}U^*_{\tau 2}$}}
\put(340,245){\makebox(0,0){$\nu 1.\nu 2$}}
\end{picture}

\begin{picture}(400,100)(20,0)
\put(70,300){\line(1,0){85}} \put(115,310){\makebox(0,0){$U_{e 1}
U^*_{\tau 1}$}} \put(70,300){\line(2,-1){73}}
\put(93,272){\makebox(0,0){$U_{e 2} U^*_{\tau 2}$}}
\put(155,300){\line(-1,-3){12}} \put(175,280){\makebox(0,0){$U_{e
3} U^*_{\tau 3}$}} \put(120,245){\makebox(0,0){$e.\tau$}}
\put(295,300){\line(1,0){90}} \put(337,310){\makebox(0,0){$U_{\tau
1} U^*_{\tau 3}$}} \put(295,300){\line(1,-3){13}}
\put(278,278){\makebox(0,0){$U_{e 1}U^*_{e 3}$}}
\put(385,300){\line(-2,-1){77}}
\put(365,268){\makebox(0,0){$U_{\mu 1}U^*_{\mu 3}$}}
\put(340,245){\makebox(0,0){$\nu 1.\nu 3$}}
\end{picture}

\begin{picture}(400,100)(20,0)
\put(70,300){\line(1,0){100}} \put(120,310){\makebox(0,0){${
U}_{\mu 3}{ U}^*_{\tau 3}$}} \put(70,300){\line(3,-2){50}}
\put(78,275){\makebox(0,0){${ U}_{\mu 1}{U}^*_{\tau 1}$}}
\put(170,300){\line(-3,-2){50}} \put(170,275){\makebox(0,0){${
U}_{\mu 2}{ U}^*_{\tau 2}$}}
\put(120,245){\makebox(0,0){$\mu$.$\tau$}}
\put(275,300){\line(1,0){112}} \put(331,310){\makebox(0,0){$U_{\mu
2}U^*_{\mu 3}$}} \put(275,300){\line(3,-1){103}}
\put(318,272){\makebox(0,0){$U_{\tau 2}U^*_{\tau 3}$}}
\put(387,300){\line(-1,-4){8.6}} \put(407,282){\makebox(0,0){$U_{e
2}U^*_{e 3}$}} \put(340,245){\makebox(0,0){$\nu 2$.$\nu 3$}}
\end{picture}
\vspace{-8.5cm} \caption{The six unitarity triangles in the
leptonic sector.} \label{leputs}
\end{figure}
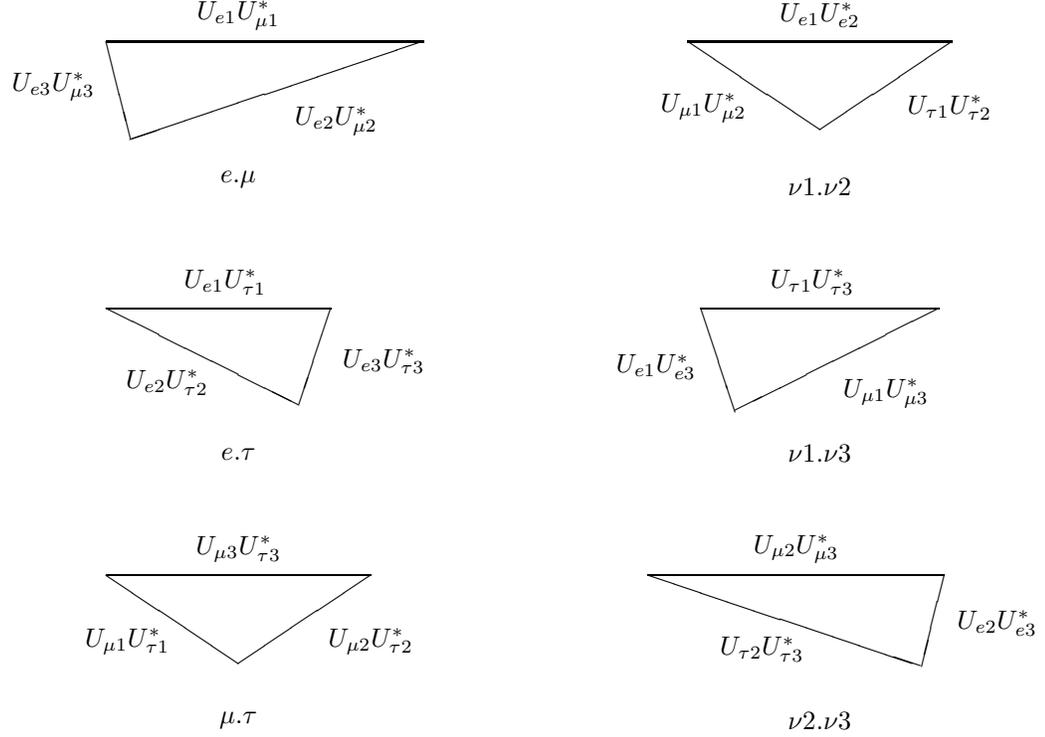

Unitarity of the PMNS matrix implies nine relations, three in
terms of normalization conditions, the other six can be defined as
 \be \sum_{i=1,2,3} U_{\alpha i} {U^*_{\beta i}} =\delta_{\alpha\beta}
 ~~~~~~~(\alpha \neq \beta) , \label{ut1} \ee
 \be \sum_{\alpha=e,\mu,\tau} U_{\alpha i}
{U^*_{\alpha j}} =\delta_{ij}  ~~~~~~~(i \neq j) , \label{ut2-4}
\ee where Latin indices run over the mass eigenstates $(1,2,3)$
and Greek ones run over the flavor eigenstates $(e,\mu,\tau)$.
These six non-diagonal relations can be expressed through six
independent unitarity triangles in the complex plane, shown in
Fig.~\ref{leputs} and can also be expressed as \beqn
  e.\mu \qquad  U_{e1}U_{\mu 1}^{*} + U_{e2}U_{\mu 2}^{*} + U_{e3}U_{\mu 3}^{*}
& = & 0 \,, \label{e-mu} \\e.\tau \qquad  U_{e1}U_{\tau 1}^{*} +
U_{e2}U_{\tau 2}^{*} + U_{e3}U_{\tau 3}^{*} & = & 0 \,,
\label{e-tau}
\\
 \mu.\tau \qquad  U_{\mu 1}U_{\tau 1}^{*} +
U_{\mu 2}U_{\tau 2}^{*} + U_{\mu 3}U_{\tau 3}^{*} & = & 0 \,,
\label{mu-tau}  \\ \nu 1.\nu 2 \qquad  U_{e 1}U_{e 2}^{*} + U_{\mu
1}U_{\mu 2}^{*} + U_{\tau 1}U_{\tau 2}^{*} & = & 0 \,, \label{nu1-
nu2 }
\\
 \nu 1.\nu 3 \qquad  U_{e 1}U_{e 3}^{*} + U_{\mu
1}U_{\mu 3}^{*} + U_{\tau 1}U_{\tau 3}^{*} & = & 0 \,, \label{nu1-
nu3 } \\
  \nu 2.\nu 3 \qquad  U_{e 2}U_{e 3}^{*} + U_{\mu
2}U_{\mu 3}^{*} + U_{\tau 2}U_{\tau 3}^{*} & = & 0 \,, \label{nu2-
nu3} \eeqn where the letters `$e.\mu$' etc. represent the
corresponding unitarity triangle.

Similar to the case of quarks, all the six unitarity triangles
shown in Fig.~\ref{leputs} are equal in area and can be used to
evaluate the CP violating parameters in the leptonic sector namely
the Jarlskog's rephasing invariant parameter in the leptonic
sector $J_l$ and the related Dirac-like CP violating phase
$\delta_l$. The parameter $J_l$ is related to area of any of the
unitarity triangle as \be |J_l| = 2 \times {\rm Area~ of~any~ of~
the~ unitarity~ triangle}
 \label{area4} \ee and yields
information about the phase $\delta_l$ through the relation
\be
J_l=s_{12}s_{23}s_{13}c_{12}c_{23}c_{13}^2 \,{\rm sin}\,\delta_l.
\label{jd}\ee Similar to Eq.~(\ref{j3}) for the case of quarks,
$J_l$ can also be defined through the relation
\begin{equation}
{\rm Im} \left (U_{\alpha i} U_{\beta j} U^*_{\alpha j} U^*_{\beta
i}\right ) \; =\; J_l \sum_{\gamma}
\varepsilon_{\alpha\beta\gamma} \sum_{k} \varepsilon_{ijk}
  \; ,
\end{equation}
where the Latin subscripts ($i,j$,$k$) and the Greek subscripts
($\alpha, \beta$, $\gamma$)run respectively over $(1,2,3)$ and
$(e,\mu,\tau)$. It may be noted that $J_l$ can also be expressed
in terms of the moduli of four independent matrix elements of $U$
as follows
\begin{eqnarray}
J_l^2 & = & |U_{\alpha i}|^2 |U_{\beta j}|^2 |U_{\alpha j}|^2
|U_{\beta i} |^2 - \frac{1}{4} \left ( 1 + |U_{\alpha i}|^2
|U_{\beta j}|^2 + |U_{\alpha j}|^2 |U_{\beta i}|^2 \right .
\nonumber \\ & & \left . - |U_{\alpha i}|^2 - |U_{\beta j}|^2 -
|U_{\alpha j}|^2 - |U_{\beta i}|^2 \right )^2 \; ,
\end{eqnarray}
in which $\alpha \neq \beta$ running over $(e, \mu, \tau)$ and $i
\neq j$ running over $(1, 2, 3)$. Therefore, the information about
leptonic CP violation can in principle be extracted from the
measured moduli of the flavor mixing matrix elements.

\subsection{Leptonic unitarity triangle in the modified
tri-bimaximal scenario \label{secuttri}} As mentioned earlier, a
parameterization of the PMNS matrix in the modified tri-bimaximal
scenario has been formulated by Bjorken {\it et
al.}\cite{bjorken}. Making use of this modified tri-bimaximal
scenario, Ref.~\refcite{ourlepuni} have made an attempt to explore
the possibility of the construction of the leptonic unitarity
triangle as well as the existence of CP violation in the leptonic
sector. To facilitate the discussion as well as understanding of
this scenario, we first reproduce here some of the essential
details of Ref.~\refcite{bjorken}.

The tri-bimaximal scenario\cite{hps1}\cdash \cite{hps7} and its
further generalization can be understood by beginning with the PDG
representation of the PMNS matrix. The representation, in terms of
the three rotations and Dirac-like CP violating phase $\delta_l$,
can be expressed as
\begin{equation}
U = \left(
\begin{array}{ccc}
1 & 0 & 0 \\ 0 & {\rm cos} \theta_{23} & {\rm sin} \theta_{23} \\
0 & - {\rm sin} \theta_{23} & {\rm cos} \theta_{23} \\
\end{array}
\right) \left(
\begin{array}{ccc}
{\rm cos} \theta_{13}  & 0 & {\rm sin} \theta_{13} e^{i \delta_l}
\\ 0 & 1  & 0 \\ - {\rm sin} \theta_{13} e^{- 1 \delta_l}  & 0  &
{\rm cos} \theta_{13} \\
\end{array}
\right) \left(
\begin{array}{ccc}
{\rm cos} \theta_{12}  & {\rm sin} \theta_{12}  & 0 \\ - {\rm sin}
\theta_{12}  & {\rm cos} \theta_{12}  & 0 \\ 0 & 0  & 1
\end{array}
\right).
\end{equation}

The atmospheric neutrino experiments\cite{atmexp} along with the
data from K2K\cite{k2k} and CHOOZ\cite{chooz} experiments give the
following values of $U_{e3}$ and $U_{\mu 3}$ at 1$\sigma$ C.L.
\begin{equation}
|U_{e3}|^2\,\lesssim\,0.013, \qquad~~~~~ |U_{\mu 3}|^2=0.50\pm
0.11. \label{um3}
\end{equation}
Using unitarity, one obtains
\begin{equation}
|U_{e3}|^2 \approx 0, \qquad~~~~~ |U_{\mu 3}|\approx |U_{\tau 3}|
\approx \frac{1}{\sqrt{2}}.
\end{equation}
The state $\nu_3$ in terms of the flavor eigenstates $\nu_{\mu}$
and $ \nu_{\tau}$, to a good approximation, can be written as
\begin{equation}
\nu_3 = \frac{1}{\sqrt{2}}(\nu_{\mu} - \nu_{\tau}). \label{tbmnu3}
\end{equation}
This approximation along with unitarity yields the approximate
$\nu_{\mu} - \nu_{\tau}$  symmetry, expressed as
\begin{equation}
|U_{\mu 2}|\approx |U_{\tau 2}|, \qquad~~~~~~ |U_{\mu 1}|\approx
|U_{\tau 1}|. \label{um2ut2}
\end{equation}
The above equation, along with the solar neutrino data at
1$\sigma$ C.L. , e.g.,
\begin{equation}
|U_{e2}|^2 = 0.31 \pm 0.04 \label{ue2}
\end{equation}
implies
\begin{equation}
|U_{e2}| \approx |U_{\mu 2}| \approx |U_{\tau 2}| \approx
\frac{1}{\sqrt{3}}. \label{nu2approx}
\end{equation}

Without the loss of generality, with appropriate phases, the state
$\nu_2$ can be written as
\begin{equation} \nu_2 =  \frac{1}{\sqrt{3}}(\nu_e +
\nu_{\mu} + \nu_{\tau}). \label{tbmnu2}
\end{equation}
Again using unitarity of the PMNS matrix and the
Eqs.~(\ref{tbmnu3}) and (\ref{tbmnu2}), the state $\nu_1$ is given
by
\begin{equation}
\nu_1 =  \frac{1}{\sqrt{6}}(2\nu_e -\nu_{\tau} - \nu_{\mu}).
\label{tbmnu1}
\end{equation}
Eqs.~(\ref{tbmnu3}), (\ref{tbmnu2}) and (\ref{tbmnu1}) together
define the tri-bimaximal mixing
texture\cite{hps1}\cdash\cite{hps7}, depicted as
\begin{equation}
U_{tbm}= \left( \ba{ccc} \sqrt{\frac{2}{3}} & \frac{1}{\sqrt 3}& 0
\\ -\frac{1}{\sqrt 6} & \frac{1}{\sqrt 3} & -\frac{1}{\sqrt
2} \\ -\frac{1}{\sqrt 6} & \frac{1}{\sqrt 3} & \frac{1}{\sqrt 2}
\ea \right). \label{eq2}
\end{equation}

The current situation for neutrino physics and the PMNS matrix
appears analogous to the earlier situation for $B$ physics and the
CKM matrix, in which the leading approximation to the matrix was
established experimentally, long before its smallest elements were
determined. In that case, the Wolfenstein
parameterization\cite{wolf} had become widely adopted. Motivated
by the phenomenological success of tri-bimaximal mixing and
considering it as a starting point, Bjorken {\it et
al.}\cite{bjorken} have proposed a simple approximation for the
PMNS mixing in the leptonic sector. The proposal can be expressed
as
\begin{eqnarray}
U&\simeq& \left( \ba{ccc}
  \frac{2}{\sqrt{6}} &  \frac{1}{\sqrt{3}}  &   0  \\
 -\frac{1}{\sqrt{6}} &  \frac{1}{\sqrt{3}}  & \frac{1}{\sqrt{2}} \\
 -\frac{1}{\sqrt{6}} &  \frac{1}{\sqrt{3}}  & -\frac{1}{\sqrt{2}}  \\
\ea \right)
 \left( \ba{ccc}
 {C} & {0} & {\sqrt{\frac{3}{2}}U_{e3}} \\
 {0} & {1} & {0} \\
 {-\sqrt{\frac{3}{2}}U^*_{e3}} & {0} & {C} \ea \right) \nonumber\\
\\
&=& \left( \ba{ccc} {\frac{2}{\sqrt{6}}C}& {\frac{1}{\sqrt{3}}} &
{U_{e3}} \\
 {-\frac{1}{\sqrt{6}}C-\frac{\sqrt{3}}{2}U^*_{e3}} &
{\frac{1}{\sqrt{3}}} & {\frac{1}{\sqrt{2}}C-\frac{{1}}{2}U_{e3}}\\
 {-\frac{1}{\sqrt{6}}C+\frac{\sqrt{3}}{2}U^*_{e3}} &
{\frac{1}{\sqrt{3}}} & {-\frac{1}{\sqrt{2}}C-\frac{{1}}{2}U_{e3}}
\\ \ea \right),  \label{exactparam}
\end{eqnarray}
where
\begin{equation}
C=\sqrt{1-\frac{3}{2}|U_{e3}|^2}\simeq 1. \label{cdef}
\end{equation}
Dropping terms of order $|U_{e3}|^2$ Eq.~(\ref{exactparam}) can be
approximated as
\begin{eqnarray}
U&\simeq&
 \left( \ba{ccc}{\frac{2}{\sqrt{6}}} & {\frac{1}{\sqrt{3}}}&
 {U_{e3}}\\
{-\frac{1}{\sqrt{6}}-\frac{\sqrt{3}}{2}U^*_{e3}}&
{\frac{1}{\sqrt{3}}} & {\frac{1}{\sqrt{2}}-\frac{{1}}{2}U_{e3}} \\
{-\frac{1}{\sqrt{6}}+\frac{\sqrt{3}}{2}U^*_{e3}}&
{\frac{1}{\sqrt{3}}} & {-\frac{1}{\sqrt{2}}-\frac{{1}}{2}U_{e3}}
\\ \ea \right). \label{bjm}
\end{eqnarray}
This parameterization is important in the sense that it does not
involve the three mixing angles, instead it enables one to
directly deal with the complex parameter $U_{e3}$ of the mixing
matrix. Therefore, in this parameterization of the PMNS matrix,
the matrix can easily be constructed in case one has knowledge
regarding the value of the matrix element $U_{e3}$.

Coming to the analysis by Ref.~\refcite{ourlepuni}, since it was
carried out before the recent announcements regarding the
measurement of $U_{e3}$, therefore for the purpose of construction
of PMNS matrix a few representative values of $U_{e3}$ which
broadly agree with those considered by Farzan and
Smirnov\cite{farsm} were chosen. In this regard, theoretical
models\cite{albright} suggest $U_{e3}$ taking values around 0.05,
0.10 and 0.15 which can be used for the construction of the
magnitudes of the PMNS matrix elements. It may be added that the
present limits on $U_{e3}$ are well within its range considered
here. To have realistic estimates of the elements of the PMNS
matrix, following Farzan and Smirnov\cite{farsm}, modest errors to
the mixing elements considered by Bjorken {\it et
al.}\cite{bjorken} can be attached. To this end,
Ref.~\refcite{ourlepuni} associate
 5$\%$ errors with the elements $U_{e1}$, $U_{e2}$,
$U_{\mu 2}$ and $U_{\tau 2}$ of the matrix given in
Eq.~(\ref{bjm}) and for $U_{e3}$ values 0.05, 0.10 and 0.15 have
been taken and considered 10$\%$ variations to these. The matrices
corresponding to $U_{e3}$ values $0.05 \pm 0.005$, $0.10 \pm 0.01$
and $0.15 \pm 0.015$ are respectively as follows
 \be U = \left(  \ba{ccc}
  0.8165\pm0.0408 & 0.5774\pm0.0289  &  0.05\pm0.005\\
0.4516\pm0.0022  &  0.5774\pm0.0289  & 0.6821\pm0.0034\\
 0.3649\pm0.0018 &0.5774\pm0.0289 &0.7321\pm0.0037
 \ea \right), \label{.05} \ee
 \be U = \left(  \ba{ccc}
  0.8165\pm0.0408 & 0.5774\pm0.0289  &  0.1\pm0.01\\
0.4948\pm0.0049  &  0.5774\pm0.0289  & 0.6571\pm0.0066\\
 0.3216\pm0.0032 &0.5774\pm0.0289 &0.7571\pm0.0076
 \ea \right), \label{.1} \ee
 \be U = \left(  \ba{ccc}
  0.8165\pm0.0408 & 0.5774\pm0.0289  &  0.15\pm0.015\\
0.5382\pm0.0081  &  0.5774\pm0.0289  & 0.6321\pm0.0095\\
 0.2783\pm0.0042 &0.5774\pm0.0289 &0.7821\pm0.0117
 \ea \right), \label{.15} \ee
wherein the magnitudes of the elements have been given , as is
usual.

Construction of the $db$ unitarity triangle in the quark sector
immediately provides a clue for exploring the probability of non
zero Dirac-like CP violating phase $\delta_l$ in the leptonic
sector, even when leptonic mixing matrix is approximately known.
Out of the six triangles defined by Eqs.~(\ref{e-mu})-(\ref{nu2-
nu3}), Bjorken {\it et al.}\cite{bjorken} have considered the
$\nu_2.\nu_3$ triangle, depicted by Eq.~(\ref{nu2- nu3}), which is
the leptonic analogue of the $db$ triangle of the quark sector.
This triangle can immediately be constructed in the scenario
considered by Bjorken {\it et al.}\cite{bjorken} in case one uses
some values of $U_{e3}$ to construct the PMNS matrix.

\begin{figure}[tbp]
\centerline{\epsfysize=2.4in\epsffile{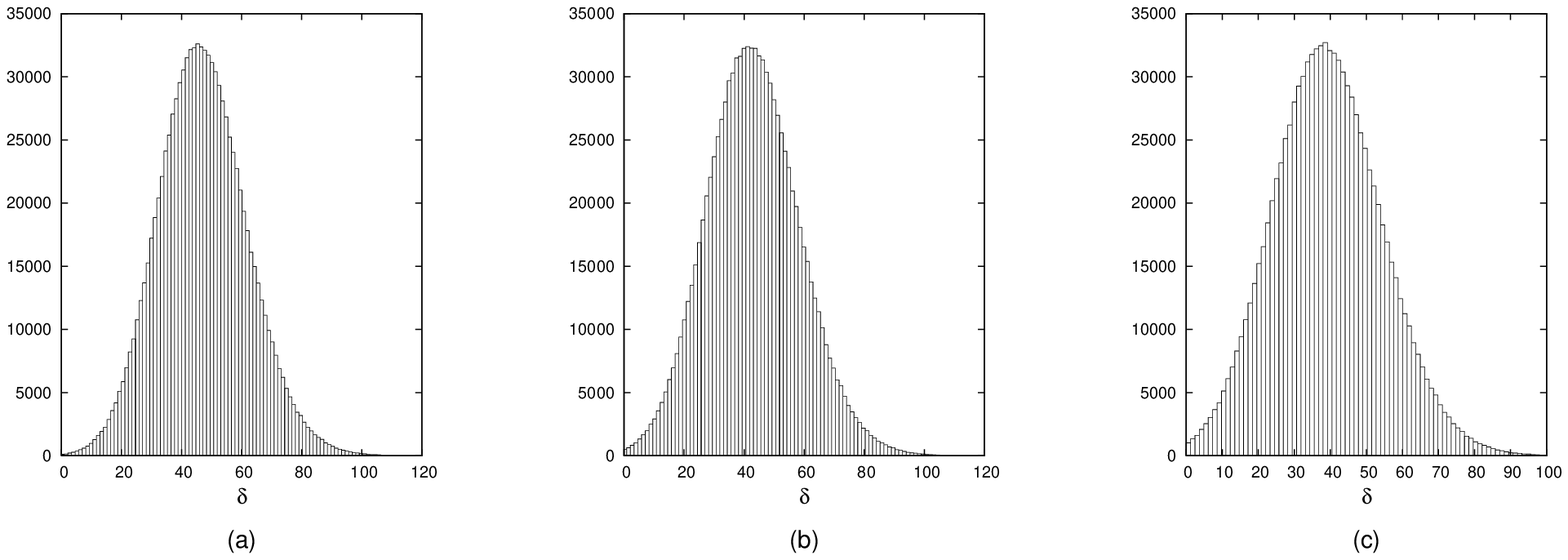}}
\vspace{0.08in} \caption{Histogram of $\delta_l$ plotted by
considering `$\nu_2.\nu_3$' triangle in the modified tri-bimaximal
scenario for (a) $U_{e3}=0.05 \pm 0.005$ (b) $U_{e3}=0.1 \pm 0.01$
(c) $U_{e3}=0.15 \pm 0.015$. }
  \label{delneut}
  \end{figure}

To this end, considering the elements of the above matrices
appearing in the $\nu_2.\nu_3$ triangle, given in Eq.~(\ref{nu2-
nu3}), to be Gaussian one can obtain the corresponding respective
values of the Jarlskog's rephasing invariant parameter in the
leptonic sector $J_l$ as
 \be J_l= 0.009 \pm 0.003, \label{j-.05}\ee
 \be J_l= 0.017 \pm 0.006, \label{j-.1}\ee
 \be J_l= 0.023 \pm 0.009. \label{j-.15}\ee
Using these values of $J_l$ and by considering various elements of
Eq.~(\ref{jd}) to be Gaussian, one can find the corresponding
distributions of $\delta_l$. Using these distributions, shown in
Fig.~\ref{delneut}, the $\delta_l$ values corresponding to
$U_{e3}$ values $0.05 \pm 0.005$, $0.10 \pm 0.01$ and $0.15 \pm
0.015$ are respectively as follows
 \be \delta_l\simeq47^{\rm o} \pm 15^{\rm o}, \ee
 \be \delta_l\simeq43^{\rm o} \pm 15^{\rm o}, \ee
 \be \delta_l\simeq39^{\rm o} \pm 15^{\rm o}. \ee
It is interesting to note that the Dirac-like CP violating phase
$\delta_l$ comes out to be around $43^{\rm o}$ and is not much
sensitive to $U_{e3}$ in the range $0.05-0.15$. Further, the above
calculated values of $\delta_l$, indicating a $2.5\sigma$
deviation from $0^{\rm o}$, in the modified tri-bimaximal scenario
for different values of $U_{e3}$, are in line with the suggestion
by several authors\cite{marciano4}\cdash\cite{giunti4}, about the
expected CP violation in the leptonic sector. Further, it is
interesting to note that this analysis carried out purely on
phenomenological inputs is very much in agreement with several
analyses based on expected outputs from different experimental
scenarios\cite{farsm,marciano4,balaji}\cdash\cite{white}.

\subsection{Leptonic CP violation after the
measurement of mixing angle $s_{13}$} As mentioned earlier, the
recent T2K\cite{t2k}, MINOS\cite{minos}, DAYA BAY\cite{dayabay}
and RENO\cite{reno} observations regarding the mixing angle
$s_{13}$ immediately provide a clue for exploring the possibility
of existence of CP violation in the leptonic sector. In the
context of neutrino oscillation phenomenology, the last few years
have seen impressive advances in fixing the neutrino mass and
mixing parameters through solar\cite{solexp1}\cdash\cite{solexp7},
atmospheric\cite{atmexp}, reactor (CHOOZ\cite{chooz},
KamLAND\cite{kamland}) and accelerator (K2K\cite{k2k},
MINOS\cite{minos,minosnew}) neutrino experiments. Adopting the
three neutrino framework, several
authors\cite{kamland,schwetztortolavalle}\cdash\cite{foglinew}
have presented updated information regarding these parameters
obtained by carrying out detailed global analyses. In particular,
incorporating the above mentioned developments regarding the angle
$s_{13}$, Fogli {\it et al}.\cite{foglinew} have carried out a
global three neutrino oscillation analysis, yielding
\be
 \Delta m_{21}^{2} = 7.58^{+0.22}_{-0.26}\times
 10^{-5}~\rm{eV}^{2},~~~~
|\Delta m_{31}^{2}| = 2.35^{+0.12}_{-0.21}\times 10^{-3}~
\rm{eV}^{2},
 \label{solatmmass}\ee
\be
{\rm sin}^2\,\theta_{12}  =  0.312^{+0.017}_{-0.016},~~~
 {\rm sin}^2\,\theta_{23}  =  0.42^{+0.08}_{-0.03},~~~
 {\rm sin}^2\,\theta_{13}  =  0.025\pm 0.007.~~~\label{s12s23} \ee

In analogy with the quark mixing phenomenon, the above value of
$s_{13}$ suggests likelihood of CP violation in the leptonic
sector. A comparison of the mixing angles in the leptonic sector
with those in the quark sector point out that the CP violation
could, in fact, be considerably large in this case. This
possibility, in turn, can have deep phenomenological implications.
As is well known, the two CP violating Majorana phases do not play
any role in the case of neutrino oscillations, therefore any hint
regarding the value of Dirac-like CP violating phase in the
leptonic sector $\delta_l$ will go a long way in the formulation
of proposals on observation of CP violation in the Long BaseLine
(LBL) experiments\cite{minos,k2k,opera}. In the absence of any
hints from the data regarding leptonic CP violation, keeping in
mind the parallelism between the neutrino mixing and the quark
mixing, an analysis of the quark mixing phenomena could provide
some viable clues regarding this issue in the leptonic sector.

It may be noted that in the context of fermion mixing phenomena,
the Pontecorvo-Maki-Nakagawa-Sakata
(PMNS)\cite{pmns1}\cdash\cite{pmns4} and the
Cabibbo-Kobayashi-Maskawa (CKM)\cite{cabibbo,kobayashi} matrices
have similar parametric structure. Also, regarding the three
mixing angles corresponding to the quark and leptonic sector, it
is interesting to note that in both the cases the mixing angle
$s_{13}$ is smaller as compared to the other two. Taking note of
these similarities of features, using the analogy of the quark
mixing case Ref.~\refcite{lepcp} have made an attempt to find the
possibility of the existence of CP violation in the leptonic
sector. Parallel to the leptonic sector wherein only the three
mixing angles or correspondingly the magnitudes of the three
elements of the mixing matrix are known, the author has first
considered a similar situation in the quark sector and examined
whether one can deduce any viable information regarding the
existence of CP violation in the quark mixing phenomena. Employing
the Particle Data Group (PDG) representation\cite{pdg10} of the
CKM matrix and making use of the fact the mixing angle $s_{13}$
($\equiv V_{ub}$) is small in comparison to both $s_{12}$ ($\equiv
V_{us}$) and $s_{23}$ ( $\equiv V_{cb}$) the approximate
magnitudes of the elements of the quark mixing matrix have been
constructed, e.g,\be V_{{\rm CKM}}= \left( \ba{ccc}
  0.97431\pm0.00021 & 0.2252\pm0.0009  &  0.00389\pm0.00044\\
  0.2250\pm0.0009  &  0.97351\pm0.00021  & 0.0406\pm0.0013\\
 0.00914\pm0.00029 &0.0396\pm0.0013 &0.999168\pm0.000053
 \ea \right). \label{ckmmag} \ee
The above matrix is in fairly good agreement with the one given by
PDG\cite{pdg10}.

 \begin{figure}[tbp]
\centerline{\epsfysize=2.6in\epsffile{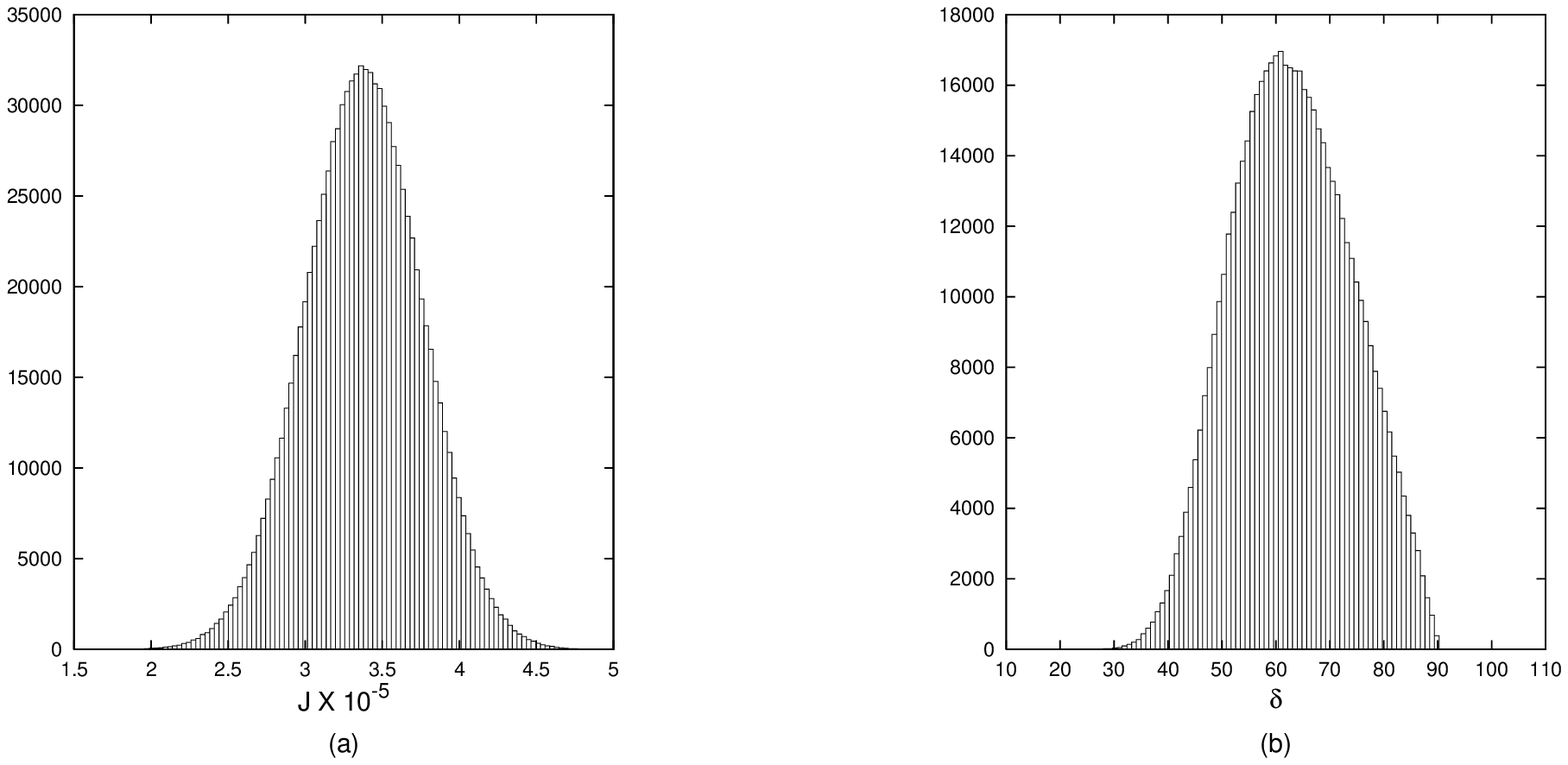}}
\vspace{0.08in}
   \caption{Histogram of $J$ and $\delta$ for `$db$'
triangle in the case of quarks}
  \label{jdelqua}
  \end{figure}

Making use of the above matrix and the usually considered `$db$'
unitarity triangle in the quark sector, the Jarlskog's rephasing
invariant parameter $J$ has been constructed using the magnitudes
of the elements of the CKM matrix. From the histogram of $J$,
shown in Fig.~\ref{jdelqua}(a), one finds
\be
J= (3.36 \pm 0.38) \times 10^{-5},\ee the corresponding histogram
of $\delta$, shown in Fig.~\ref{jdelqua}(b), yields \be
\delta=62.60^{\rm o} \pm 10.98^{\rm o}. \label{delq} \ee
Interestingly, the above mentioned $J$ and $\delta$ values are
compatible with those given by PDG 2010\cite{pdg10}.

For the case of existence of CP violation in the leptonic sector,
analogous to the construction of the CKM matrix presented in
Eq.~(\ref{ckmmag}) and using the inputs given in
Eq.~(\ref{s12s23}),the approximate magnitudes of the elements of
the PMNS matrix have been given as
 \be U = \left(  \ba{ccc}
  0.8190\pm0.0105 & 0.5516\pm0.0151  &  0.1581\pm0.0221\\
0.4254\pm0.0315  &  0.6317\pm0.0442 & 0.6399\pm0.0610\\
0.3620\pm0.0358  &  0.5376\pm0.0516 & 0.7520\pm0.0519
 \ea \right). \label{r2} \ee
It is interesting to note that this mixing matrix is compatible
with those given by Refs.~\refcite{othersmm1}
$-$\refcite{othersmm4}.

\begin{figure}[tbp]
\centerline{\epsfysize=2.6in\epsffile{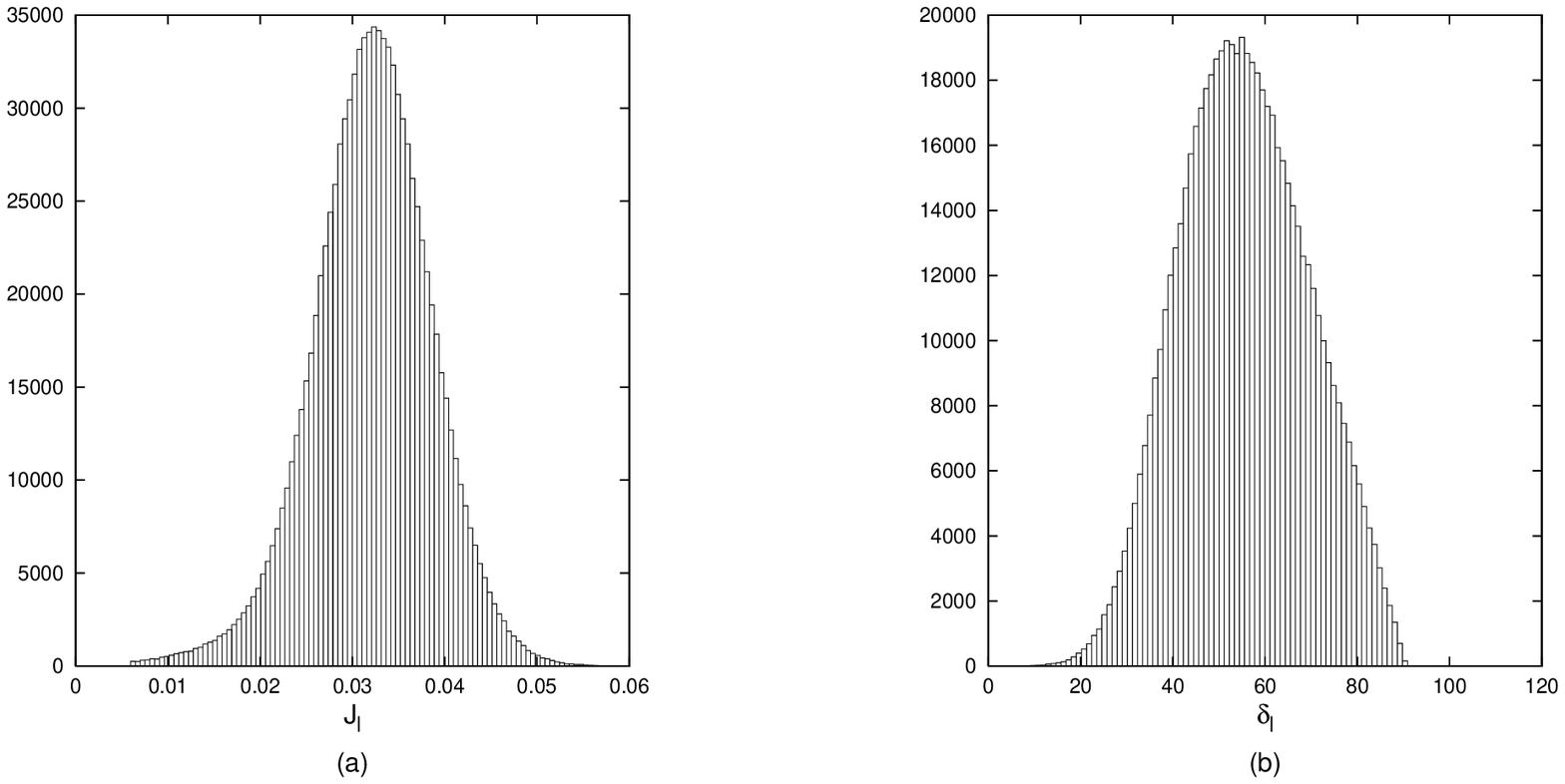}} \vspace{0.08in}
   \caption{Histogram of $J_l$ and $\delta_l$ for `$\nu_1.\nu_3$'
triangle in the case of neutrinos}
  \label{jdeln}
  \end{figure}

Analogous to the `$db$' triangle in the quark sector, considering
the `$\nu_1.\nu_3$' unitarity triangle, expressed as \be U_{e
1}U_{e 3}^{*} + U_{\mu 1}U_{\mu 3}^{*} + U_{\tau 1}U_{\tau 3}^{*}
 = 0 \,, \ee and using the
matrix given in Eq.~(\ref{r2}), the Jarlskog's rephasing invariant
parameter in the leptonic sector $J_l$ comes out to be \be J_l=
0.0318 \pm 0.0065, \label{jval}\ee corresponding distribution
plotted in Fig.~\ref{jdeln}(a).  Also, the corresponding phase
$\delta_l$ from the histogram given in Fig.~\ref{jdeln}(b) is
given by \be \delta_l=54.98^{\rm o} \pm 13.81^{\rm o}.
\label{deltanu} \ee Interestingly, this likely value of the phase
$\delta_l$ has a good overlap with several phenomenological
analyses\cite{farsm,ourlepuni,marciano4}\cdash\cite{white}.

\section{Fermion mass matrices \label{tsmm}}
Coming to the issue of fermion masses, as is well known, along
with fermion mixings, these provide a good opportunity to hunt for
physics beyond the SM. In view of the relationship of fermion
mixing phenomena with that of fermion mass matrices, understanding
flavor physics essentially implies formulating fermion mass
matrices. As mentioned earlier, the lack of a viable approach from
the `top down' perspective brings up the need for formulating
fermion mass matrices from a `bottom up' approach. In this
context, initially several {\it
ans\"{a}tze}\cite{frzans1,frzans2,ansatze1,ansatze2} were
suggested for quark mass matrices. One of the successful {\it
ans\"{a}tze} incorporating the `texture zero' approach was
initiated by Fritzsch\cite{frzans1,frzans2}. {\it A particular
texture structure is said to be texture $n$ zero, if it has $n$
number of non-trivial zeros, for example, if the sum of the number
of diagonal zeros and half the number of the symmetrically placed
off diagonal zeros is $n$}.

Further, related to the Fritzsch {\it ans\"{a}tze}, Branco {\it et
al.}\cite{branconni1} have formulated mass matrices in the
Nearest-Neighbour Interaction (NNI) basis. In particular, it as
been shown that in the SM, starting with arbitrary Yukawa
couplings, it is possible to find a weak basis where the quark
mass matrices have the NNI form. It has also been shown that NNI
textures for the quark mass matrices can be obtained through the
introduction of an Abelian family symmetry. In the context of two
Higgs doublet extension of the SM, the authors point out that the
NNI form for the quark mass matrices can be obtained through the
introduction of a $Z_4$ symmetry. The authors also point
out\cite{branconni2} that the NNI scheme, with small deviations
from hermiticity, can correctly reproduce the experimentally
allowed values for quark masses and CKM mixings. In the present
review, our emphasis will be essentially on the texture zero
formulation of mass matrices, related to the Fritzsch {\it
ans\"{a}tze}.

\subsection{Quark mass matrices}
The mass matrices, having their origin in the Higgs fermion
couplings, are arbitrary in the SM, therefore the number of free
parameters available with a general mass matrix is larger than the
physical observables. For example, if no restrictions are imposed,
there are $36$ real free parameters in the two $3 \times 3$
general complex mass matrices, $M_U$ and $M_D$, which in the quark
sector need to describe ten physical observables, i.e., six quark
masses, three mixing angles and one CP violating phase. Similarly,
in the leptonic sector, physical observables described by lepton
mass matrices are six lepton masses, three mixing angles and one
CP violating phase for Dirac neutrinos (two additional phases in
case neutrinos are Majorana particles). Therefore, to develop
viable phenomenological fermion mass matrices one has to limit the
number of free parameters in the mass matrices.

In this context, it is well known that in the SM and its
extensions wherein the right handed fields in the Lagrangian are
SU(2) singlets, without loss of generality, the mass matrices can
be considered as hermitian\cite{9912358}. This immediately brings
down the number of real free parameters from 36 to 18, which
however, is still a large number compared to the number of
observables. To this end, Weinberg\cite{weinberg} implicitly and
Fritzsch\cite{frzans1,frzans2} explicitly initiated the idea of
texture specific mass matrices which on the one hand imparted
predictability to mass matrices while on the other hand, it paved
the way for the phenomenology of texture specific mass matrices.
To define the various texture specific cases, we present the
typical Fritzsch-like texture specific hermitian quark mass
matrices, e.g.,
\be
 M_{U}=\left( \ba{ccc}
0 & A _{U} & 0      \\ A_{U}^{*} & D_{U} &  B_{U}     \\
 0 &     B_{U}^{*}  &  C_{U} \ea \right), \qquad
M_{D}=\left( \ba{ccc} 0 & A _{D} & 0      \\ A_{D}^{*} & D_{D} &
B_{D}     \\
 0 &     B_{D}^{*}  &  C_{D} \ea \right),
\label{nf2zero}\ee
 where $M_{U}$ and $M_{D}$ correspond to up and
down mass matrices respectively. It may be noted that each of the
above matrix is texture 2 zero type with $A_{i}
=|A_i|e^{i\alpha_i}$
 and $B_{i} = |B_i|e^{i\beta_i}$, where $i= U,D$.

The texture 6 zero Fritzsch mass matrices can be obtained from the
above mentioned matrices by taking both $D_{U}$ and $D_{D}$ to be
zero, which reduces the matrices $M_{U}$ and $M_{D}$ each to
texture 3 zero type. This Fritzsch {\it
ans\"{a}tze}\cite{frzans1,frzans2} as well as some other {\it
ans\"{a}tze}\cite{ansatze1,ansatze2} were ruled out because of the
large value predicted by these for $|V_{cb}|$ due to the high `t'
quark mass, in disagreement with the experimental data. Further, a
few other texture 6 zero mass matrices were analyzed by Ramond,
Roberts and Ross\cite{rrr} revealing that these matrices were
again ruled out because of the large predicted value of
$|V_{cb}|$. They also explored the question of connection between
phenomenological quark mass matrices considered at low energies
and the possible mass patterns at the GUT scale and showed that
the texture structure of mass matrices is maintained as we come
down from GUT scale to $M_Z$ scale. This important conclusion also
leads to the fact that the texture zeros of fermion mass matrices
can be considered as phenomenological zeros, thereby implying that
at all energy scales the corresponding matrix elements are
sufficiently suppressed in comparison with their neighboring
counterparts. This, therefore, opens the possibility of
considering lesser number of texture zeros.

Besides Ramond, Roberts and Ross\cite{rrr}, several
authors\cite{9912358,group5zero}\cdash\cite{5zero3} have explored
the texture 5 zero quark mass matrices. Fritzsch-like texture 5
zero matrices can be obtained by taking either $D_{U}$ = 0 and
$D_{D} \neq 0$ or $D_{U} \neq 0$ and $D_{D}$ = 0 in
Eq.~(\ref{nf2zero}), thereby giving rise to two possible cases of
texture 5 zero mass matrices pertaining to either $M_U$ or $M_D$
being texture 3 zero type while the other being texture 2 zero
type. These analyses reveal that texture 5 zero mass matrices
although not ruled out unambiguously yet are not able to reproduce
the entire range of data.

As an extension of texture 5 zero mass matrices, several
authors\cite{9912358,group4zero1}\cdash\cite{group4zero5} carried
out the study of the implications of the Fritzsch-like texture 4
zero mass matrices. It may be noted that Fritzsch-like texture 4
zero mass matrices can be obtained by considering both $M_U$ and
$M_D$, with non zero $D_i (i=U,D)$ in Eq.~(\ref{nf2zero}), to be
texture 2 zero type. Although from the above mentioned analyses
one finds that texture 4 zero mass matrices were able to
accommodate the quark mixing data quite well, however it may be
noted that these analyses assumed `strong hierarchy', to be
defined later, of the elements of the mass matrices as well as
explored only their limited domains. Further, in the absence of
any precise information about CP violating phase ${\delta}$,
sin$\,2\beta$ and related parameters, adequate attention was not
given to the phases of the mass matrices.

Recent refinements in quark mixing data as well as information
about the CP violating phase motivated several
authors\cite{hallraisin}\cdash\cite{cps} to have a re-look at the
compatibility of Fritzsch-like texture 4 zero mass matrices with
the quark mixing data. In particular, using assumption of `strong
hierarchy' of the elements of the mass matrix defined as $D_i <
|B_{i}|
< C_i, (i=U,D)$, having its motivation in the
hierarchy of the quark mixing angles several
attempts\cite{hallraisin}\cdash\cite{branco} were made to predict
the value of precisely known parameter sin$\,2\beta$.
Unfortunately, the value of sin$\,2\beta$ predicted by these
analyses came out to be in quite disagreement with its precisely
known value. A somewhat detailed and comprehensive analyses of
texture 4 zero quark mass matrices was carried out by Xing and
Zhang\cite{xingzhang}, in particular they attempted to find the
parameter space available to the elements of mass matrices. Their
analysis has also given valuable clues about the phase structure
of the mass matrices, in particular for the strong hierarchy case
they conclude that only one of the two phase parameters plays a
dominant role. Subsequently, attempts have also been made by Verma
{\it et al.}~\cite{s2b,cps} to update and broaden the scope of the
analysis carried out by Xing and Zhang\cite{xingzhang}, in
particular regarding the structural features of the mass matrices
having implications for the value of parameter sin$\,2\beta$ and
the CP violating phase $\delta$. Further, recently, an exhaustive
analysis of texture 6 and 5 zero non Fritzsch-like texture
specific mass matrices have also been carried
out\cite{neelu56zeroquarks} leading to some interesting
conclusions. In view of the large parameter space available for
fitting the data and very large number of possibilities of non
Fritzsch-like texture 4 zero mass matrices, an exhaustive analysis
in this regard is yet to be carried out. In the sequel, we briefly
discuss the analyses of
Refs.~\refcite{s2b}-\refcite{neelu56zeroquarks} regarding the
texture specific mass matrices.

\subsubsection{Relationship of quark mass matrices and mixing
matrix} Before detailing the analyses of texture specific mass
matrices, for the sake of completeness, we present essentials
regarding the relationship between the quark mass matrices and the
CKM mixing matrix. In the SM, the quark mass terms for three
generations of quarks can be expressed as
\be
{\overline q}_{U_L} M_U ~ {q}_{U_R}  + {\overline q}_{D_L} M_D ~
{q}_{D_R}\,,
 \label{mc} \ee
where ${q}_{U_{L(R)}}$ and  ${q}_{D_{L(R)}}$ are the left handed
(right handed)
 quark fields for the  up sector
$(u,c,t)$ and down sector $(d,s,b)$ respectively. $M_U$ and $M_D$
are the mass matrices for the up and the down sector  of quarks.
 In order to re-express above equation in terms of the physical quark
fields, one can diagonalize the mass matrices by the following
bi-unitary transformations \be V_{U_L}^{\dagger}M_U V_{U_R} =
M_U^{diag} \equiv {\rm Diag}\, (m_u,m_c,m_t)\,, \label{v11} \ee
\be V_{D_L}^{\dagger}M_D V_{D_R} = M_D^{diag} \equiv {\rm Diag}
\,(m_d,m_s,m_b)\,, \label{v21} \ee
 where $M_{U,D}^{diag}$ are real and diagonal, while $V_{U_L}$ and
  $V_{U_R}$ etc. are complex unitary matrices. The quantities $m_u, m_d$, etc. denote the
  eigenvalues of the mass matrices, i.e. the physical
quark masses. Using Eqs.~(\ref{v11}) and (\ref{v21}), one can
rewrite (\ref{mc}) as
 \be  \overline{q}_{U_L}V_{U_L}M_U^{diag}
V_{U_R}^{\dagger} q_{U_R}  + \overline{q}_{D_L} V_{D_L} M_D^{diag}
V_{D_R}^{\dagger} q_{D_R}\, , \ee which can be re-expressed in
terms of physical quark fields as
\be
 \overline{q}^{phys}_{{U}_L} M_U^{diag}
q^{phys}_{U_R}  + \overline{q}^{phys}_{{D}_L}M_D^{diag}
 q^{phys}_{D_R}\, ,\ee
where $q^{phys}_{U_L}\,=\,V_{U_L}^{\dagger} q_{U_L}$ and
 $q^{phys}_{D_L}\,=\,V_{D_L}^{\dagger} q_{D_L}$ and so on.

The mismatch of diagonalizations of up and down quark mass
matrices leads to the quark mixing matrix $V_{{\rm CKM}}$,
referred to as the Cabibbo-Kobayashi-Maskawa
(CKM)\cite{cabibbo,kobayashi} matrix given as \be V_{\rm CKM} =
V_{U_L}^{\dagger} V_{D_L}.
 \label{1mix}
\ee The CKM matrix expresses the relationship between quark mass
eigenstates $d,\,s,\,b$ which participate in the strong q$-$q and
q$-\overline{\rm{q}}$ interactions and the interaction eigenstates
or flavor eigenstates $d', s', b'$ which participate in the weak
interactions and are the linear combinations of mass eigenstates,
e.g., \be  \left( \ba{c} d'
\\ s'
\\ b' \ea \right) = \left( \ba {lll}
V_{ud} & V_{us} & V_{ub} \\ V_{cd} & V_{cs} & V_{cb} \\ V_{td} &
V_{ts} & V_{tb} \\ \ea \right) \left( \ba {c} d\\ s \\ b \ea
\right), \label {1vckm}\ee where $V_{ud}$, $V_{us}$, etc. describe
the transition of $u$ to $d$, $u$ to $s$ respectively, and so on.

In view of the relationship of the mixing matrix with the mass
matrix, a knowledge of the CKM matrix elements would have
important implications for the mass matrices. The most commonly
used parameterization of the quark mixing matrix, the standard
parameterization given by Particle Data Group (PDG)\cite{pdg10},
has already been presented in Eq.~(\ref{1ckm}). Keeping in mind
the ever increasing precision in the measurement of CKM
phenomenological parameters, it is desirable to keep updating the
analyses of the mass matrices for their compatibility with the
mixing data.

\subsubsection{Texture 6 zero quark mass matrices} To begin with, we
first consider texture 6 zero Fritzsch quark mass matrices given
by
 \be
 M_{U}=\left( \ba{ccc}
0 & A _{U} & 0      \\ A_{U}^{*} & 0 &  B_{U}     \\
 0 &     B_{U}^{*}  &  C_{U} \ea \right), \qquad
M_{D}=\left( \ba{ccc} 0 & A _{D} & 0      \\ A_{D}^{*} & 0 & B_{D}
\\
 0 &     B_{D}^{*}  &  C_{D} \ea \right),
\label{2zero}\ee
 where $M_{U}$ and $M_{D}$ correspond to up and
down mass matrices respectively . The non Fritzsch-like mass
matrices differ from the above mentioned Fritzsch-like mass
matrices in regard to the position of `zeros' in the structure of
the mass matrices. One can get non Fritzsch-like mass matrices by
shifting the position of $C_i$ ($i=U, D$) on the diagonal as well
as by shifting the position of zeros among the non diagonal
elements. For example, a non Fritzsch-like texture 3 zero matrix
is obtained if (1,1) element is non zero, the other diagonal
elements are zero leaving the non diagonal elements unchanged,
with (1,1) referring to the element corresponding to the first row
and first column of the matrix. Similarly, by considering (1,2)
and (2,1) element to be zero and (1,3) and (3,1) element to be non
zero, without disturbing other elements, we again get texture 3
zero non Fritzsch-like mass matrix. This results into a total of
20 different possible texture patterns, out of which 8 are easily
ruled out by imposing the following conditions
\be
{\rm Trace}~ M_{U,D}  \neq 0 \qquad {\rm and} \qquad {\rm Det}~
M_{U,D} \neq 0, \label{trace}\ee corresponding to non zero, non
degenerate quark masses.

These possible patterns of texture specific mass matrices can be
limited further by considering the constraints imposed by
diagonalization procedure of mass matrices in up and down sector
to obtain CKM matrix, details of diagonalization can be looked up
in Refs.~\refcite{detaildiag} and \refcite{Monika}. An essential
step in this process is to consider the invariants trace $M$,
trace $M^{2}$ and determinant $M$ which yield the relations
involving elements of mass matrices and mass eigenvalues $m_1$,
$-m_2$ and $m_3$\cite{9912358,Monika}, taking the second
eigenvalue as $-m_2$ facilitates the diagonalization procedure
without affecting the consequences\cite{9912358}. Following
Ref.~\refcite{neelu56zeroquarks}, it is interesting to note that
the 12 possible textures break into two classes as shown in
Table~\ref{t1} depending upon the equations these matrices
satisfy. For example, six matrices of class I, mentioned in
Table~\ref{t1}, satisfy the following equations
\be
C = m_1- m_2+ m_3, \quad A^2 + B^2 = m_1m_2 + m_2m_3 -m_1m_3,
\quad A^2 C = m_1m_2m_3. \label{classI}\ee
 Similarly, in case of class II
all six matrices satisfy the following equations
\be
C + D= m_1 -m_2 +m_3,\quad A^2 -C D = m_1m_2 + m_2m_3 - m_1m_3,
\quad A^2 C = m_1m_2m_3. \label{classII}\ee
 The subscripts U and D have not been
used as these are valid
 for both kind of mass matrices.

\begin{table}
 \tbl{Twelve possibilities of
texture 3 zero hermitian mass matrices categorized into two
classes I and II.} {\bt{|c|c|c|} \hline
  & Class I  & Class II \\ \hline
a & $\left ( \ba{ccc} {\bf 0} & Ae^{i\alpha} & {\bf 0} \\
Ae^{-i\alpha}  & {\bf 0} & Be^{i\beta} \\ {\bf 0} & Be^{-i\beta} &
C \ea \right )$  & $\left ( \ba{ccc} {\bf 0} & Ae^{i\alpha} & {\bf
0} \\ Ae^{-i\alpha}  & D & {\bf 0} \\ {\bf 0} & {\bf 0}  & C \ea
\right )$ \\ b &  $\left ( \ba{ccc} {\bf 0} &{\bf 0} &
Ae^{i\alpha} \\ {\bf 0}  & C & Be^{i\beta} \\Ae^{-i\alpha} &
B^{-i\beta}  & {\bf 0} \ea \right )$  &
 $\left ( \ba{ccc} {\bf 0} & {\bf 0} & Ae^{i\alpha}
 \\ {\bf 0}  & C & {\bf 0} \\ Ae^{-i\alpha}  & {\bf 0}  &
D \ea \right )$ \\ c &  $\left ( \ba{ccc} {\bf 0} & Ae^{i\alpha} &
Be^{i\beta} \\ Ae^{-i\alpha}  & {\bf 0} & {\bf 0} \\ Be^{-i\beta}
& {\bf 0}  & C \ea \right )$  &
 $\left ( \ba{ccc} D & Ae^{i\alpha} &
{\bf 0} \\ Ae^{-i\alpha}  & {\bf 0} & {\bf 0} \\ {\bf 0} & {\bf 0}
& C \ea \right )$ \\ d &  $\left ( \ba{ccc} C & Be^{i\beta} & {\bf
0}
\\ Be^{-i\beta}  & {\bf 0} & Ae^{i\alpha}\\ {\bf 0}  & Ae^{-i\alpha} &
{\bf 0} \ea \right )$  &
 $\left ( \ba{ccc} C & {\bf 0} & {\bf 0}
 \\ {\bf 0}  & D & Ae^{i\alpha} \\ {\bf 0} & Ae^{-i\alpha}  &
{\bf 0} \ea \right )$ \\ e &  $\left ( \ba{ccc} {\bf 0} &
Be^{i\beta}  & Ae^{i\alpha} \\ Be^{-i\beta}  & C & {\bf 0} \\
Ae^{-i\alpha}  & {\bf 0}  &
 {\bf 0} \ea \right )$  &
 $\left ( \ba{ccc} D & {\bf 0} & Ae^{i\alpha}
\\ {\bf 0} & C &  {\bf 0} \\ Ae^{-i\alpha}  & {\bf 0}  &
{\bf 0}  \ea \right )$ \\ f & $\left ( \ba{ccc} C & {\bf 0} &
Be^{i\beta}
 \\ {\bf 0}  & {\bf 0}  & Ae^{i\alpha} \\Be^{-i\beta} & Ae^{-i\alpha}  &
{\bf 0} \ea \right )$  &
 $\left ( \ba{ccc} C & {\bf 0} &{\bf 0}
 \\ {\bf 0}  & {\bf 0} & Ae^{i\alpha} \\ {\bf 0} & Ae^{-i\alpha}  &
D \ea \right )$ \\    \hline \et} \label{t1}
\end{table}

The matrices $M_{U}$ and $M_{D}$ each can correspond to any of the
12 possibilities, therefore yielding 144 possible combinations
which in principle can yield 144 quark mixing matrices. These 144
combinations can be put into 4 different categories, e.g., if
$M_{U}$ is any of the 6 matrices from class I, then $M_{D}$ can be
either from class I or class II yielding 2 categories of 36
matrices each. Similarly, we obtain 2 more categories of 36
matrices each when $M_{U}$ is from class II and $M_{D}$ is either
from class I or class II. The 36 combinations in each category
further can be shown to be reduced to groups of six combinations
of mass matrices, each yielding same CKM matrix. For example, six
of the 36 combinations belonging to first category, when both up
and down sectors mass matrices are of the form $M_{U_{i}}$ and
$M_{D_{i}}$ (i = a,b,c,d,e,f), yield the same CKM matrices.
Similarly, the remaining 30 matrices in category one yield five
groups of six matrices each corresponding to five independent
mixing matrices. A similar simplification can be achieved in other
three categories.

Keeping in mind the hierarchical nature of the CKM matrix, an
analytical analysis of categories as mentioned above yields only 4
groups of $M_{U_{i}}$ and $M_{D_{i}}$ corresponding to 6
combinations each of which yield 4 CKM matrices. To illustrate
this point one can consider first matrix $M_{U}$ to be of type (a)
from class I and similarly $M_{D}$ to be of type (b) from the same
class. The corresponding CKM matrix is expressed as \beqn
 V_{{\rm CKM}} &=& \left( \ba {lll}
V_{ud} & V_{us} & V_{ub} \\ V_{cd} & V_{cs} & V_{cb} \\ V_{td} &
V_{ts} & V_{tb} \\ \ea \right)  \label{km0}  \eeqn

\be
=\left[ \ba {ccc}  -e^{i(\alpha_U - \alpha_D)} &
(\sqrt{\frac{m_d}{m_s}}) e^{i(\alpha_U - \alpha_D)} &
\sqrt{\frac{m_u}{m_c}}e^{i\beta_D}  \\
\sqrt{\frac{m_u}{m_c}}e^{i(\alpha_U - \alpha_D)}
+\sqrt{\frac{m_d}{m_b}}e^{i\beta_D} &
  -\sqrt{\frac{m_s}{m_b}} e^{i\beta_D}-\sqrt{\frac{m_c}{m_t}} e^{i\beta_U} &
e^{-i\beta_D}\\ \sqrt{\frac{m_d}{m_s}}e^{-i\beta_U}  & -1 &
\sqrt{\frac{m_s}{m_b}}e^{-i\beta_U} + \sqrt{\frac{m_c}{m_t}}
e^{-i\beta_D}   \ea \right], \label{np}\ee
 where $\alpha_i$ and
$\beta_i$, $i=U,D$ are related to the phases of the elements $A_i$
and $B_i$ of the mass matrices given in Eq.~(\ref{2zero}). From
the above structure of CKM matrix one can easily find out that off
diagonal elements, e.g., $|V_{cb}|$ and $|V_{ts}|$ are of the
order of unity whereas diagonal elements $|V_{cs}|$ and $|V_{tb}|$
are smaller than unity which is in complete contrast to the
structure of CKM matrix. In a similar manner, one can conclude
that remaining indistinguishable combinations also lead to such
non-physical mixing matrices. In case we apply the above criteria,
interestingly we are left with only four groups of mass matrices
as mentioned in Table~\ref{t3}. Thus the problem of exploring the
compatibility of 144 phenomenological allowed texture 6 zero
combinations with the recent low energy data is reduced only to an
examination of 4 groups each having 6 combinations of mass
matrices corresponding to the same CKM matrix.

The compatibility of these 4 groups of texture 6 zero mass
matrices with the quark mixing data has been examined in
Ref.~\refcite{neelu56zeroquarks}. Considering the quark masses and
mass ratios at $M_z$ scale(GeV)\cite{leut},
\be
m_{u}=0.002- 0.003, m_{c}=0.6- 0.7, m_{t}=169.5- 175.5,
\label{uct}\ee
\be
 m_{d}=0.0037- 0.0052, m_{s}= 0.072- 0.097, m_{b}= 2.8- 3.0,
\label{dsb}\ee
\be
 m_{u}/m_{d} = 0.51- 0.60, m_{s}/m_{d} = 18.1-19.7 .
\label{ratio}\ee and by giving full variation to phases $\phi_{1}$
and $\phi_{2}$, some of the CKM parameters have been reproduced
and compared with the following data,
\be
|V_{us}| = 0.2236-0.2274, |V_{cb}|= 0.0401- 0.0423,
\label{vckm}\ee
\be
{\rm sin\,}2\beta=0.656 -0.706,~~ J=(2.85-3.24)\times 10^{-5},~~
\delta=45^{\circ} - 107^{\circ}. \label{sin}\ee From
Table~\ref{t3} one can immediately find that all possible
combinations of texture 6 zero are ruled out as these are not able
to reproduce the CKM element $|V_{cb}|$. Thus, none of the texture
6 zero combinations, Fritzsch-like as well as non Fritzsch-like,
is found to be compatible with the recent quark mixing data,
ruling out the existence of these mass matrices.

\begin{table}
\tbl{Predicted values of $|V_{cb}|$, sin$\,2\beta$, $\delta$ and J
for 4 independent texture 6 zero combinations where i =
a,b,c,d,e,f.}
 {\bt{|cc|c|c|c|c|c|} \hline $M_{U}$ & $M_{D}$ & $|V_{cb}|$ &
$sin\,2\beta$ & $\delta^{\circ}$ & $J\times10^{-5}$  \\ \hline
$I_{i}$ & $I_{i}$ & 0.09-0.24 & 0.48-0.57 & 78-100 & 12-76  \\
$II_{i}$ & $II_{i}$  & 0 & not defined & not defined & 0 \\
$I_{i}$ & $II_{i}$  & 0.055-0.065 & 0.48-0.55 & 74-90 & 4.7-5.3 \\
$II_{i}$ & $I_{i}$  & 0.15-0.18 & 0.50-0.54 & 80-95 & 32-44 \\
\hline \et} \label{t3} \end{table}

\subsubsection{Texture 5 zero quark mass matrices} As already mentioned,
several authors\cite{9912358,rrr}\cdash\cite{5zero3} have explored
the case of texture 5 zero quark mass matrices, however, recently
a detailed and comprehensive analysis of all possible,
Fritzsch-like as well as non Fritzsch-like texture 5 zero mass
matrices has been carried out\cite{neelu56zeroquarks}. A brief
discussion of this analysis is perhaps in order. As mentioned
earlier, texture 5 zero mass matrices can be obtained by either
considering $M_{U}$ being 2 zero and $M_{D}$ being 3 zero type or
vice versa. Texture 3 zero possibilities have already been
enumerated and after taking into consideration the conditions
mentioned in Eq.~(\ref{trace}) one can check that there are 18
possible texture 2 zero patterns. These textures further break
into three classes detailed in Table~\ref{t2}. It can be easily
shown that while constructing the CKM matrix, the element F in
type `a' matrix of class V is much smaller than the other elements
of considered mass matrix, therefore it can be considered as a
very small perturbation on corresponding texture 3 zero pattern.
Similar conclusion can be drawn for other matrices of the same
class as well as it can be shown that matrices of class IV also
reduce to texture 3 zero patterns. Therefore, one can conclude
that the matrices in class IV and V effectively reduce to the
corresponding 3 zero patterns, leaving only one, class III of
texture 2 zero matrices, that needs to be explored for texture 5
zero combinations. All matrices of this class satisfy the
following equation
\be
 C + D = m_1- m_2+ m_3,\quad A^2 + B^2 - C D = m_1m_2 + m_2m_3 -m_1m_3,\quad A^2 C = m_1m_2m_3.
\label{classIII}\ee

\begin{table}
\tbl{Texture 2 zero possibilities categorized into three classes
III, IV and V.} {\bt{|c|c|c|c|} \hline
 & Class III  & Class IV  & Class V   \\  \hline
a & $\left ( \ba{ccc} {\bf 0} & Ae^{i\alpha} & {\bf 0} \\
Ae^{-i\alpha}  & D & Be^{i\beta} \\ {\bf 0} & Be^{-i\beta}  & C
\ea \right )$  &
 $\left ( \ba{ccc} D & Ae^{i\alpha} &
{\bf 0} \\ Ae^{-i\alpha}  & {\bf 0} &  Be^{i\beta} \\ {\bf 0} &
Be^{-i\beta}  & C \ea \right )$     &
 $\left ( \ba{ccc} {\bf 0} & Ae^{i\alpha} &
 Fe^{i\gamma}\\ Ae^{-i\alpha}  &{\bf 0}  & Be^{i\beta} \\ Fe^{-i\gamma} & Be^{-i\beta}  &
C \ea \right )$ \\ b &  $\left ( \ba{ccc} {\bf 0} & {\bf 0}  &
Ae^{i\alpha} \\ {\bf 0}  & C & Be^{i\beta} \\  Ae^{-i\alpha} &
Be^{-i\beta}  & D \ea \right )$  &
 $\left ( \ba{ccc} D & {\bf 0} & Ae^{i\alpha}
 \\ {\bf 0} & C & Be^{i\beta} \\  Ae^{-i\alpha} & Be^{-i\beta}  &
{\bf 0} \ea \right )$     & $\left ( \ba{ccc} {\bf 0} &
Fe^{i\gamma} & Ae^{i\alpha}
\\  Fe^{-i\gamma}  & C  & Be^{i\beta} \\  Ae^{-i\alpha} & Be^{-i\beta}  &
 {\bf 0}\ea \right )$ \\
c &  $\left ( \ba{ccc} D & Ae^{i\alpha} &
  Be^{i\beta}\\ Ae^{-i\alpha}  & {\bf 0}  & {\bf 0} \\  Be^{-i\beta} & {\bf 0}
& C \ea \right )$  & $\left ( \ba{ccc} {\bf 0} & Ae^{i\alpha} &
 Be^{i\beta} \\ Ae^{-i\alpha}  & D & {\bf 0}  \\  Be^{-i\beta}  & {\bf 0} &
C \ea \right )$     & $\left ( \ba{ccc} {\bf 0} & Ae^{i\alpha} &
Be^{i\beta}
\\ Ae^{-i\alpha} & {\bf 0} & Fe^{i\gamma} \\ Be^{-i\beta}  & Fe^{-i\gamma} &
C \ea \right )$ \\ d &  $\left ( \ba{ccc} C & Be^{i\beta} & {\bf
0}
 \\ Be^{-i\beta} & D & Ae^{i\alpha}  \\ {\bf 0} & Ae^{-i\alpha} & {\bf 0}
 \ea \right )$  &
 $\left ( \ba{ccc} C & Be^{i\beta} &  {\bf 0}
 \\ Be^{-i\beta} & {\bf 0} & Ae^{i\alpha} \\ {\bf 0} & Ae^{-i\alpha}  &
D \ea \right )$     &
 $\left ( \ba{ccc} C & Be^{i\beta}  & Fe^{i\gamma}
\\ Be^{-i\beta}   & {\bf 0}  &  Ae^{i\alpha} \\ Fe^{-i\gamma} & Ae^{-i\alpha}  &
 {\bf 0}\ea \right )$ \\
e &  $\left ( \ba{ccc} D & Be^{i\beta} & Ae^{i\alpha}
  \\ Be^{-i\beta}  & C & {\bf 0} \\ Ae^{-i\alpha} & {\bf 0}
& {\bf 0} \ea \right )$  &
 $\left ( \ba{ccc} {\bf 0} & Be^{i\beta} & Ae^{i\alpha}
  \\  Be^{-i\beta}  & C & {\bf 0}  \\ Ae^{-i\alpha}   & {\bf 0} &
D \ea \right )$     & $\left ( \ba{ccc} {\bf 0}  & Be^{i\beta} &
Ae^{i\alpha}
\\ Be^{-i\beta}  & C & Fe^{i\gamma} \\ Ae^{-i\alpha} & Fe^{-i\gamma}&
{\bf 0}  \ea \right )$ \\ f &  $\left ( \ba{ccc} C & {\bf 0} &
Be^{i\beta}
 \\ {\bf 0} & {\bf 0} & Ae^{i\alpha}  \\ Be^{-i\beta} & Ae^{-i\alpha} & D
 \ea \right )$  &
$\left ( \ba{ccc} C  &  {\bf 0} & Be^{i\beta}
 \\ {\bf 0} & D & Ae^{i\alpha} \\Be^{-i\beta}  & Ae^{-i\alpha}  &
0 \ea \right )$     &
 $\left ( \ba{ccc} C & Fe^{i\gamma} & Be^{i\beta}
\\  Fe^{-i\gamma} & {\bf 0}  &  Ae^{i\alpha} \\ Be^{i\beta} & Ae^{-i\alpha}  &
 {\bf 0}\ea \right )$ \\  \hline
\et} \label{t2}
\end{table}

Considering class III of texture 2 zero mass matrices along with
different patterns of class I and II of texture 3 zero mass
matrices we find a total of 144 possibilities of texture 5 zero
mass matrices, in sharp contrast to the case if we had considered
class IV and V also yielding 432 possibilities. Keeping in mind
the hierarchy of CKM matrix, out of 144 cases, one is again left
with only 4 groups of texture 5 zero mass matrices detailed in
Table~\ref{t4}. A general look at the table immediately reveals
that as compared to texture 6 zero case $|V_{cb}|$ predictions are
quite different. However, interestingly only in one case
$|V_{cb}|$ as well as other CKM elements are found to be
compatible with their experimentally measured values.
Interestingly, this possibility corresponds to the usual
Fritzsch-like texture 5 zero mass matrix where $M_{U}$ is of
texture 2 zero and $M_{D}$ is of texture 3 zero type. It may be
noted that the possibility corresponding to $M_{U}$ being texture
3 zero and $M_{D}$ being texture 2 zero is not viable which is due
to the fact that pertaining to this case the hierarchy of elements
of mass matrices comes out be very strong, therefore making it
difficult to fit the data.

\begin{table}
\tbl{Texture 5 zero combinations and their corresponding predicted
values pertaining to $|V_{cb}|$, $|V_{ub}|$, sin$\,2\beta$,
$\delta$ and $J$.} {\bt{|cc|c|c|c|c|c|c|} \hline $M_{U}$ & $M_{D}$
& $|V_{cb}|$ & $|V_{ub}|$ & sin$\,2\beta$ & $\delta^{\circ}$ &
$J\times10^{-5}$
\\ \hline $I_{i}$ & $III_{i}$ & 0.09-0.28 & 0.005-0.02 & 0.44-0.60
& 60-100 & 10-110 \\ $II_{i}$ & $III_{i}$  & 0.14-0.26 &
0.008-0.02 & 0.46-0.59 & 65-93 & 0.27-0.97 \\ $III_{i}$ & $I_{i}$
& 0.0401-0.0423 & 0.0032-0.0041 & 0.656-0.701 & 55-100 & 2.4-3.8
\\ $III_{i}$ & $II_{i}$  & 0.06-0.29 & 0.003-0.02 & 0.51-0.54 &
45-88 & 4.8-110   \\ \hline \et} \label{t4}
\end{table}

\begin{figure}
\psfig{file=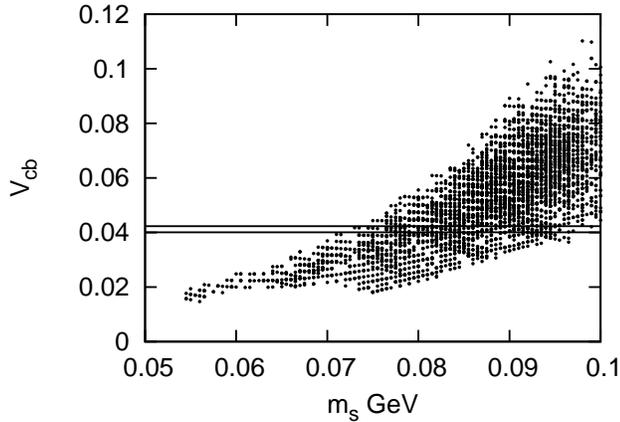, width=3.5in} \caption{Plot showing the
allowed range of $|V_{cb}|$ versus $m_s$}
  \label{vcbms}
  \end{figure}

Further, Ref.~\refcite{neelu56zeroquarks} notes that agreement
pertaining to texture 5 zero case mentioned above is valid only
when Leutwyler\cite{leut} quark masses are used. This agreement
gets ruled out in case one use the latest quark masses proposed by
Xing {\it et al.}~\cite{xingmass}. This can be understood from a
study of the dependence of the elements $|V_{cb}|$ and $|V_{ub}|$
on the mass $m_s$. For example, by giving full variation to all
other parameters within their allowed ranges, Fig.~\ref{vcbms}
depicts the plot showing the allowed range of $|V_{cb}|$ versus
$m_s$. From the figure, it immediately becomes clear that when
$m_s$ is lower than 0.075, then one is not able to obtain
$|V_{cb}|$ in the allowed range, thereby leading to the conclusion
that viability of texture 5 zero Fritzsch-like case is very much
dependent on light quark masses used.

\subsubsection{Texture 4 zero quark mass matrices}
Fritzsch-like texture 4 zero quark mass matrices are known to be
compatible with specific models of GUTs\cite{9912358,bando},
Abelian family symmetries\cite{abelian}, as well as describe the
quark mixing data quite well. In the present work our emphasis
will be to discuss the attempts\cite{s2b,cps} pertaining to the
compatibility of these mass matrices with the precisely measured
parameter sin$\,2\beta$, the issue whether one can consider
`weakly' hierarchical mass matrices to reproduce `strongly'
hierarchical mixing angles and the parameter space available to
their elements in explaining the quark mixing data. Also, it may
be mentioned that we have not discussed non Fritzsch-like texture
4 zero mass matrices as the texture two zero possibilities,
presented in Table~\ref{t2}, result into 324 texture 4 zero
possibilities making it a difficult and an elaborate task, yet to
be carried out.

It may be noted that keeping in mind that the parameter
sin$\,2\beta$ provides vital clues to the structural features of
texture specific mass matrices, comprising of hierarchy and phases
of the elements of the mass matrices, several
authors\cite{hallraisin}\cdash\cite{s2b} have explored its
implications for these. In particular, using assumption of `strong
hierarchy' of the elements of the mass matrix, having its
motivation in the hierarchy of the quark mixing angles, the
following leading order relationships between the various elements
of the mixing matrix and quark masses have been obtained in
Refs.~\refcite{hallraisin} -\refcite{branco}, \beqn \left|
{\frac{V_{ub}}{V_{cb}}} \right| =
\sqrt{\frac{m_u}{m_c}},~~~~~~~~\left| {\frac{V_{td}}{V_{ts}}}
\right| = \sqrt{\frac{m_d}{m_s}},~~~~~~~\left| V_{us} \right| =
\left| {\sqrt{\frac{m_d}{m_s}}}e^{i \phi} -
\sqrt{{\frac{m_u}{m_c}}} \right|. \label{ratios} \eeqn  Following
Particle Data Group (PDG)\cite{pdg10} definition, these further
give the expression for $\beta$ in the `strong hierarchy' case,
e.g.,
\begin{eqnarray} \beta\equiv{\rm arg}\left[-\frac{V_{cd} V_{cb}^*}{V_{td}
 V_{tb}^*}\right]={\rm arg}\left[ 1- \sqrt{\frac{m_u m_s}{m_c m_d}} e^{-i \phi}
  \right].\label{betaothers} \end{eqnarray}
Unfortunately, the value of sin$\,2\beta$ predicted by the above
formula is in quite disagreement with its present precisely known
value. In particular, with the present values of input quark
masses and by giving full variation to phase $\phi$, the maximum
value of sin$\,2\beta$ comes out to be 0.5, which is in sharp
conflict with its present PDG 2010 value $0.673 \pm
0.023$\cite{pdg10}. However, recently a detailed and comprehensive
analysis by Verma {\it et al.} has been carried out\cite{s2b}
wherein this conflict seems to have been resolved.

To begin with, the authors\cite{s2b,cps} define the `weak' and
`strong' hierarchy of the elements of the mass matrices. For
example, in case $D_i < |B_{i}| < C_i$ it would lead to a strongly
hierarchical mass matrix whereas a weaker hierarchy of the mass
matrix implies $D_{i} \lesssim |B_i| \lesssim C_i$. It may also be
added that for the purpose of numerical work, the ratio $D_i/C_i
\sim 0.01$ characterizes strong hierarchy whereas $D_i/C_i \gtrsim
0.2$ implies weak hierarchy of the elements of the mass matrices.
This can be understood by expressing these parameters in terms of
the quark masses, in particular $D_U/C_U \sim 0.01$ implies $C_U
\sim m_t$ and $D_D/C_D \sim 0.01$ leads to $C_D \sim m_b$.

The mass matrices $M_U$ and $M_D$ can be exactly diagonalized. To
facilitate the understanding of modification of the formula of
sin$\,2\beta$ given in Eq.~(\ref{betaothers}), we present some
essential details. The texture 4 zero hermitian mass matrix $M_i$
$(i = U, D)$ can be expressed as
\be
M_i= Q_i M_i^r P_i \,  \label{mk} \ee or  \be M_i^r= Q_i^{\dagger}
M_i P_i^{\dagger}\,, \label{mkr} \ee where $M_i^r$ is a real
symmetric matrix with real eigenvalues and $Q_i$ and $P_i$ are
diagonal phase matrices. The matrix $M_i^r$ can be diagonalized by
the orthogonal transformation, e.g., \be M_i^{\rm diag} = O_i^T
M_i^{r} O_i
 \,,   \label{o2}\ee
where \be M_i^{\rm diag} = {\rm diag}(m_1,\,-m_2,\,m_3)\,, \ee the
subscripts 1, 2 and 3 referring respectively to $u,\, c$ and $t$
for the up sector as well as $d,\,s$ and $b$ for the down sector.
Using the invariants, tr$M_i^r$, tr ${M_i^r}^2$ and det$M_i^r$,
the values of the elements of the mass matrices $A_{i}$, $B_{i}$
and $C_{i}$, in terms of the free parameter $D_{i}$ and the quark
masses are given as
 \beqn
  C_i& = &(m_1-m_2+m_3-D_i)\,, \\
   |A_i| &=&(m_1 m_2 m_3/C_i)^{1/2}\,, \\
 |B_i| &= &
 [(m_3-m_2-D_i)(m_3+m_1-D_i)(m_2-m_1+D_i)/C_i]^{1/2}\,.
\label{elements} \eeqn The exact diagonalizing transformation
$O_i$ is expressed as
 \be O_i
= \left( {\renewcommand{\arraystretch}{1.7}
 \ba{ccc}
  \pm {\sqrt \frac{m_2 m_3 (C_i-m_1)}{(m_3-m_1)(m_2+m_1)C_i} } &
   \pm  {\sqrt \frac{m_1 m_3 (C_i+m_2)}{C_i (m_2+m_1) (m_3+m_2)}} &
  \pm{\sqrt \frac{m_1 m_2 (m_3-C_i)}{C_i (m_3+m_2)(m_3-m_1)}}\\
 \pm{\sqrt \frac{m_1 (C_i-m_1)}{(m_3-m_1)(m_2+m_1)} } &
 \mp{\sqrt \frac{m_2 (C_i+m_2)}{(m_3+m_2)(m_2+m_1)} }&
 \pm{\sqrt \frac{m_3(m_3-C_i)}{(m_3+m_2)(m_3-m_1)} } \\
 \mp{\sqrt \frac{m_1 (m_3-C_i)(C_i+m_2)}{C_i(m_3-m_1)(m_2+m_1)} } &
 \pm{\sqrt \frac{m_2 (C_i-m_1) (m_3-C_i)}{C_i
(m_3+m_2)(m_2+m_1)} } &
  \pm{\sqrt \frac{m_3 (C_i-m_1)(C_i+m_2)}{C_i
(m_3+m_2)(m_3-m_1)}}  \ea} \right). \label{ou1} \ee \vskip 0.5cm

It may be noted that while finding the diagonalizing
transformation $O_i$, one has the freedom to choose several
equivalent possibilities of phases. Similarly, while normalizing
the diagonalized matrix to quark masses, one again has the freedom
to choose the phases for the quark masses. This is due to the fact
that the diagonalizing transformations of $M_U$ and $M_D$ occur in
a particular manner in the weak charge current interactions of
quarks to give the CKM mixing matrix. As is usual, the phase of
$m_2$ has been chosen to be negative facilitating the
diagonalization process as well as the construction of the CKM
matrix. This is one of the possibilities considered by Xing and
Zhang\cite{xingzhang}, the others are related and are all
equivalent, these only redefining the phases $\phi_1$ and $\phi_2$
which in any case are arbitrary. In the work being discussed, the
authors have chosen the possibility \be O_i= \left( \ba{ccc}
~~O_i(11)& ~~O_i(12)& ~O_i(13)
\\
 ~~O_i(21)& -O_i(22)& ~O_i(23)\\
     -O_i(31) & ~~O_i(32) & ~O_i(33) \ea \right). \ee

As already mentioned, the CKM mixing matrix can be obtained using
$O_{U(D)}$ through the relation $V_{\rm CKM}= O_U^T (P_U
P_D^{\dagger}) O_D$. Explicitly, the elements of the CKM mixing
matrix can be expressed as
\be
V_{l m} = O_{1 l}^U O_{1 m}^D e^{-i \phi_1} + O_{2 l}^U O_{2 m}^D
 + O_{3 l}^U O_{3 m}^D e^{i \phi_2},
\label{vckmelement1} \ee
 where the subscripts $l$ and $m$ run respectively over $u,\, c$, $t$  and $d,\,s$, $b$
and $\phi_1=  \alpha_U- \alpha_D$, $\phi_2= \beta_U- \beta_D$.

To evaluate the parameter sin$\,2\beta$, the elements $V_{cd}$, $
V_{cb}$, $V_{td}$ and $V_{tb}$ are required to be known. Using the
above equation, these elements can be easily found, e.g.,
 \beqn
  V_{cd}=\sqrt{\frac{m_u m_t (-D_u + m_u +
m_t)}{C_u(m_u+m_c)(m_c+m_t)}} \sqrt{\frac{m_s m_b(-D_d + m_b -
m_s)}{C_d(m_b-m_d)(m_s+m_d)}}~e^{-i \phi_1}  \nonum
\\- \sqrt{\frac{m_c (-D_{u} + m_u +
m_t)}{(m_u+m_c)(m_c+m_t)}} \sqrt{\frac{m_d (-D_d + m_b -
m_s)}{(m_b-m_d)(m_s+m_d)}}~~~~~~~~~~~~~~~~~~~ \nonum
\\ - \sqrt{\frac{m_c (-D_u + m_t - m_c)(D_u - m_u +
m_c)}{C_u(m_u+m_c)(m_t+m_c)}} \times
~~~~~~~~~~~~~~~~~~~~~~~~~~\nonum
\\ \sqrt{\frac{m_d (D_d - m_d + m_s)(-D_d + m_d +
m_b)}{C_d(m_b-m_d)(m_s+m_d)}}~e^{i
\phi_2}.~~~~~~~~~~~~~~~~~~~~~~~~~\label{fvcd}
  \eeqn
In case one uses the above complicated expression for $V_{cd}$ as
well as similar expressions of the other elements to evaluate
sin$\,2\beta$, one can find that these would yield a long and
complicated formula from which it would be difficult to understand
the implications on the phases and other parameters of the mass
matrices. To derive a simple and informative formula, one needs to
first rewrite the diagonalizing transformation $O_i$ keeping in
mind $m_3 \gg m_2 \gg m_1$ and the element of the mass matrix $C_i
\gg m_1$, which is always valid without any dependence on the
hierarchy of the elements of the mass matrices. It may be
mentioned that this approximation induces less than a fraction of
a percentage error in the numerical results. The structure of
$O_i$ can be simplified and expressed as
 \be O_i
= \left( {\renewcommand{\arraystretch}{1.7}
 \ba{ccc}
  1&
   \zeta_{1 i} {\sqrt \frac{m_1}{m_2}} &
 {\frac{\zeta_{2 i}}{\zeta_{3 i}}} {\sqrt \frac{m_1 m_2}{m_3^2}} \\
  \zeta_{3 i} {\sqrt \frac{m_1}{m_2}} &
  - \zeta_{1 i} \zeta_{3 i}&
 \zeta_{2 i} \\
  - \zeta_{1 i} \zeta_{2 i} {\sqrt \frac{m_1}{m_2}} &
  \zeta_{2 i} &
  \zeta_{1 i} \zeta_{3 i} \ea} \right), \label{oisimple2} \ee
  where the three parameters $\zeta_{1 i}$, $\zeta_{2 i}$, $\zeta_{3
i}$, with $i$ denoting $U$ and $D$ are given by \be \zeta_{1 i}=
\sqrt{1 + \frac{m_2}{C_i}},~~~~\zeta_{2 i}= \sqrt{1 -
\frac{C_i}{m_3}},~~~~\zeta_{3 i}= \sqrt{\frac{C_i}{m_3}}. \ee
Making use of this equation, along with relation
(\ref{vckmelement1}), the following elements needed to evaluate
$\beta$ are obtained as
 \be V_{cd} = \zeta_{1 U} \sqrt{\frac{m_u}{m_c}}
e^{-i \phi_1} - \sqrt{\frac{m_d}{m_s}}~ \left[ \zeta_{1 U}\,
\zeta_{3 U}\, \zeta_{3 D} + \zeta_{2 U}\, \zeta_{1 D}\, \zeta_{2
D}\, e^{i \phi_2} \right], \label{cvcd}\ee

 \be V_{cb} = \frac{\zeta_{1 U} \zeta_{2 D}}{\zeta_{3 D}} \sqrt{\frac{m_u m_d m_s}{m_c m_b^2}}
e^{-i \phi_1} - \left[ \zeta_{1 U}\, \zeta_{3 U}\, \zeta_{2 D} -
\zeta_{2 U}\, \zeta_{1 D}\, \zeta_{3 D}\, e^{i \phi_2} \right],
\label{cvcb}\ee

 \be V_{td} = \frac{\zeta_{2 U}}{\zeta_{3 U}} \sqrt{\frac{m_u m_c}{m_t^2}}
e^{-i \phi_1} + \sqrt{\frac{m_d}{m_s}}~\left[ \zeta_{2 U}\,
\zeta_{3 D} - \zeta_{1 U}\, \zeta_{3 U}\, \zeta_{1 D}\, \zeta_{2
D}\, e^{i \phi_2} \right], \label{cvtd}\ee

\be V_{tb} = \frac{\zeta_{2 U} \zeta_{2 D}}{\zeta_{3 U} \zeta_{3
D}} \sqrt{\frac{m_u m_c m_d m_s}{m_t^2 m_b^2}} e^{-i \phi_1} +
\left[ \zeta_{2 U}\, \zeta_{2 D} + \zeta_{1 U}\, \zeta_{3 U}\,
\zeta_{1 D}\, \zeta_{3 D}\, e^{i \phi_2} \right]. \label{cvtb}\ee
A general look on the above elements clearly shows that the above
relations are not only more compact but also more useful to view
the dependence of these CKM matrix elements on the quark masses
and phases. Using these elements, after some non trivial algebra,
one arrives at the following expression of $\beta$, wherein its
dependence on the quark masses and the elements of the quark mass
matrices is visible in a simple and clear manner, e.g.,
\begin{eqnarray} \beta\equiv{\rm arg}\left[-\frac{V_{cd}
V_{cb}^*}{V_{td}
 V_{tb}^*}\right]={\rm arg}\left[ \left( 1- \sqrt{\frac{m_u m_s}{m_c m_d}} e^{-i (\phi_1 + \phi_2)} \right)
 \left( \frac{1-r_2 e^{i \phi_2}}{1-r_1 e^{i \phi_2}} \right)  \right],
 \label{beta} \end{eqnarray} where the parameters $r_1$ and $r_2$
can be expressed in terms of the quark masses and the elements of
the quark mass matrices via the relations,
 \be r_1 = \frac{\zeta_{1 U} \,
\zeta_{3 U}\, \zeta_{1 D}\, \zeta_{2 D}}{\zeta_{2 U}\, \zeta_{3
D}}~~~~~~~~~{\rm and}~~~~~~~~~ r_2 = \frac{\zeta_{1 U} \, \zeta_{3
U}\, \zeta_{2 D}}{\zeta_{2 U}\, \zeta_{1 D}\, \zeta_{3 D}}. \ee

The relationship given in Eq.~(\ref{beta}) is an `exact' formula
emanating from texture 4 zero mass matrices, incorporating both
the phases. This formula has several interesting aspects. Apart
from clearly underlying the dependence of small quark masses and
the phases $\phi_1$ and $\phi_2$, it also clearly establishes the
modification of the earlier formula. It is very easy to check that
the earlier formula can be easily deduced from the present one by
using the strong hierarchy assumption which essentially translates
to $\zeta_{1 D} \simeq \zeta_{1 U} \simeq 1$, further implying
$r_{1}$=$r_{2}$ and $\left( \frac{1-r_2 e^{i \phi_2}}{1-r_1 e^{i
\phi_2}} \right)=1$. The phase of the earlier formula can be
obtained by identifying $\phi_1 + \phi_2$ as $\phi$, taking values
from 0 to 2$\pi$.

It also needs to be re-emphasized that while arriving at the above
`exact' relationship, the hierarchy of the quark masses, e.g.,
$m_t \gg m_u$ and $m_b \gg m_d$ has been considered as well as
$m_3 \gg m_2 \gg m_1$ and the element of the mass matrix $C_i \gg
m_1$ have been used, these being valid in both weak and strong
hierarchy cases. It may also be added that the formula remains
valid for both the weak hierarchy of the elements of the mass
matrices as well as for the strong hierarchy assumption.
Interestingly, the modification to the earlier formula contributes
only when $\phi_2 \neq 0$, implying thereby that both the phases
of the mass matrices might play an important role in achieving the
agreement with data.

In order to investigate the implications of the formula given in
Eq.~(\ref{beta}) on the structural features of the mass matrices
and the CKM parameters, as a first step one can find the range of
sin$\,2\beta$ predicted by the above formula. In this regard, the
following ranges of quark masses at the energy scale of $M_z$ have
been adopted by Refs.~\refcite{s2b} and \refcite{cps}, e.g., \beqn
m_u=0.8 -1.8\, {\rm MeV},~~ m_d=1.7 -4.2\, {\rm MeV},~~ m_s=40.0
-71.0\, {\rm MeV},~~~~~~~~\nonumber\\ ~~~~~~~m_c=0.6- 0.7\, {\rm
GeV},~~ m_b=2.8- 3.0\, {\rm GeV},~~ m_t=169.5- 175.5\, {\rm GeV}.
~~~~~~~~\label{qmasses1} \eeqn With these inputs and the `exact'
formula given in Eq.~(\ref{beta}), one can evaluate sin$\,2\beta$
by giving full variation to the phases $\phi_1$ and $\phi_2$, the
parameters $D_U$ and $D_D$ have been given wide variation in
conformity with the natural hierarchy of the elements of the mass
matrices, e.g., $D_i
< C_i$ for $i=U, D$. The sin$\,2\beta$ so evaluated comes out to
be
\be
\\{\rm sin}\,2\beta = 0.4105 - 0.7331.\ee Interestingly, one finds that
the above value is inclusive of its experimental range $0.681\pm
0.025$. This clearly indicates that the formula which includes
weak hierarchy as well as additional phase factors plays a crucial
role in bringing out reconciliation between texture 4 zero mass
matrices and the present precise value of sin$\,2\beta$.

 \begin{figure}
\psfig{file=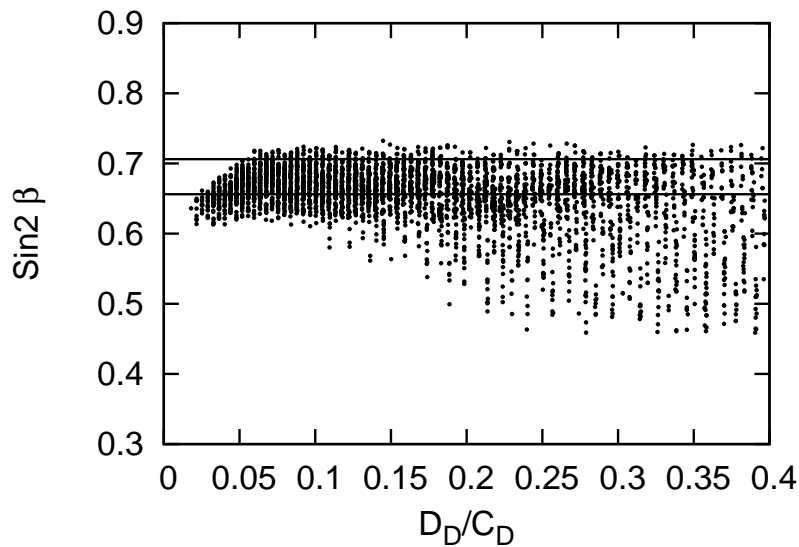, width=4.5in} \caption{Plot showing
variation of CP violating parameter sin$\,2\beta$ versus hierarchy
characterizing ratio $D_D/C_D$}
  \label{s2bddcd}
  \end{figure}

After having shown the compatibility of texture 4 zero mass
matrices with sin$\,2\beta$, Ref. \refcite{s2b} examines the
constraints imposed on the ratio $D_i/C_i$ for $i=U, D$,
characterizing hierarchy, as well as on the phases $\phi_1$ and
$\phi_2$ of the mass matrices. As a first step, using the relation
obtained earlier, one can investigate the role of hierarchy by
plotting sin$\,2\beta$ against the ratio $D_D/C_D$, shown in
Fig.~\ref{s2bddcd}. Several interesting conclusions follow from
the graph. It can be easily noted that when $D_D/C_D$ $<$ 0.02,
one is not able to reproduce any point within the 1$\sigma$ range
of sin$\,2\beta$, even after giving full variation to all the
other parameters. It may be of interest to mention that the
earlier attempts\cite{hallraisin}\cdash\cite{branco} had
considered a value of $D_D/C_D$ $\lesssim$ 0.02, thereby resulting
in the incompatibility of texture 4 zero mass matrices with
sin$\,2\beta$. From the figure it can be easily checked that only
for $D_D/C_D$ $>$ 0.05, full range of sin$\,2\beta$ is reproduced.
This clearly shows that as one deviates from strong hierarchy
characterized by the ratio $D_D/C_D \sim 0.01$ towards weak
hierarchy given by $D_D/C_D \gtrsim 0.1$, one is able to reproduce
the results. It may be mentioned that although the graph has been
plotted for $D_D/C_D$ up to 0.4, however the same pattern is
followed up to $D_D/C_D \sim 0.6$, beyond which the basic
structure of the mass matrix is changed. It may be added that the
corresponding graph of $D_U/C_U$ is also very much similar. One
would also like to emphasize that the agreement between
sin$\,2\beta$ and higher values of $D_i/C_i$ does not spoil the
overall agreement of texture 4 zero mass matrices with the CKM
matrix derived earlier. This brings out an extremely important
point as the conventional belief was that the hierarchical quark
mixing angles can be reproduced only by strong hierarchy mass
matrices.

  \begin{figure}
\psfig{file=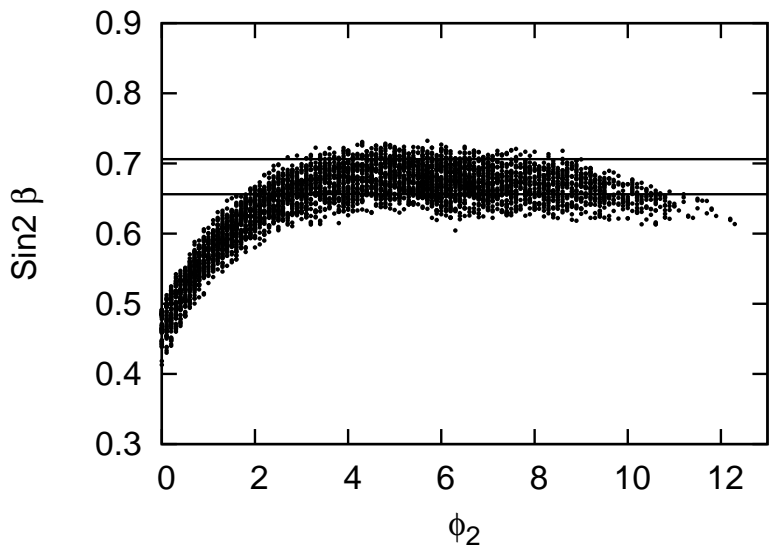, width=4.5in} \caption{Plot showing
variation of CP violating parameter sin$\,2\beta$ versus the phase
$\phi_2$}
  \label{s2bph2}
  \end{figure}

Regarding the phases $\phi_1$ and $\phi_2$ of the mass matrices,
in Fig.~\ref{s2bph2} graph of sin$\,2\beta$ versus angle $\phi_2$
has been plotted. The graph clearly illustrates the crucial role
played by the phase $\phi_2$ in bringing out agreement of texture
4 zero mass matrices and the precisely known sin$\,2\beta$. It is
interesting to emphasize that despite giving full variation to
other parameters, for $\phi_2=0^{\rm o}$ one is not able to
reproduce sin$\,2\beta$ within the experimental range.

Coming to the issue whether `weakly hierarchical' quark mass
matrices are able to reproduce the mixing data which involves
strongly hierarchical parameters. In this context, it may be
mentioned that in the case of quark mass matrices, usually the
elements are assumed to follow `strong hierarchy', whereas there
is no such compulsion for the leptonic mass matrices. Therefore,
in case one needs to invoke quark-lepton
unification\cite{qlepuni}, this issue becomes interesting. This is
all the more important as the texture 4 zero mass matrices perhaps
provide the simplest parallel structure for quark and lepton mass
matrices which are compatible with the low energy data.

Realizing the importance of Fritzsch-like hermitian texture 4 zero
mass matrices in the context of quarks, a few years back Xing and
Zhang\cite{xingzhang} have attempted to find the parameter space
of the elements of these mass matrices. Their analysis has
provided good deal of information regarding the space available to
various parameters as well as have provided valuable insight into
the `structural features' of texture 4 zero mass matrices. In this
context, it may be noted that the hierarchy of the elements of the
mass matrices is largely governed by the (2,2) element of the
matrix. In their analysis, attempt has been made to go somewhat
beyond the minimal values of this element, corresponding to the
`strong hierarchy' case, however in case one has to consider the
`weak hierarchy' case as well then there seems a further need to
consider a still larger range for this element.

Recently, an interesting analysis has been carried out by Verma
{\it et al.}~\cite{cps} which updates and broadens the scope of
the analysis carried out by Xing and Zhang\cite{xingzhang} as well
as examines the implications of recent precision measurements on
the structural features of texture 4 zero mass matrices. The
analysis incorporates the latest values of quark masses and their
ratios and full variation has been given to the phases associated
with the mass matrices $\phi_1$ and $\phi_2$, the parameters $D_U$
and $D_D$ have been given wide variation in conformity with the
hierarchy of the elements of the mass matrices e.g., $D_i < C_i$
for $i=U, D$. The extended range of these parameters allows the
calculations for the case of weak hierarchy of the elements of the
mass matrices as well.

\begin{figure}[hbt]
\vspace{0.12in}
\centerline{\epsfysize=1.55in\epsffile{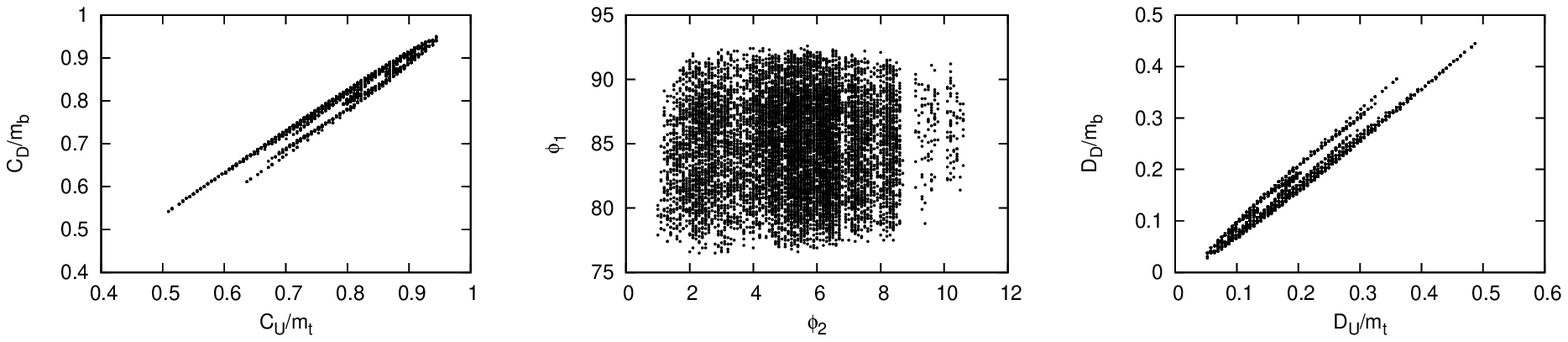}}
\vspace{0.08in}
\caption{Plots showing the allowed ranges of (a)
$C_{\rm U}/m_t$ versus $C_{\rm D}/m_b$, (b) $\phi_1$ versus
$\phi_2$ and (c) $D_{\rm U}/m_t$ versus $D_{\rm D}/m_b$}
  \label{cpsfigs}
  \end{figure}

The key findings of Ref.~\refcite{cps} can be understood from the
following figures, reproduced here for the readers' convenience.
In Fig.~\ref{cpsfigs}(a), $C_{U }/m_t$ versus $C_{ D}/m_b$ has
been plotted revealing that both $C_{U}/m_t$ as well as $C_{
D}/m_b$ take values from $\sim 0.55-0.95$, which interestingly
indicates the ratios being almost proportional. Also, the figure
gives interesting clues regarding the role of strong and weak
hierarchy. In particular, one finds that in case one restricts to
the assumption of strong hierarchy then these ratios take large
values around $0.95$. However, for the case of weak hierarchy, the
ratios $C_{U }/m_t$ and $C_{ D}/m_b$ take much larger number of
values, in fact almost the entire range mentioned above, which are
compatible with the data.

In Fig.~\ref{cpsfigs}(b), the plot of $\phi_1$ versus $\phi_2$ has
been presented. Interestingly, the present refined inputs limit
the ranges of the two phases to $\phi_1 \sim 76^{\rm o} - 92^{\rm
o}$ and $\phi_2 \sim 1^{\rm o} - 11^{\rm o}$. Keeping in mind that
full variation has been given to the free parameters $D_U$ and
$D_D$, corresponding to both strong as well as weak hierarchy
cases, it may be noted that the allowed ranges of the two phases
come out to be rather narrow. In particular, for the strong
hierarchy case one gets $\phi_2 \sim 10^{\rm o}$, whereas for the
case of weak hierarchy $\phi_2$ takes almost its entire range
mentioned above. Also, the analysis indicates that although
$\phi_1 \gg \phi_2$, still both the phases are required for
fitting the mixing data.

As a next step, in Fig.~\ref{cpsfigs}(c) $D_{U }/m_t$ versus $D_{
D}/m_b$ has been given, representing an extended range of the
parameters $D_U$ and $D_D$. A closer look at the figure reveals
both $D_{U}/m_t$ as well as $D_{ D}/m_b$ take values $\sim
0.05-0.5$. The lower limit of the range i.e. when the ratios
$D_{U}/m_t$ and $D_{ D}/m_b$ are around $0.05$ corresponds to
strong hierarchy amongst the elements of the mass matrices,
whereas when the elements have weak hierarchy then these ratios
take a much larger range of values. From this one may conclude
that in the case of strongly hierarchical elements of the texture
4 zero mass matrices, one has limited compatibility of these
matrices with the quark mixing data, whereas the weakly
hierarchical ones indicate the compatibility for much broader
range of the elements.

The above discussion can also be understood by the construction of
the mass matrices. However, as the phases of the elements of the
mass matrices can be separated out, as can be seen from
Eq.~(\ref{mkr}), one needs to consider $M_i^r$ ($i=U, D$) instead
of $M_i$. The ranges of the elements of these matrices $M_U^r$ and
$M_D^r$ are as follows \be M_U^r = m_t \left( \ba {ccc} 0 &
0.000174-0.000252 & 0
\\ 0.000174-0.000252 &0.0464-0.4870 & 0.2184-0.5017
\\0 & 0.2184-0.5017 & 0.5094-0.9500 \ea \right), \label{mu} \ee

\be M_D^r = m_b \left( \ba {ccc} 0 & 0.003555-0.006154 & 0
\\ 0.003555-0.006154 &0.0276-0.4448 & 0.2194-0.5044
\\0 & 0.2194-0.5044 & 0.5418-0.9505 \ea \right). \label{md} \ee
It may be noted that the elements of the mass matrices $A_i$,
$B_i$, $C_i$ and $D_i$ satisfy the relation $|B_i|^2-C_i D_i
\simeq m_2 m_3$ for both the strong and the weak hierarchy cases
characterized respectively by $D_i < |B_{i}|
< C_i$ and $D_{i} \lesssim |B_i| \lesssim C_i$. This relation can
be numerically checked from the above mentioned mass matrices in
Eqs.~(\ref{mu}) and (\ref{md}). The above constraint on the
elements of the mass matrices as well as the ranges of various
ratios , particularly in the case of weak hierarchy, provide an
interesting possibility for checking the viability of various mass
matrices formulated at the GUTs scale or obtained using horizontal
symmetries. From a different point of view, this can also provide
vital clues to the formulation of mass matrices which are in
agreement with the low energy data.

After constructing the mass matrices, the authors have also
constructed the corresponding CKM mixing matrix and compared it
with the one arrived through global analysis. The CKM mixing
matrix obtained by the authors is as follows
 \be V_{{\rm CKM}} = \left( \ba{ccc}
  0.9738-0.9747 &~~~~   0.2236-0.2274 &~~~~  0.00357-0.00429 \\
 0.2234-0.2274  &~~~~   0.9729-0.9739    &~~~~  0.0401-0.0423\\
0.0057-0.0114  &~~~~  0.0388-0.0420 &~~~~  0.9991-0.9992 \ea
\right). \label{3sm} \ee A general look at the matrix reveals that
the ranges of CKM elements obtained here are quite compatible with
those obtained by recent global
analyses\cite{pdg10}\cdash\cite{hfag}. The Jarlskog's rephasing
invariant parameter $J$ has also been evaluated which comes out to
be \be J=(1.807 - 3.977) 10^{-5} \label{j2}. \ee Further, using
this value of $J$ one can obtain the following range of the CP
violating phase $\delta$
 \be \delta=28.8.8^{\rm o}- 110.4^{\rm o}. \label{delta2}\ee
The above mentioned ranges of the parameters $J$ and the phase
$\delta$ are inclusive of the values given by latest PDG
data\cite{pdg10}.

\subsection{Lepton mass matrices} In the leptonic sector, the observation of neutrino oscillations has
added another dimension to the issue of fermion masses and mixing.
In fact, the pattern of neutrino masses and mixings seems to be
vastly different from that of quarks. At present, the available
neutrino oscillation data does not throw any light on the neutrino
mass hierarchy, which may be normal/ inverted and may even be
degenerate. Further, the situation becomes complicated when one
realizes that neutrino masses are much smaller than charged
fermion masses as well as it is not clear whether neutrinos are
Dirac or Majorana particles. The situation becomes more
complicated in case one has to understand the quark and neutrino
mixing phenomena in a unified manner.

In the literature, several attempts have been made to formulate
the phenomenological mass matrices considering charged leptons to
be diagonal, usually referred to as the flavor basis
case\cite{0307359}. In case the PMNS matrix is known, then, in
principle, the mass matrices for Dirac and Majorana case can be
expressed as
\be
m_{\nu}^{\alpha \beta} = \sum_{i} (U^*)_{\alpha i} m_i (U^\dag)_{
i \beta }\,. \ee It is clear from the above equation that the
elements of the neutrino mass matrix in the flavor basis can be
determined in case we have complete knowledge about the mass
eigenvalues as well as the elements of the mixing matrix. In the
case of Dirac neutrinos, one would require knowledge of the three
mixing angles and one Dirac-like CP violating phase $\delta_l$.
For Majorana neutrinos, things become more complicated as one
needs to have additional information about the two Majorana phases
also. It is likely in the near future the precision regarding the
measurement of the three mixing angles would increase and also
some information about the phase $\delta_l$ becomes available.
Similarly, it is also likely that one may be able to get some
clues about the overall scale of neutrino masses. However,
extracting information about Majorana phases from neutrinoless
double $\beta$ decay experiments would be a challenging task.

To understand the pattern of neutrino masses and mixings, texture
zero approach has been tried with good deal of
success\cite{0307359,leptex1}\cdash\cite{neelu6zerolep}. In the
light of quark-lepton unification hypothesis, advocated by
Smirnov\cite{qlepuni}, in the sequel, parallel to the case of
quarks, we would restrict ourselves to the Fritzsch-like as well
as non Fritzsch-like texture specific lepton mass matrices. An
important attempt to formulate texture specific lepton mass
matrices, however in the flavor basis, was carried out by
Frampton, Glashow and Marfatia\cite{framp}, wherein assuming a
complex symmetric Majorana mass matrix and considering seven
possible texture 2 zero cases, they carried out the implications
of these for the neutrino oscillation data. Further, without
considering the flavor basis, for normal hierarchy of neutrino
masses, Fukugita, Tanimoto and Yanagida\cite{fuku} carried out an
analysis of Fritzsch-like texture 6 zero mass matrices. Similarly,
Zhou and Xing\cite{zhou} also carried out a systematic analysis of
all possible texture 6 zero mass matrices for Majorana neutrinos
with an emphasis on normal hierarchy of neutrino masses.
Recently,\cite{ourneut6zero,ourneut4zero,ourneut6zero2} for all
possible hierarchies of neutrino masses, for both Majorana as well
as Dirac neutrinos, detailed analyses of Fritzsch-like texture 6,
5 and 4 zero mass matrices was carried out. Further, in view of
absence of any theoretical justification for Fritzsch-like mass
matrices, recent attempts\cite{neelu6zerolep} were made to
consider non Fritzsch-like mass matrices for neutrinos. Before we
present the essential details of some of these analyses, in the
following section, we first present the relation between lepton
mass matrices and mixing matrix.

\subsubsection{Relationship of lepton mass matrices and mixing
matrix} The observation of neutrino oscillation phenomenon which
essentially implies the flavor conversion of neutrinos is similar
to the quark mixing phenomenon. This possibility of flavor
conversion was originally examined by B. Pontecorvo and further
generalized by Maki, Nakagawa and
Sakata\cite{pmns1}\cdash\cite{pmns4}. The emerging picture that
neutrinos are massive and therefore mix has been proved beyond any
doubt and provides an unambiguous signal of NP.

In the case of neutrinos, the generation of masses is not
straight-forward as they may have either the Dirac masses or the
more general Dirac-Majorana masses. A Dirac mass term can be
generated by the Higgs mechanism with the standard Higgs doublet
by introducing singlet right handed neutrinos in the SM. In this
case, the neutrino mass term can be written as
\begin{equation}
 \overline{\nu}_{a_{L}} M_{\nu D} {\nu}_{a_{R}} + h.c.,
\end{equation}
where $a$ = $e$, $\mu$, $\tau$ and $\nu_e$, $\nu_\mu$, $\nu_\tau$
are the flavor eigenstates. $M_{\nu D}$ is a complex $3\times 3$
Dirac mass matrix. The mass term mentioned above would also be
characterized by the same symmetry breaking scale such as that of
charged leptons or quarks, therefore, in this case very small
masses of neutrinos would be very unnatural from the theory point
of view. On the other hand, the neutrino might be a Majorana
particle which is defined as is its own anti-particle and is
characterized by only two independent particle states of the same
mass ($\nu^{~}_{\rm L}$ and $\bar{\nu}^{~}_{\rm R}$ or
$\nu^{~}_{\rm R}$ and $\bar{\nu}^{~}_{\rm L}$). A Majorana mass
term, which violates both the law of total lepton number
conservation and that of individual lepton flavor conservation,
can be written either as
\begin{equation}
 \frac{1}{2} \overline{\nu}_{a_{L}} M_L {\nu}^c_{a_R}
+ h.c.~~~~~~~~~{\rm or~as}~~~~~~~~~\frac{1}{2}
\overline{\nu}^c_{a_L} M_R {\nu}_{a_R} + h.c.,
\end{equation}
where $M_l$ and $M_R$ are complex symmetric matrices.

A simple extension of the SM is to include one right handed
neutrino in each of the three lepton families, while the
Lagrangian of the electroweak interactions is kept invariant under
$SU(2)_L \times U(1)_Y$ gauge transformations. This can be shown
to lead to Dirac-Majorana mass terms which further lead to the
famous seesaw mechanism\cite{seesaw1}\cdash\cite{seesaw5} for the
generation of small neutrino masses, e.g.,
 \be M_{\nu}=-M_{\nu D}^T\,(M_R)^{-1}\,M_{\nu D},
\label{seesaweq2} \ee \noindent where $M_{\nu D}$ and $M_R$ are
respectively the Dirac neutrino mass matrix and the right handed
Majorana neutrino mass matrix. The seesaw mechanism is based on
the assumption that, in addition to the standard Higgs mechanism
of generation of the Dirac mass term, there exists a beyond the SM
mechanism of generation of the right handed Majorana mass term,
which changes the lepton number by two and is characterized by a
mass $M \gg m$. The Dirac mass term mixes the left handed field
$\nu_L$, the component of a doublet, with a single field
$(\nu^c)_R$. As a result of this mixing the neutrino acquires
Majorana mass, which is much smaller than the masses of leptons or
quarks.

Similar to the quark sector, the lepton mass matrices can be
diagonalized by bi-unitary transformations and the corresponding
mixing matrix obtained, known as Pontecorvo-Maki-Nakagawa-Sakata
(PMNS) or lepton mixing matrix\cite{pmns1}\cdash\cite{pmns4} is
given as
\be
V_{\rm PMNS}= V^{\dagger}_{l_{L}} V_{\nu_{L}}. \ee

The PMNS matrix expresses the relationship between the neutrino
mass eigenstates and the flavor eigenstates, e.g.,
  \be
\left( \ba{c} \nu_e
\\ \nu_{\mu}
\\ \nu_{\tau} \ea \right)
  = \left( \ba{ccc} V_{e1} & V_{e2} & V_{e3} \\ V_{\mu 1} & V_{\mu 2} &
  V_{\mu 3} \\ V_{\tau 1} & V_{\tau 2} & V_{\tau 3} \ea \right)
 \left( \ba {c} \nu_1\\ \nu_2 \\ \nu_3 \ea \right),  \label{nm1}  \ee
where $ \nu_{e}$, $ \nu_{\mu}$, $\nu_{\tau}$ are the flavor
eigenstates, $ \nu_1$, $ \nu_2$, $ \nu_3$ are the mass eigenstates
and the $3 \times 3$ mixing matrix is the leptonic mixing
matrix\cite{pmns1}\cdash\cite{pmns4}. For the case of three Dirac
neutrinos, in the standard PDG parameterization\cite{pdg10},
involving three angles $\theta_{12}$, $\theta_{23}$, $\theta_{13}$
and the Dirac-like CP violating phase ${\delta}_l$ the mixing
matrix has the form
\begin{eqnarray}
V_{\rm PMNS}=   \left (
  \begin{array}{ccc}
    c_{12} c_{13} & s_{12} c_{13} & s_{13} e^{-i {\delta}_l} \\
    -s_{12} c_{23} - c_{12} s_{23} s_{13} e^{i {\delta}_l} & c_{12} c_{23} - s_{12}
    s_{23} s_{13}e^{i {\delta}_l} & s_{23} c_{13} \\
    s_{12} s_{23} - c_{12} c_{23} s_{13}e^{i {\delta}_l} & -c_{12} s_{23} - s_{12}
    c_{23} s_{13}e^{i {\delta}_l} & c_{23} c_{13}
  \end{array}
  \right ),
\label{ch1pmns2}
\end{eqnarray}
with $s_{ij} = {\rm sin}\theta_{ij}$, $c_{ij} = {\rm
cos}\theta_{ij}$. In the case of the Majorana neutrinos, there are
extra phases which cannot be removed. Therefore, the above matrix
takes the following form \beqn {\left( \ba{ccc} c_{12} c_{13} &
s_{12} c_{13} & s_{13}e^{-i {\delta}_l} \\ - s_{12} c_{23} -
c_{12} s_{23} s_{13} e^{i {\delta}_l} & c_{12} c_{23} - s_{12}
s_{23} s_{13} e^{i {\delta}_l} & s_{23} c_{13}
\\ s_{12} s_{23} - c_{12} c_{23} s_{13} e^{i \delta_l} & - c_{12}
s_{23} - s_{12} c_{23} s_{13} e^{i { \delta}_l} & c_{23} c_{13}
\ea \right)} \left( \ba{ccc} e^{i \alpha_1/2} & 0 & 0 \\ 0 &e^{i
\alpha_2/2} & 0 \\ 0 & 0  & 1 \ea \right),\nonumber \\ \eeqn where
$\alpha_1$ and $\alpha_2$ are the Majorana phases which do not
play any role in neutrino oscillations.

\subsubsection{Texture 6 zero lepton mass matrices \label{lepmm}}
Parallel to the case of quarks, there are a total number of 144
possible cases of texture 6 zero mass matrices. In the case of
lepton mass matrices, for each of the 144 combinations, there are
6 cases each corresponding to normal/ inverted hierarchy and
degenerate scenario of neutrino masses for Majorana neutrinos as
well as Dirac neutrinos, leading to a total of 864 cases.
Recently, a detailed analysis of these cases was carried out by
Mahajan {\it et al.}~\cite{neelu6zerolep}, revealing several
interesting points. In particular, their investigations for Dirac
neutrinos show that there are no viable texture 6 zero lepton mass
matrices for normal/ inverted hierarchy as well as degenerate
scenario of neutrino masses. For the case of Majorana neutrinos,
again all the cases pertaining to inverted hierarchy and
degenerate scenario of neutrino masses are ruled out. For normal
hierarchy of Majorana neutrinos, the analysis reveals that out of
144, only 16 combinations are compatible with current neutrino
oscillation data at $3\sigma$ C.L.. For a detailed discussion, we
refer the readers to Ref.~\refcite{neelu6zerolep}.

Keeping in mind the important role played by Fritzsch-like mass
matrices in understanding the quark and neutrino mixing
 phenomenology, in the present work, we would like to present the
details regarding this case. Recently, using Fritzsch-like texture
6 zero lepton mass matrices, detailed predictions for cases
pertaining to normal/inverted hierarchy as well as degenerate
scenario of neutrino masses have been carried out for the case of
Majorana neutrinos by Randhawa {\it et al.}~\cite{ourneut6zero}
and for Dirac neutrinos by Ahuja {\it et
al.}~\cite{ourneut6zero2}.

\begin{figure}[hbt]
\centerline{\epsfysize=4.in\epsffile{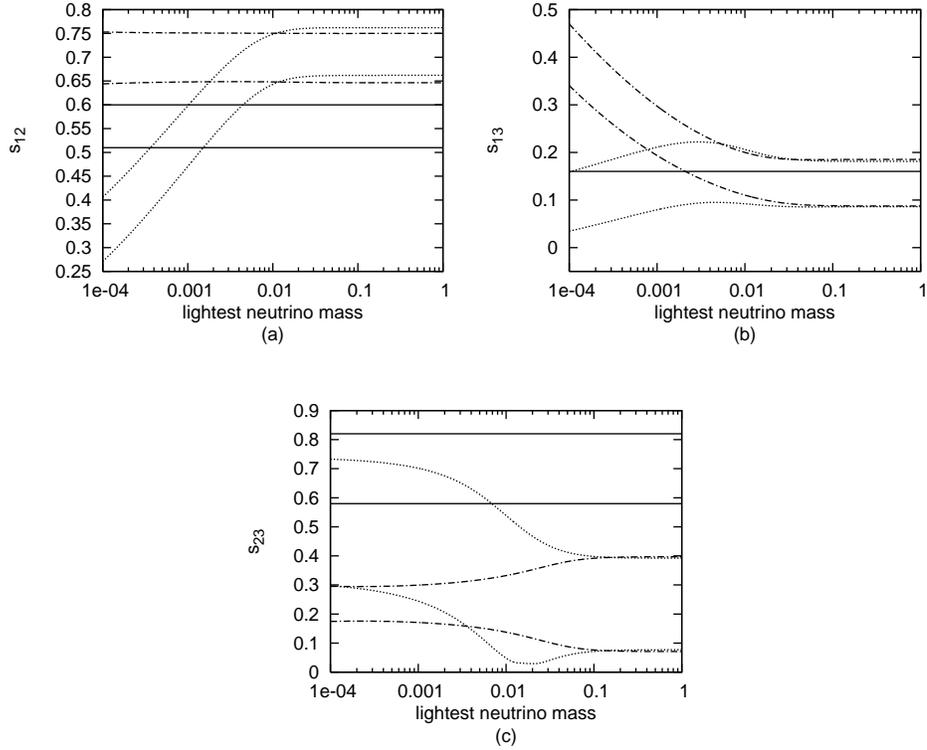}}
 \caption{Plots showing variation of mixing angles
$s_{12}$, $s_{13}$ and $s_{23}$ with the lightest neutrino mass
for texture 6 zero case of Majorana neutrinos. The dotted lines
and the dot-dashed lines depict the limits obtained assuming
normal and inverted hierarchy respectively, the solid horizontal
lines show the 3$\sigma$ limits of the mixing angles.}
\label{nhih6z}
\end{figure}

Without getting into the detailed methodology and inputs used in
these analyses\cite{ourneut6zero,ourneut6zero2}, it is instructive
to present some of the discussions leading to key conclusions. In
particular, for both Majorana and Dirac neutrinos, all the cases
pertaining to inverted hierarchy and degenerate scenario of
neutrino masses seem to be ruled out. For the case of Majorana
neutrinos, following Ref.~\refcite{ourneut6zero}, the ruling out
of inverted hierarchy can be understood from the graphs shown in
Fig.~\ref{nhih6z}, wherein by giving full variations to other
parameters, plots of the mixing angles $s_{12}$, $s_{23}$ and
$s_{13}$ against the lightest neutrino mass have been given. The
dotted lines and the dot-dashed lines depict the limits obtained
assuming normal and inverted hierarchy respectively, the solid
horizontal lines show the 3$\sigma$ limits of the plotted mixing
angle. Interestingly, it is clear from Figs.~\ref{nhih6z}(a) and
\ref{nhih6z}(c) that inverted hierarchy is ruled out by the
experimental limits on $s_{12}$ and $s_{23}$ respectively,
however, in the case of Fig.~ \ref{nhih6z}(b) inverted hierarchy
seems to be allowed for $m_{\nu_1} \gtrsim 0.001$. To this end, it
may be noted that to rule out inverted hierarchy any single plot
of Fig.~\ref{nhih6z} is sufficient.

Coming to the degenerate scenarios of Majorana neutrinos
characterized by either $m_{\nu_1} \lesssim m_{\nu_2} \sim
m_{\nu_3} \sim 0.1~\rm{eV}$ or $m_{\nu_3} \sim m_{\nu_1} \lesssim
m_{\nu_2} \sim 0.1~\rm{eV}$ corresponding to normal hierarchy and
inverted hierarchy respectively. For both these degenerate
scenarios Figs.~\ref{nhih6z}(a) and \ref{nhih6z}(c) can again be
used to rule them out at 3$\sigma$ C.L.. It needs to be mentioned
that while plotting these figures the range of the lightest
neutrino mass is taken to be $10^{-5}\,\rm{eV}-10^{-1}\,\rm{eV}$,
which includes the neutrino masses corresponding to degenerate
scenario, therefore by discussion similar to the one given for
ruling out inverted hierarchy, degenerate scenarios of neutrino
masses are ruled out as well.

\begin{figure}[hbt]
\centerline{\epsfysize=3.2 in\epsffile{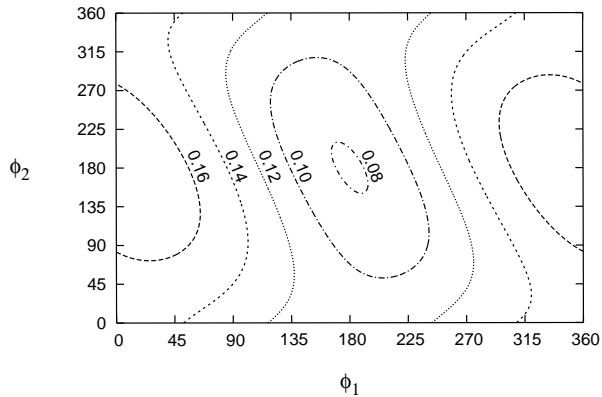}}
 \caption{The contours of $s_{13}$ in $\phi_1 - \phi_2$ plane for
 texture 6 zero matrices for the normal hierarchy case of Majorana
neutrinos.} \label{s13cont6z}
\end{figure}

After ruling out the cases pertaining to inverted hierarchy and
degenerate scenarios, the authors discuss the normal hierarchy
cases. Interestingly, the possibility when charged leptons are in
flavor basis is also completely ruled out. For the case of $M_l$
being texture specific, the viable ranges of neutrino masses,
mixing angle $s_{13}$, Jarlskog's rephasing invariant parameter in
the leptonic sector $J_l$, Dirac-like CP violating phase in the
leptonic sector $\delta_l$ and effective neutrino mass $ \langle
m_{ee} \rangle$ related to neutrinoless double beta decay
$(\beta\beta)_{0 \, \nu}$ have been evaluated. One finds that the
calculated values of parameters $m_{\nu_1}$, $s_{13}$, $J_l$ and $
\langle m_{ee} \rangle$ are well within the ranges obtained by
other similar approaches. It may be added that the predicted lower
limit on $s_{13}$ is in agreement with recent measurements of
$s_{13}$. Further, a measurement of effective mass $\langle m_{ee}
\rangle $, through the $(\beta\beta)_{0 \, \nu}$ decay
experiments, would also have implications for these kind of mass
matrices. Besides the above mentioned parameters, the implications
of $s_{13}$ on the phases $\phi_1$ and $\phi_2$ have also been
considered. To this end, in Fig.~\ref{s13cont6z} the contours for
$s_{13}$ in $\phi_1 - \phi_2$ plane have been shown. From the
figure it is clear that $s_{13}$ plays an important role in
constraining the phases, in particular if lower limit of $s_{13}$
is on the higher side, then $\phi_1$ is restricted to I or IV
quadrant.

\begin{figure}[hbt]
\centerline{\epsfysize=4.in\epsffile{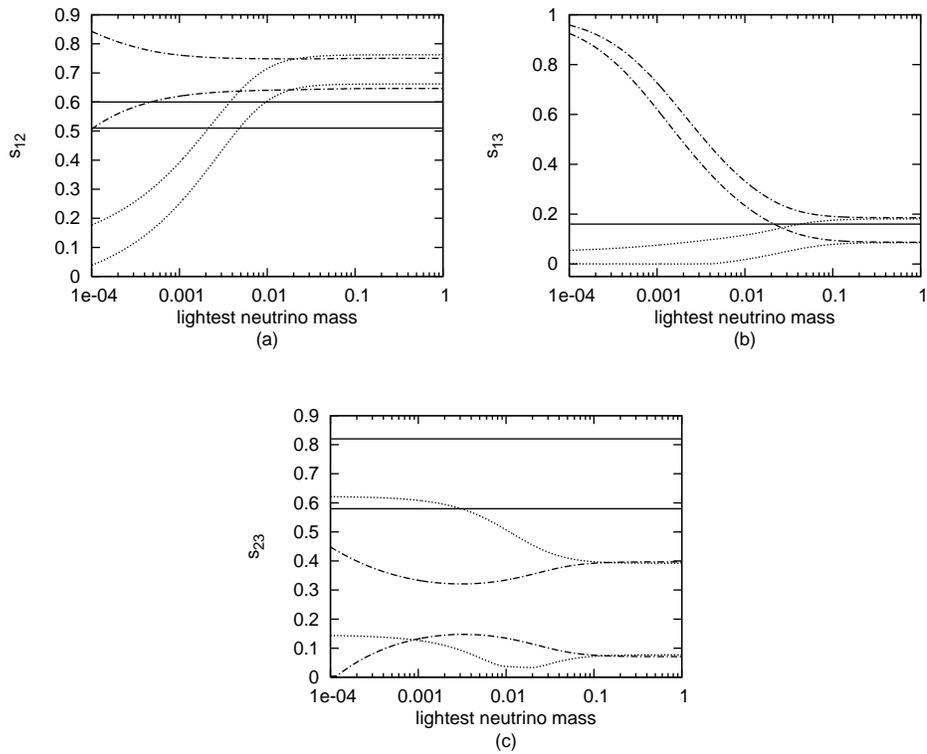}}
 \caption{Plots showing variation of mixing angles
$s_{12}$, $s_{13}$ and $s_{23}$ with lightest neutrino mass for
texture 6 zero case of Dirac neutrinos. The dotted lines and the
dot-dashed lines depict the limits obtained assuming normal and
inverted hierarchy respectively, the solid horizontal lines show
the 3$\sigma$ limits of the mixing angles.} \label{nhih6zdir}
\end{figure}

After studying the implications of texture 6 zero neutrino mass
matrices on the various hierarchies of neutrino masses for the
Majorana neutrinos, it becomes desirable to carry out similar
investigations for Dirac neutrinos. As expected, similar to the
Majorana case, all the cases pertaining to inverted hierarchy and
degenerate scenarios of neutrino masses seem to be ruled out.
Parallel to the case of Majorana neutrinos, in
Fig.~\ref{nhih6zdir}, by giving full variations to other
parameters, the mixing angles against the lightest neutrino mass
have been plotted. The dotted lines and the dot-dashed lines
depict the limits obtained assuming normal and inverted hierarchy
respectively, the solid horizontal lines show the 3$\sigma$ limits
of the plotted mixing angle. A general comparison of these plots
with those for Majorana neutrinos, shown in Fig.~\ref{nhih6z},
suggests that the variation of the mixing angles $s_{12}$ and
$s_{23}$ with the lightest neutrino mass does not depict much
change  for the two different types of neutrinos. Also, it is
easily evident from the Fig.~\ref{nhih6zdir}(c) that inverted
hierarchy is ruled out by the experimental limits on the mixing
angle $s_{23}$.

One can easily check that degenerate scenarios characterized by
either $m_{\nu_1} \lesssim m_{\nu_2} \sim m_{\nu_3} \sim
0.1~\rm{eV}$ or $m_{\nu_3} \sim m_{\nu_1} \lesssim m_{\nu_2} \sim
0.1~\rm{eV}$ are clearly ruled out from Figs.~\ref{nhih6zdir}(a)
and \ref{nhih6zdir}(c). This can be understood by noting that
around $0.1~\rm{eV}$, the limits obtained assuming normal and
inverted hierarchies have no overlap with the experimental limits
of angles $s_{12}$ and $s_{23}$.

For the normal hierarchy of neutrino masses, similar to the
texture 6 zero mass matrices for Majorana neutrinos, the
possibility of charged leptons being in the flavor basis is again
completely ruled out. Corresponding to the case when $M_l$ is
considered texture specific, the viable ranges of neutrino masses,
mixing angle $s_{13}$, Jarlskog's rephasing invariant parameter in
the leptonic sector $J_l$, Dirac-like CP violating phase in the
leptonic sector $\delta_l$ have again been evaluated.
Interestingly, the viable ranges of masses $m_{\nu_1}$,
$m_{\nu_2}$, and $m_{\nu_3}$ as well as the range of $s_{13}$ for
Dirac neutrinos are much narrower compared to the Majorana case.
Further, compared to the Majorana neutrinos, the upper limit of
$J_l$ now cones out to be considerably lower.

\begin{figure}[hbt]
\centerline{\epsfysize=3.2 in\epsffile{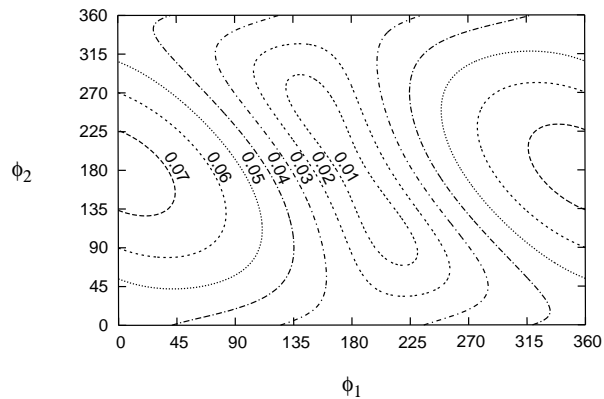}}
 \caption{The contours of $s_{13}$ in $\phi_1 - \phi_2$ plane for
texture 6 zero matrices for the normal hierarchy case of Dirac
neutrinos.} \label{s13cont6zdir}
\end{figure}

For Dirac neutrinos also, the implications of the mixing angle
$s_{13}$ on the phases $\phi_1$ and $\phi_2$ have been examined.
In this context, in Fig.~\ref{s13cont6zdir} the contours for
$s_{13}$ in $\phi_1 - \phi_2$ plane have been plotted. These
contours indicate that the mixing angle $s_{13}$ constrains both
the phases $\phi_1$ and $\phi_2$. For example, if the lower limit
of $s_{13}$ is around 0.07, then $\phi_1$ lies in either the I or
the IV quadrant and $\phi_2$ lies between 135$^{\circ}$ -
225$^{\circ}$.

\subsubsection{Texture 5 zero lepton mass matrices} Similar to the
case of texture 6 zero lepton mass matrices, the implications for
different hierarchies in the case of Fritzsch-like and non
Fritzsch-like texture 5 zero lepton mass matrices have also been
investigated for both Majorana and Dirac
neutrinos\cite{ourneut6zero,ourneut6zero2,lepnonfritex5zero}.
Following Gupta\cite{lepnonfritex5zero}, one finds that for the
two types of neutrinos, corresponding to normal/ inverted
hierarchy and degenerate scenario of neutrino masses 360 cases
each have been considered for carrying out the analysis, making it
a total of 2160 cases. For Majorana neutrinos with normal
hierarchy of neutrino masses, out of the 360 combinations, 67 are
compatible with the neutrino mixing data. Most of the
phenomenological implications of combinations of different
categories are similar, however, still these can be experimentally
distinguished with more precise measurements of $\theta_{13}$ and
$\theta_{23}$. Interestingly, degenerate scenario of Majorana
neutrinos is completely ruled out by the existing data. In the
case of inverted hierarchy, 24 combinations out of 360 are
compatible with the neutrino mixing data. For Dirac neutrinos with
normal hierarchy of neutrino masses, as compared to Majorana
cases, out of 360 only 44 combinations are compatible with
neutrino mixing data. Interestingly, 6 combinations out of 44 can
accommodate degenerate Dirac neutrinos. For inverted hierarchy, 24
combinations are compatible with the existing data. Again, similar
to the discussion presented for the case of texture 6 zero lepton
mass matrices, for the case of texture 5 zero matrices also, we
would like to present some of the details of the
analyses\cite{ourneut6zero,ourneut6zero2} carried out for the
Fritzsch-like case.

\begin{figure}[hbt]
\centerline{\epsfysize=4.in\epsffile{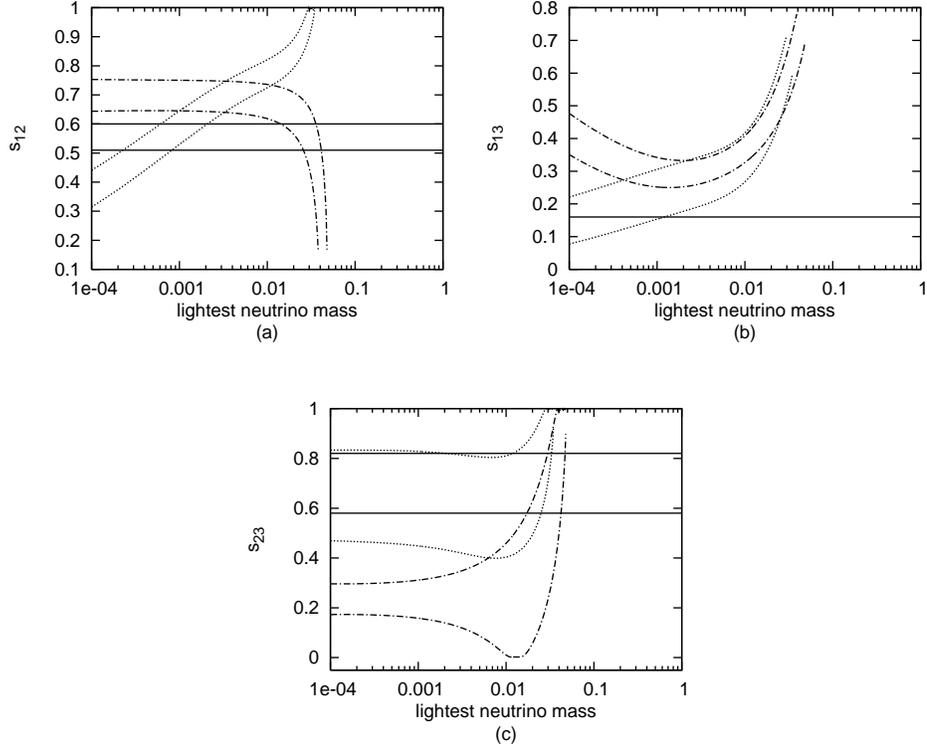}}
 \caption{Plots showing variation of mixing angles
$s_{12}$, $s_{13}$ and $s_{23}$ with lightest neutrino mass for
texture 5 zero $D_l=0$ case of Majorana neutrinos, with a value
$D_{\nu}= \sqrt{m_{\nu_3}}$. The representations of the curves
remain the same as in Fig.~\ref{nhih6z}.} \label{nhih5zdn}
\end{figure}

Following Ref.~\refcite{ourneut6zero}, for texture 5 zero cases,
we first discuss the case when $D_l=0$ and $D_{\nu} \neq 0$. In
Fig.~\ref{nhih5zdn} the plots of the mixing angles against the
lightest neutrino mass for both normal and inverted hierarchy for
a particular value of $D_{\nu}= \sqrt{m_{\nu_3}}$ have been given.
Interestingly, texture 5 zero $D_l=0$ case shows a big change in
the behaviour of the mixing angles versus the lightest neutrino
mass as compared to the texture 6 zero case shown in
Fig.~\ref{nhih6z}. A look at Fig.~\ref{nhih5zdn}(b) clearly shows
that inverted hierarchy as well as degenerate scenario
corresponding to it are ruled out by the experimental limits on
$s_{13}$. A closer look at Figs.~\ref{nhih5zdn}(a) and
\ref{nhih5zdn}(c) reveals that the region pertaining to inverted
hierarchy, depicted by dot-dashed lines, shows an overlap with the
experimental limits on $s_{12}$ and $s_{23}$ respectively,
depicted by solid horizontal lines, around the region when
neutrino masses are almost degenerate. This suggests that, for
these angles, in case the degenerate scenario is ruled out
inverted hierarchy is also ruled out.

Coming to the texture 5 zero $D_{\nu}=0$ and $D_l \neq 0$ case,
interestingly the plots of mixing angles against the lightest
neutrino mass are very similar to those in Fig.~\ref{nhih6z}
pertaining to the texture 6 zero case. By similar arguments, this
case is also ruled out for inverted hierarchy as well as for the
two degenerate scenarios.

\begin{table}
 \tbl{Calculated ranges for neutrino mass and mixing
parameters obtained by varying $\phi_1$ and $\phi_2$ from 0 to
2$\pi$ for the normal hierarchy cases of Majorana neutrinos. All
masses are in $\rm{eV}$.} {\begin{tabular} {|c|c|c|c|} \hline
 Parameter & 5 zero $D_l=0$ & 5 zero $D_{\nu}=0$\\

 & ($M_l$ 3 zero, $M_{\nu D}$ 2 zero) &  ($M_l$ 2 zero, $M_{\nu D}$ 3
 zero)
 \\ \hline

$m_{\nu_1}$ & 0.00020 - 0.0020  & 0.0005 - 0.0032
\\
 $m_{\nu_2}$ &  0.0086 - 0.0094  & 0.0086 -
0.0097
\\
$m_{\nu_3}$ & 0.0421 - 0.0547 & 0.0421 - 0.055
\\
$s_{13}$ & 0.076 - 0.160 &0.055 - 0.160\\

 $J_l$ & $\sim$ 0 - 0.025 & $\sim$ 0 - 0.037 \\

$\delta_l$ & $0^{\circ}$ - 50.0$^{\circ}$
 &$0^{\circ}$ - 90.0$^{\circ}$ \\

$\langle m_{ee} \rangle$ & 0.0029 - 0.0059 & 0.0028 - 0.0068
\\ \hline
\end{tabular}}
\label{tab-5n}
\end{table}

Coming to the normal hierarchy cases, for the two cases of texture
5 zero mass matrices, in Table~\ref{tab-5n} the viable ranges of
neutrino masses, mixing angle $s_{13}$, Jarlskog's rephasing
invariant parameter in the leptonic sector $J_l$, Dirac-like CP
violating phase in the leptonic sector $\delta_l$ and effective
neutrino mass $ \langle m_{ee} \rangle$ related to neutrinoless
double beta decay $(\beta\beta)_{0 \, \nu}$ have been presented.
Considering the $D_l=0$ case, interestingly, results are obtained
for both the possibilities of $M_l$ having Fritzsch-like structure
as well as $M_l$ being in the flavor basis. When $M_l$ is assumed
to have Fritzsch-like structure, the possibility of $D_{\nu} \neq
0$ considerably affects the viable range of $m_{\nu_1}$ as well as
of $s_{13}$. The Pontecorvo-Maki-Nakagawa-Sakata (PMNS) mixing
matrix\cite{pmns1}\cdash\cite{pmns4} constructed for this case is
as follows
 \be U=\left( \ba{ccc}
 0.7898  -  0.8571   &    0.5035  -  0.5971 &      0.0761  -  0.1600 \\
  0.1845  -  0.4413   &    0.5349  -  0.7459 &      0.5725  -  0.8135 \\
  0.3546  -  0.5615   &    0.3926  -  0.6689 &      0.5652  -
  0.8107
 \ea \right). \ee
 When $M_l$ is considered in the flavor basis, one gets a very narrow
range of masses, $m_{\nu_1} \sim 0.00063$,
$m_{\nu_2}=0.0086-0.0088$ and $m_{\nu_3}=0.0534-0.0546$, for which
5 zero matrices are viable. Also for this case, there is a limited
overlap of  with its recent measurements. In case the value of
$s_{13}$ gets constrained further, then it may have implications
for this case.

For the texture 5 zero $D_{\nu}=0$ case, when $M_l$ is considered
in the flavor basis, one does not find any viable solution,
however when it has Fritzsch-like structure there are a few
important observations. The range of $m_{\nu_1}$ gets extended as
compared to the texture 6 zero case, whereas compared to the
texture 5 zero $D_l=0$ case, both the lower and upper limits of
$m_{\nu_1}$ have higher values. Interestingly, this case has the
widest $s_{13}$ range among all the cases considered here. The
PMNS matrix corresponding to this case does not show any major
variation compared to the earlier case, except that the ranges of
some of the elements like $U_{\mu 1}$, $U_{\mu 2}$, $U_{\tau 1}$
and $U_{\tau 2}$ become little wider. This can be understood when
one realizes that $D_l$ can take much wider variation compared to
$D_{\nu}$.

\begin{figure}[hbt]
\centerline{\epsfysize=4.in\epsffile{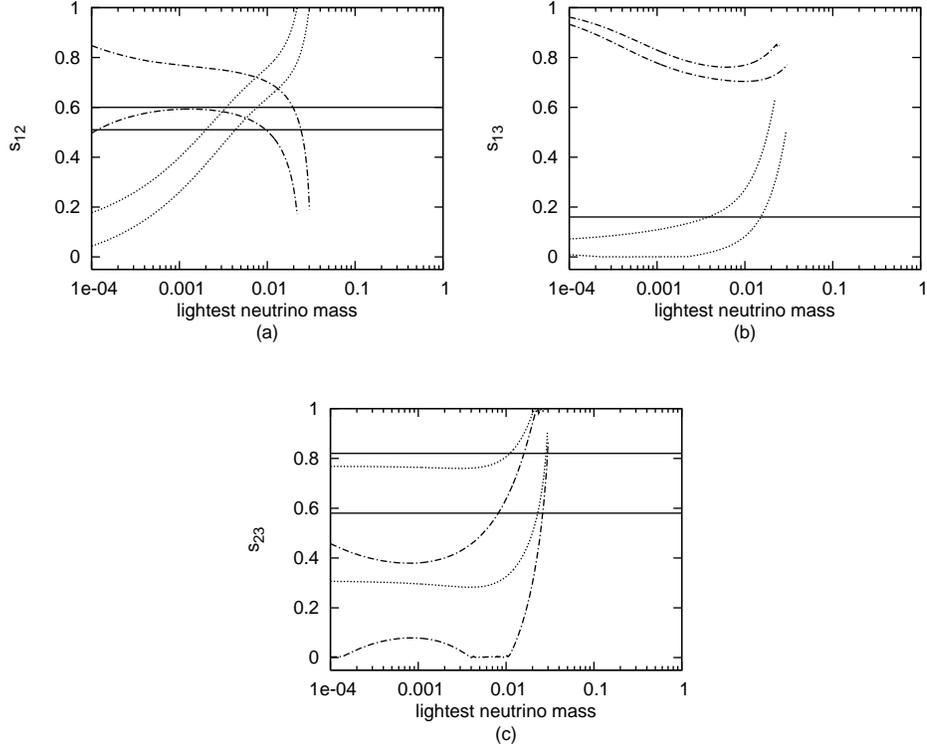}}
 \caption{Plots showing variation of mixing angles
$s_{12}$, $s_{13}$ and $s_{23}$ with lightest neutrino mass for
texture 5 zero $D_l=0$ case of Dirac neutrinos, with a value
$D_{\nu}= \sqrt{m_{\nu_3}}$. The representations of the curves
remain the same as in Fig.~\ref{nhih6zdir}.} \label{nhih5zdndir}
\end{figure}

Coming to the case of Dirac neutrinos\cite{ourneut6zero2}, we
first discuss the texture 5 zero case when $D_l=0$ and $D_{\nu}
\neq 0$. Again to facilitate comparison with the earlier texture 5
zero $D_l=0$ case for Majorana neutrinos, in
Fig.~\ref{nhih5zdndir} the mixing angles against the lightest
neutrino mass for both normal and inverted hierarchy for a
particular value of $D_{\nu}= \sqrt{m_{\nu_3}}$ have been plotted.
Interestingly, we find that similar to the texture 6 zero case,
the texture 5 zero $D_l=0$ case for Dirac neutrinos also shows a
big change in the behaviour of $s_{13}$ versus the lightest
neutrino mass as compared to the corresponding case of texture 5
zero matrices shown in Fig.~\ref{nhih5zdn}. This again suggests
that the variation of the other two mixing angles, $s_{12}$ and
$s_{23}$, with the lightest neutrinos is not much dependent on the
type of neutrinos for the texture 5 zero $D_l=0$ case also. The
graph of $s_{13}$ versus the lightest neutrino mass, shown in
Fig.~\ref{nhih5zdndir}(b) immediately rules out inverted hierarchy
by experimental limits on angle $s_{13}$.

For the texture 5 zero $D_{\nu}=0$ and $D_l \neq 0$ case, similar
to the Majorana case,  the plots of mixing angles against the
lightest neutrino mass are very similar to Fig.~\ref{nhih6zdir}
pertaining to the texture 6 zero case of Dirac neutrinos.
Therefore, arguments similar to the ones for the texture 6 zero
case lead us to conclude that both inverted hierarchy as well as
degenerate scenarios of neutrino masses are ruled out for this
case as well. It may be mentioned that similarities observed in
the mixing angles variation with the lightest neutrino mass for
the texture 5 zero $D_{\nu}=0$ and the texture 6 zero cases of
both Majorana and Dirac neutrinos can be understood by noting that
a very strong hierarchy in the case of charged leptons reduces the
texture 5 zero $D_{\nu}=0$ case essentially to the texture 6 zero
case only.

\begin{table}
\tbl{Calculated ranges for neutrino mass and mixing parameters
obtained by varying $\phi_1$ and $\phi_2$ from 0 to 2$\pi$ for the
normal hierarchy cases of Dirac neutrinos. All masses are in
$\rm{eV}$. } {\begin{tabular}{|c|c|c|c|} \hline
 Parameter & 5 zero $D_l=0$ & 5 zero $D_{\nu}=0$\\

 & ($M_l$ 3 zero, $M_{\nu D}$ 2 zero) &  ($M_l$ 2 zero, $M_{\nu D}$ 3
 zero)
 \\ \hline

$m_{\nu_1}$ & 0.00032 - 0.0063  & 0.0025 - 0.0079
\\
 $m_{\nu_2}$ & 0.0086 - 0.0112  & 0.0089 -
0.0122
\\
$m_{\nu_3}$ & 0.0421 - 0.0550 & 0.0422 - 0.0552
\\
$s_{13}$ & 0.005 - 0.160 &0.0001 - 0.135\\

 $J_l$ & $\sim$ 0 - 0.027 & $\sim$ 0 - 0.028 \\

$\delta_l$ &$0^{\circ}$ - 80.2$^{\circ}$
 &$0^{\circ}$ - 90.0$^{\circ}$

\\ \hline
\end{tabular}}
\label{tab1dir2}
\end{table}

Coming to the normal hierarchy cases for the texture 5 zero mass
matrices corresponding to Dirac neutrinos, we first consider the
$D_l=0$ case. Analogous to the corresponding case of Majorana
neutrinos, both the possibilities of $M_l$ having Fritzsch-like
structure as well as $M_l$ being in the flavor basis yield viable
ranges for the various phenomenological quantities presented in
Table~\ref{tab1dir2}. One finds that going from texture 6 zero to
texture 5 zero $D_l=0$ case, the viable ranges of $m_{\nu_1}$ and
$m_{\nu_3}$ get much broader. Also, the upper limit of $s_{13}$ is
pushed considerably higher which can be understood by noting that
$s_{13}$ is quite sensitive to variations in $D_{\nu}$. The
Pontecorvo-Maki-Nakagawa-Sakata (PMNS) mixing
matrix\cite{pmns1}\cdash\cite{pmns4} obtained for the texture 5
zero $D_l=0$ case is given by
 \be U=\left( \ba{ccc}
 0.7897  -  0.8600   &    0.5036  -  0.5998 &      0.0054  -  0.1600 \\
  0.1838  -  0.4748   &    0.4859  -  0.7438 &      0.5726  -  0.8194 \\
  0.3107  -  0.5633   &    0.3974  -  0.6890 &      0.5650  -
  0.8142
 \ea \right). \label{dirmat}\ee

For the case of $M_l$ being in the flavor basis, the range of
masses so obtained are $m_{\nu_1}=0.0020-0.0040$,
$m_{\nu_2}=0.0088-0.0100$ and $m_{\nu_3}=0.0422-0.0548$. The range
of the mixing angle $s_{13}$ is $0.0892-0.1594$, indicating that
the lower limit of $s_{13}$ is considerably high which implies
that refinements in the measurement of this angle would have
consequences for this case of texture 5 zero mass matrices for
Dirac neutrinos.

A comparison of the texture 5 zero $D_l=0$ case of Dirac neutrino
with the corresponding case of Majorana neutrinos, presented in
Table~\ref{tab1dir2} indicates that now there is an expansion of
the range of $m_{\nu_1}$, in particular its upper limit is pushed
higher. There is also a significant lowering down of the lower
limit of the angle $s_{13}$ and a considerable increase in the
upper limit of the Dirac-like CP violating phase $\delta_l$ for
the Dirac neutrinos as compared to the Majorana neutrinos. This
difference in the viable ranges of these quantities could provide
clues to neutrinos being Dirac or Majorana.

Considering the texture 5 zero $D_{\nu}=0$ case for Dirac
neutrinos, again the possibility of $M_l$ being in the flavor
basis does not yield any results, however the case of $M_l$ having
Fritzsch-like structure reveals several interesting facts. For
Dirac neutrinos, a comparison of the texture 5 zero $D_{\nu}=0$
case with the texture 5 zero $D_l=0$ case shows both the lower and
upper limits of $m_{\nu_1}$ have higher values. Interestingly, for
this case the lower limit of $s_{13}$ becomes almost 0. Also, it
may be noted that for this case one gets a wide range of the
Dirac-like CP violating phase $\delta_l$. The PMNS matrix
corresponding to this case is quite similar to the one presented
in Eq.~(\ref{dirmat}), except for somewhat wider ranges of the
elements $U_{e 3}$, $U_{\mu 2}$, $U_{\mu 2}$, $U_{\tau 1}$ and
$U_{\tau 2}$.

\subsubsection{Texture 4 zero lepton mass matrices} Like the case
of quarks, the number of viable possibilities for the case of
texture 4 zero lepton mass matrices is also quite large. Recently,
an interesting analysis has been carried out by Branco {\it et
al.}\cite{brancolep4z} wherein they have classified and analyzed
the texture 4 zero {\it ans\"{a}tze} for the charged lepton mass
matrix and the neutrino mass matrix with a parallel structure. It
has been pointed out that these {\it ans\"{a}tze} do have physical
implications, since not all the zeros can be obtained
simultaneously, just by making weak basis (WB) transformations.
Further, it has been shown that these texture 4 zero {\it
ans\"{a}tze} can be classified in four classes, one of which is
not compatible with the experimental data. For the remaining three
classes, the authors have presented a summary of predictions
pertaining to the lightest neutrino mass and the effective
Majorana mass.

Similar, to the discussion presented for the texture 4 zero quark
mass matrices, for the case of leptons also we have presented
details pertaining to only the Fritzsch-like texture 4 zero mass
matrices. Before proceeding further, for ready reference as well
as to facilitate subsequent discussions, we would again like to
mention that most of the attempts to understand the pattern of
neutrino masses and mixings have been carried out using the seesaw
mechanism\cite{seesaw1}\cdash\cite{seesaw5} given by
 \be M_{\nu}=-M_{\nu D}^T\,(M_R)^{-1}\,M_{\nu D},
\label{seesaweq6} \ee \noindent where $M_{\nu D}$ and $M_R$ are
respectively the Dirac neutrino mass matrix and the right handed
Majorana neutrino mass matrix. Similarly, we would like to
reiterate that the predictions are quite different, on the one
hand when texture is imposed only on $M_{\nu D}$ and $M_R$ and on
the other hand when $M_{\nu}$ and $M_{\nu D}$ have the same
texture by imposing `texture invariant
conditions'\cite{0307359,matsuda}.

We begin by discussing briefly the
attempts\cite{0307359,matsuda,ourneut4zero} made by a few authors
regarding the texture 4 zero mass matrices. Assuming normal
hierarchy of masses as well as imposing texture 4 zero structure
on $M_{\nu D}$ and charged lepton mass matrices,
Xing\cite{0307359} has not only shown the compatibility of these
with neutrino oscillation phenomenology but have also shown the
seesaw invariance of these structures under certain conditions.
Very recently, for normal hierarchy, Matsuda {\it et
al.}\cite{matsuda} have shown that texture 4 zero lepton mass
matrices can accommodate large values of mixing angle $s_{13}$. In
particular, by imposing texture invariant conditions they have
shown that $M_{\nu}$ can be texture 2 zero when one assumes
Fritzsch-like texture 2 zero structure for $M_{\nu D}$, $M_R$ as
well as for charged lepton mass matrix. Further, for both Majorana
and Dirac neutrinos, Ahuja {\it et al.}\cite{ourneut4zero} have
carried out detailed and comprehensive investigation regarding the
compatibility of texture 4 zero lepton mass matrices with the
normal/inverted hierarchy, degenerate scenario of neutrino masses
and have arrived at some very interesting conclusions. In the
present work, we have presented essential details of the analysis
by Ref.~\refcite{ourneut4zero}.

For the sake of the convenience of the reader as well as for
better understanding of the conclusions by
Ref.~\refcite{ourneut4zero}, we first define the Fritzsch-like
texture 4 zero mass matrices, e.g.,
 \be
 M_{l}=\left( \ba{ccc}
0 & A _{l} & 0      \\ A_{l}^{*} & D_{l} &  B_{l}     \\
 0 &     B_{l}^{*}  &  C_{l} \ea \right), \qquad
M_{\nu D}=\left( \ba{ccc} 0 &A _{\nu} & 0      \\ A_{\nu}^{*} &
D_{\nu} &  B_{\nu}     \\
 0 &     B_{\nu}^{*}  &  C_{\nu} \ea \right),
 \label{frzmm6}
 \ee
$M_{l}$ and $M_{\nu D}$ respectively corresponding to charged
lepton and Dirac neutrino mass matrices. The matrices $M_{l}$ and
$M_{\nu D}$ together are referred to as the texture 4 zero mass
matrices, each being texture 2 zero type with $D_l$ and $D_{\nu}$
being non zero. Corresponding to these mass matrices, the PMNS
matrix for Dirac neutrinos is expressed as
\be
 U = O_l^{\dagger} Q_l P_{\nu D} O_{\nu D} \,, \ee
where $Q_l$, $P_{\nu D}$ are the diagonal phase matrices and
$O_l$, $O_{\nu D}$ correspond to the orthogonal transformations
used for diagonalizing the matrices $M_{l}$ and $M_{\nu D}$. For
the two types of neutrinos, the details of diagonalization
procedure of the mass matrices and the elements of the PMNS
matrices have been presented in Appendices A and B respectively.

The analysis by Ref.~\refcite{ourneut4zero} incorporates the
following inputs at 3$\sigma$ C.L.,
\be
 \Delta m_{12}^{2} = (7.1 - 8.9)\times
 10^{-5}~\rm{eV}^{2},~~~~
 \Delta m_{23}^{2} = (2.0 - 3.2)\times 10^{-3}~ \rm{eV}^{2},
 \label{solatmmass6}\ee
\be
{\rm sin}^2\,\theta_{12}  =  0.24 - 0.40,~~~
 {\rm sin}^2\,\theta_{23}  =  0.34 - 0.68,~~~
 {\rm sin}^2\,\theta_{13} \leq 0.040. \label{s136}
\ee For the purpose of calculations, the lightest neutrino mass,
the phases $\phi_1$, $\phi_2$ and the elements of the mass
matrices $D_{l, \nu}$ are considered as free parameters, the other
two masses are constrained by $\Delta m_{12}^2 = m_{\nu_2}^2 -
m_{\nu_1}^2 $ and $\Delta m_{23}^2 = m_{\nu_3}^2 - m_{\nu_2}^2 $
in the normal hierarchy case and by $\Delta m_{23}^2 = m_{\nu_2}^2
- m_{\nu_3}^2$ in the inverted hierarchy case. For all the three
hierarchies, the explored range of the lightest neutrino mass is
$10^{-8}\,\rm{eV}-10^{-1}\,\rm{eV}$. In the absence of any
constraint on the phases, $\phi_1$ and $\phi_2$ have again been
given full variation from 0 to $2\pi$.

\begin{figure}[hbt]
\vspace{0.12in}
\centerline{\epsfysize=2.8in\epsffile{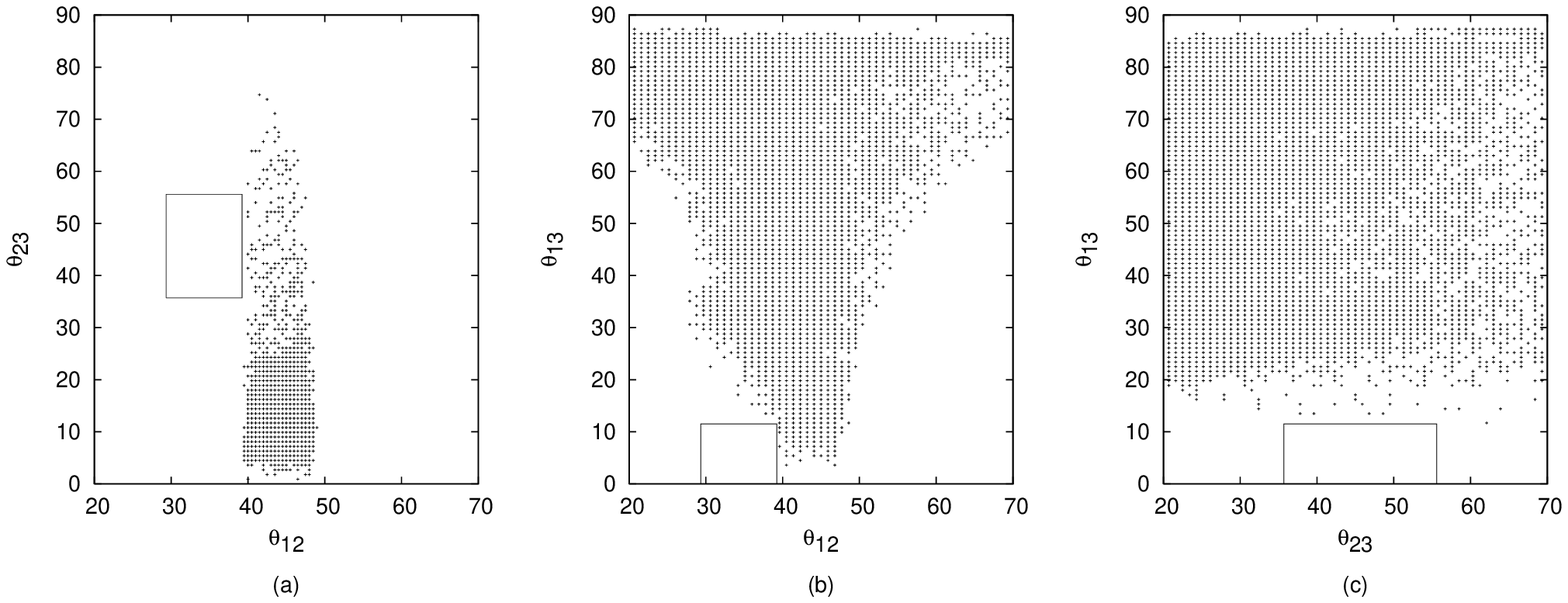}}
\vspace{0.08in}
   \caption{Plots showing the parameter space
corresponding to any of the two mixing angles by constraining the
third angle by its experimental limits and giving full allowed
variation to other parameters for Majorana neutrinos. The blank
rectangular region indicates the experimentally allowed $3\sigma$
region of the plotted angles.}
  \label{fig1}
  \end{figure}
To begin with, the authors consider the inverted hierarchy case
for both Majorana and Dirac neutrinos. In this context, it may be
mentioned that for both the possibilities texture is imposed only
on $M_{\nu D}$, with no such restriction on $M_{\nu}$ for the
Majorana case. For Majorana neutrinos, Figs.~\ref{fig1}(a),
\ref{fig1}(b) and \ref{fig1}(c) present the plots pertaining to
the parameter space corresponding to any of the two mixing angles
by constraining the third angle by its values given in
Eq.~(\ref{s136}) while giving full allowed variation to other
parameters. Also included in the figures are blank rectangular
regions indicating the experimentally allowed $3\sigma$ region of
the plotted angles. Interestingly, a general look at these figures
reveals that the case of inverted hierarchy seems to be ruled out.
From Fig.~\ref{fig1}(a) showing the plot of angles $\theta_{12}$
versus $\theta_{23}$, one can immediately conclude that the
plotted parameter space includes the experimentally allowed range
of $\theta_{23}$, however it excludes the experimentally allowed
range of $\theta_{12}$. This clearly indicates that at 3$\sigma$
C.L. inverted hierarchy is not viable. The conclusions arrived
above can be further checked from Figs.~ \ref{fig1}(b) and
\ref{fig1}(c) wherein $\theta_{12}$ versus $\theta_{13}$ and
$\theta_{23}$ versus $\theta_{13}$ have been plotted respectively
by constraining angles $\theta_{23}$ and $\theta_{12}$. Both the
figures indicate that the plotted parameter space does not include
simultaneously the experimental bounds of the plotted angles,
e.g., $\theta_{12}$ in the case of Fig.~\ref{fig1}(b) and
$\theta_{13}$ in Fig.~\ref{fig1}(c).

For Dirac neutrinos, again inverted hierarchy seems to be ruled
out using graphs which can be plotted in a manner similar to the
Majorana case by constraining one mixing angle by its experimental
limits and plotting the parameter space for the other two angles.
Again, the plotted parameter space does not overlap with the
experimental limits of at least one of the plotted angles, thereby
indicating that inverted hierarchy is ruled out at 3$\sigma$ C.L.
for Dirac neutrinos as well.

For Majorana or Dirac neutrinos the cases of neutrino masses being
degenerate, characterized by either $m_{\nu_1} \lesssim m_{\nu_2}
\sim m_{\nu_3} \lesssim 0.1~\rm{eV}$ or $m_{\nu_3} \sim m_{\nu_1}
\lesssim m_{\nu_2} \lesssim 0.1~\rm{eV}$ corresponding to normal
and inverted hierarchy respectively, are again ruled out.
Considering degenerate scenario corresponding to inverted
hierarchy, Figs.~\ref{fig1} can again be used to rule out
degenerate scenario at 3$\sigma$ C.L. for Majorana neutrinos
respectively. It needs to be mentioned that while plotting these
figures the range of the lightest neutrino mass is taken to be
$10^{-8}\,\rm{eV}-10^{-1}\,\rm{eV}$, which includes the neutrino
masses corresponding to degenerate scenario, therefore by
discussion similar to the one given for ruling out inverted
hierarchy, degenerate scenario of neutrino masses is ruled out as
well.

\begin{figure}[tbp]
\centerline{\epsfysize=2.8in\epsffile{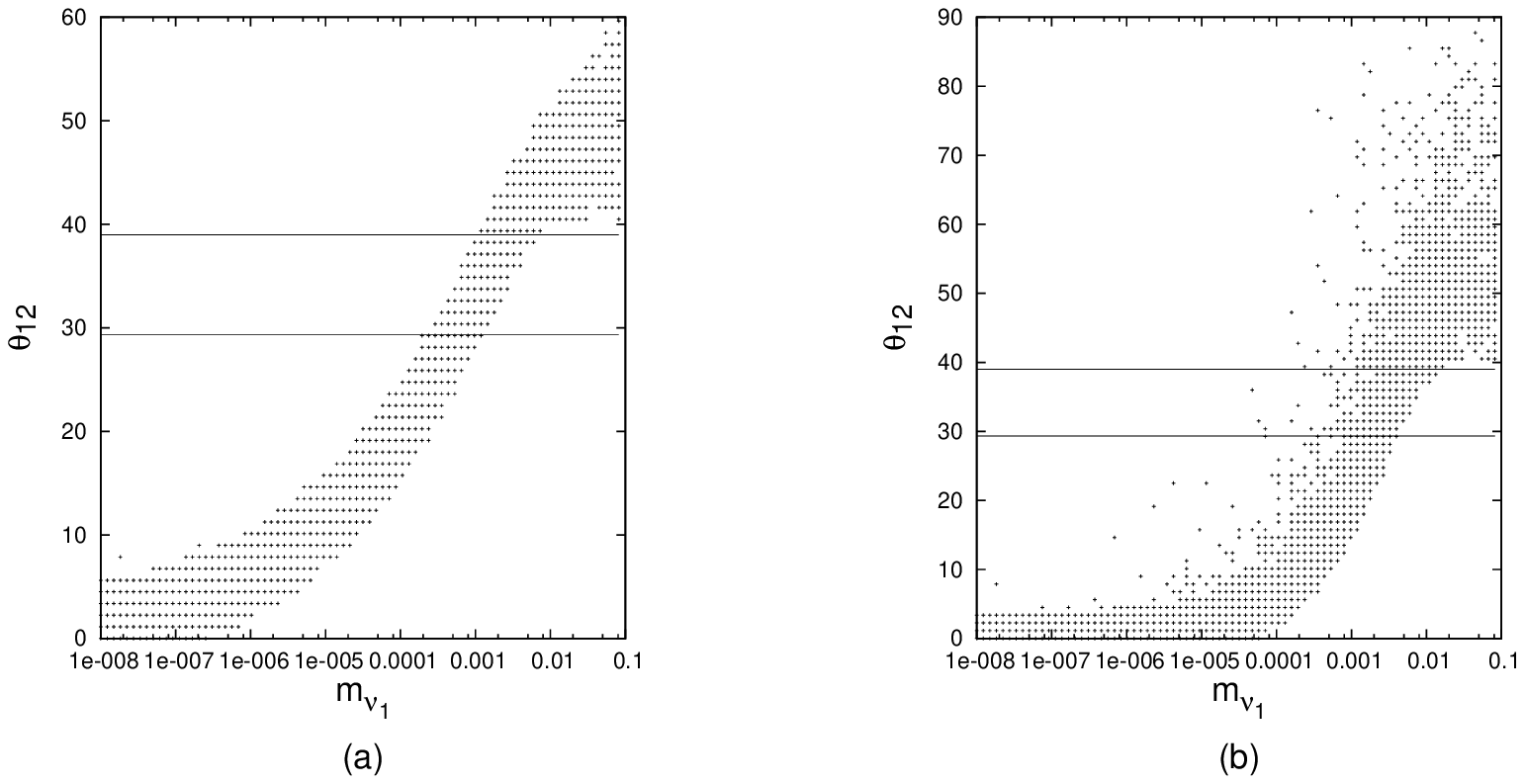}}
\vspace{0.08in}
 \caption{Plots showing variation of mixing angle
$\theta_{12}$ with lightest neutrino mass $m_{\nu_1}$ by giving
full variation to other parameters for (a) Majorana neutrinos and
(b) Dirac neutrinos. The parallel lines indicate the $3\sigma$
limits of angle $\theta_{12}$.}
  \label{th12vm-md}
  \end{figure}

Coming to degenerate scenario corresponding to normal hierarchy,
one can easily show that this is ruled out again. To this end, in
Fig.~\ref{th12vm-md}, by giving full variation to other
parameters, the mixing angle $\theta_{12}$ against the lightest
neutrino mass $m_{\nu_1}$ has been plotted.
Fig.~\ref{th12vm-md}(a) corresponds to the case of Majorana
neutrinos and Fig.~\ref{th12vm-md}(b) to the case of Dirac
neutrinos. From the figures one can immediately find that the
values of $\theta_{12}$ corresponding to $m_{\nu_1} \lesssim
0.1~\rm{eV}$ lie outside the experimentally allowed range, thereby
ruling out degenerate scenario for Majorana as well as Dirac
neutrinos at 3$\sigma$ C.L..

It may also be added that in the case when charged leptons are in
the flavor basis, one can easily check that inverted hierarchy and
degenerate scenarios for the texture 4 zero mass matrices are
again ruled out, in agreement with the conclusions of
Ref.~\refcite{leptex35}. The results pertaining to this case can
easily be derived from the earlier cases.

\begin{table}
\tbl{Calculated ranges for neutrino mass and mixing parameters for
the normal hierarchy case of texture 4 zero lepton mass matrices.
All masses are in $\rm{eV}$.} {\begin{tabular}{|c|cc|cc|} \hline &
\multicolumn{2}{c|} {$M_l$ being Fritzsch-like} &
\multicolumn{2}{c|} {$M_l$ being in the flavor basis} \\ \hline
&Dirac case &Majorana case & Dirac case &Majorana case\\ \hline

$m_{\nu_1}$ & 5.73 $\times 10^{-5}$  - 0.012 & 2.47 $\times
10^{-4}$ - 0.006 & (1.63 - 6.28) $\times 10^{-3}$ & (0.402 - 2.06)
$\times 10^{-3}$
\\
$m_{\nu_2}$ & 0.0084 - 0.0149 & 0.0084 - 0.0108 & 0.0086 - 0.0113
& 0.0084 - 0.0096
\\
$m_{\nu_3}$ & 0.0456 - 0.0577 & 0.0455 - 0.0575 & 0.0446 - 0.0576
& 0.0455 - 0.0573
\\
$\theta_{12}$ & 29.30$^{\circ}$ - 39.20$^{\circ}$  &
29.30$^{\circ}$ - 39.20$^{\circ}$ & 29.30$^{\circ}$ -
39.20$^{\circ}$  & 29.30$^{\circ}$ - 39.04$^{\circ}$
\\
$\theta_{23}$ & 35.70$^{\circ}$ - 55.60$^{\circ}$  &
35.70$^{\circ}$ - 55.60$^{\circ}$ & 35.70$^{\circ}$ -
55.59$^{\circ}$  & 35.70$^{\circ}$ - 40.15$^{\circ}$
 \\
$\theta_{13}$ & 0.084$^{\circ}$ - 11.50$^{\circ}$  &
1.14$^{\circ}$ - 11.50$^{\circ}$ & 3.60$^{\circ}$ -
11.15$^{\circ}$ & 8.43$^{\circ}$ - 11.50$^{\circ}$
 \\
$J_l$ & $-$0.0462 - 0.0448  & $-$0.0459 - 0.0463 & $\sim 0$ &
$\sim 0$
 \\
$\delta_l$ & $-90^{\circ}$ - 90.0$^{\circ}$  &$-90^{\circ}$ -
90.0$^{\circ}$ & $\sim 0$  &$\sim 0$
\\
$\langle m_{ee} \rangle$ & - & 0.00086 - 0.0173 & - & 0.0032 -
0.0075
 \\ \hline
\end{tabular}}
\label{tab1}
\end{table}

After ruling out the cases pertaining to inverted hierarchy and
degenerate scenarios, the authors then discuss the normal
hierarchy cases. For the charged lepton mass matrix $M_l$ being
Fritzsch-like or in the flavor basis, for Majorana as well as
Dirac neutrinos, Table~\ref{tab1} presents the viable ranges of
neutrino masses, mixing angles $\theta_{12}$, $\theta_{23}$ and
$\theta_{13}$, Jarlskog's rephasing invariant parameter in the
leptonic sector $J_l$, Dirac-like CP violating phase in the
leptonic sector $\delta_l$ and effective neutrino mass $ \langle
m_{ee} \rangle$. For both Dirac or Majorana neutrinos, the viable
range of the lightest neutrino mass $m_{\nu_1}$ is quite
different, in particular the range corresponding to Dirac
neutrinos is much wider at both the ends as compared to the
Majorana neutrinos. Similar conclusions can be arrived at by
studying the implications of the well known mixing angle
$\theta_{12}$ on the lightest neutrino mass $m_{\nu_1}$ through a
closer look at the Figs.~\ref{th12vm-md}(a) and
\ref{th12vm-md}(b). Therefore, a measurement of $m_{\nu_1}$ could
have important implications for the nature of neutrinos. Somewhat
constrained range of $m_{\nu_2}$ for the Majorana case as compared
to the Dirac case is also due to the constrained range of
$m_{\nu_1}$ for the Majorana case. Also, from the table one finds
that the lower limit on $\theta_{13}$ for both the Dirac case and
the Majorana case are quite small as compared to the recently
measured value of angle $\theta_{13}$. It must be noted that the
calculated values of $\langle m_{ee} \rangle$ are much less
compared to the present limits of $\langle m_{ee}
\rangle$\cite{heidel1,heidel2}, therefore, these do not have any
implications for the texture 4 zero cases considered here.
However, the future experiments with considerably higher
sensitivities, aiming to measure $\langle m_{ee} \rangle \simeq
3.6\times
 10^{-2}~\rm{eV}$ (MOON\cite{moon}) and $\langle m_{ee} \rangle \simeq 2.7\times
 10^{-2}~\rm{eV}$ (CUORE\cite{cuore}), would have implications on
the cases considered here. The different cases of Dirac and
Majorana neutrinos do not show any divergence for the ranges of
Jarlskog's rephasing invariant parameter.

In Table~\ref{tab1} the results when charged leptons are in the
flavor basis which can be easily deduced from the case when $M_l$
is Fritzsch-like have also been presented. Interestingly, in this
case both $J_l$ and $\delta_l$ are vanishingly small for the wide
range of parameters considered here, which can easily be
understood by examining the corresponding mixing matrix. Also, the
range of angle $\theta_{13}$ is much narrower compared to the case
when $M_l$ is Fritzsch-like, particularly for the Majorana case
the predicted range is very narrow, however being very much in
agreement with the recent measurement of $\theta_{13}$. It may
also be added that for the Majorana case, the range of
$\theta_{23}$ is compatible only with the lower part of the
present admissible range, however for the Dirac case there is no
such restriction. These conclusions are broadly in agreement with
those of Ref.~\refcite{leptex35}.

Further, the authors have also constructed the
Pontecorvo-Maki-Nakagawa-Sakata (PMNS) mixing
matrix\cite{pmns1}\cdash\cite{pmns4} which for Majorana neutrinos
is
 \be U=\left( \ba{ccc}
 0.7599  -  0.8701   &    0.4797  -  0.6294 &      0.0199  -  0.1994 \\
  0.1673 -  0.5715   &    0.3948  -  0.7606 &      0.5720  -  0.8224 \\
  0.1854  -  0.5912  &    0.3549  -  0.7363 &      0.5540  -
  0.8094
 \ea \right), \label{mmaj} \ee
 wherein the magnitude of the matrix elements have been given.
Similarly, for Dirac neutrinos, the PMNS matrix is
  \be U=\left( \ba{ccc}
 0.7604  -  0.9213  &    0.3887 -  0.6317&      0.0015  -  0.1993 \\
  0.1475  -  0.5552   &    0.4049  -  0.8170 &      0.4154  -  0.8244 \\
  0.1830  -  0.6022   &    0.3648  -  0.7441 &      0.5546  -
  0.9095
 \ea \right). \label{mdir} \ee
A general look at the two matrices reveals that the ranges of the
matrix elements are more wider in the case of Dirac neutrinos as
compared to those in the case of Majorana neutrinos. A comparison
of the two matrices shows that the lower limit of the element
$U_{\mu 3}$ show an appreciable difference, which seems to be due
to the nature of neutrinos, hence, a further precision of $U_{\mu
3}$ would have important implications. Also, it may be mentioned
that both the above mentioned matrices are fully compatible with a
very recent construction of a mixing matrix by Bjorken {\it et
al.}\cite{bjorken} assuming democratic trimaximally mixed $\nu_2$
mass eigenstate as well as with the one presented by
Giunti\cite{othersmm2}.

\section{Textures and general mass matrices \label{nmmwb}}
From the previous sections, one finds that texture specific mass
matrices are able to accommodate the quark as well as neutrino
mixing data. This immediately brings in focus the question whether
textures can be derived from more fundamental considerations. In
this context, several ideas have been advocated in the literature,
e.g., Peccei and Wang\cite{nmm} have imposed the conditions of
naturalness which makes certain elements of the texture specific
mass matrices very small relative to the others, indicating that
the idea of naturalness  is in accordance with the texture {\it
ans\"{a}tze}. Similarly, Branco and
collaborators\cite{brancolep4z,brancowb1,brancowb2} have used the
facility of weak basis transformations to derive textures. In the
sequel, we present the essentials pertaining to these ideas.

\subsection{Natural mass matrices} Peccei and Wang\cite{nmm}
advocated the concept of natural mass matrices which makes it
possible to restrict the arbitrariness in the mass pattern
construction, thereby allowing a search for viable GUT patterns
more systematically and efficiently. The essential idea consists
of avoidance of fine tuning in the elements of the mass matrices
so as to reproduce the hierarchical structure of the CKM matrix.
To illustrate their idea, Peccei and Wang first discuss the two
generation quark case. For this case, the natural mass matrices
take the following form
 \begin{equation}
\tilde{M}_{u} \simeq \left( \begin{array}{cc}
\alpha'_{u}\lambda^{4} &  \alpha_{u}\lambda^{2}   \\
\alpha_{u}\lambda^{2}     & 1
\end{array} \right), \tilde{M}_{d}  \simeq  \left( \begin{array}{cc}
\alpha'_{d}\lambda^{2} &  \alpha_{d}\lambda   \\ \alpha_{d}\lambda
& 1
\end{array} \right),
\end{equation} where $\lambda$ is a small parameter of the order
of Cabibbo angle and
\begin{equation}
\sin \theta_{u} =  \alpha_{u} \lambda^{2},~~ \alpha'_{u}-
\alpha^{2}_{u} = \xi_{uc},~~ \sin \theta_{d} = \alpha_{d}
\lambda,~~\alpha'_{d}- \alpha^{2}_{d} = \xi_{ds},~~\alpha_{d} -
\alpha_{u}\lambda \simeq 1.
 \end{equation}  With the
parameters $\alpha$ and $\alpha'$ of $O(1)$, one gets the observed
hierarchy (i.e. the $\xi$'s being of $O(1)$) without any need for
fine tunings. This also illustrates that even if the element (1,1)
is taken to be zero, reducing both $\tilde{M}_{u}$ and
$\tilde{M}_{d}$ to texture 1 zero mass matrices, the matrix will
continue to remain natural.

Further, they have also considered natural mass matrices for the
three generation case. Again keeping in mind the structure of the
CKM matrix, a convenient parameterization of the mass matrices
based on a perturbative expansion in `$\lambda$' has been
introduced. For $\theta_{1d} \sim \lambda \; , \; \theta_{1u} \sim
\theta_{2d} \sim \lambda^{2} \; , \; \theta_{2u} \sim \theta_{3u}
\sim \lambda^{4} \; \mbox{and} \; \theta_{3d} \sim \lambda^{5}$
one gets a particular mass pattern, e.g.,
\begin{eqnarray} \tilde{M}_{u} & \simeq & \left(
\begin{array}{ccc}
 u_{11}\lambda^{7} &  u_{12}\lambda^{6}  &  e^{-i\delta_{u}}u_{13}\lambda^{4}
\\
 u_{12}\lambda^{6} &  u_{22}\lambda^{4}  &  u_{23}\lambda^{4} \\
e^{i\delta_{u}} u_{13}\lambda^{4} &  u_{23}\lambda^{4}  &  1
 \end{array} \right), \nonumber \\ \nonumber \\
\tilde{M}_{d} & \simeq &  \left( \begin{array}{ccc}
 d_{11}\lambda^{4} &  d_{12}\lambda^{3}  &  e^{-i\delta_{d}}d_{13} \lambda^{5}
\\
 d_{12}\lambda^{3} &  d_{22}\lambda^{2}  &  d_{23}\lambda^{2} \\
 e^{i\delta_{d}}d_{13} \lambda^{5} &  d_{23}\lambda^{2}  &  1
 \end{array} \right), \label{matrix_pattern}
\end{eqnarray} where the real coefficients $u_{ij}$'s, $d_{ij}$'s are functions
of the $O(1)$ parameters. A closer look at the structure reveals
that some of the elements of the mass matrices can be considered
to be highly suppressed compared with others, again in consonance
with the concept of texture specific mass matrices. It may also be
mentioned that the elements of a general mass matrix, following
the hierarchy \bc (1,1), (1,3), (3,1) $ \lesssim $ (1,2), (2,1) $
\lesssim $ (2,3), (3,2), (2,2) $ \lesssim $ (3,3), \ec can be
considered to be natural mass matrices.

\subsection{Texture and weak basis transformations}
The study of viable texture zeros is further complicated by the
fact that some sets of these zeroes have by themselves no physical
significance. This is due to the fact that these can be obtained
starting from arbitrary fermion mass matrices by making
appropriate weak basis (WB) transformations under which quark mass
matrices change but the gauge currents remain real and diagonal.
In this context, several
attempts\cite{9912358,brancolep4z,brancowb1}\cdash\cite{fxwb2}
have been made wherein the above freedom is exploited to introduce
WB zeros in the fermion mass matrices $M_u$, $M_d$. If we consider
$ M_{u} $ and $ M_{d} $ hermitian mass matrices, then most general
WB transformations that leave the mass matrices hermitian is
\be
M_{d}\rightarrow M^{'}_{d} \equiv W^{T} M_{d}W, \ee
\be
M_{u}\rightarrow M^{'}_{u} \equiv W^{T} M_{u}W. \ee where W is
arbitrary unitary matrix.

The WB transformations broadly lead to two possibilities for the
texture zero fermion mass matrices. In the first possibility, as
observed by Branco {\it et al.}\cite{brancowb1,brancowb2}, one
ends up with texture 3 zero fermion mass matrices $M_u$, $M_d$
wherein one of the matrix among these pairs is a texture 2 zero
Fritzsch-like hermitian mass matrix given by \be M_{q} = \left(
\ba {lll} 0 & * & 0 \\
* & * & * \\ 0 & * & * \\ \ea \right),
\ee where $q=U/D$, while the other mass matrix is a texture 1 zero
hermitian mass matrix of the following form
 \be M^{'}_{q} = \left( \ba {lll}
0 & * & * \\ * & * & * \\ * & * & * \\ \ea \right). \label{formp}
\ee In the second possibility, as observed by Fritzsch and
Xing\cite{fxwb1,fxwb2}, one ends up with texture 2 zero fermion
mass matrices, wherein both the fermion mass matrices assume a
texture 1 zero hermitian structure of the following form
 \be M^{'}_{q} = \left( \ba {lll}
* & * & 0 \\ * & * & * \\ 0 & * &
* \\ \ea \right). \label{tex1z} \ee In addition, the position of the WB
zeros in these mass matrices can be varied by the use of parallel
$S_3$ or permutation transformations\cite{brancolep4z} on the
fermion mass matrices. However, it is
known\cite{neelu56zeroquarks} that such permutation
transformations generally lead to the re-positioning of the
various elements of these mass matrices. For details in this
regard, we refer the reader to Ref.~\refcite{brancolep4z}.

It is important to emphasize here that in the first approach,
without any loss of generality, one is able to embed at the most
three WB zeros in the fermion mass matrices, whereas the maximum
number of such WB zeros that can be introduced using the
Fritzsch-Xing approach is only two. Further, one can
conclude\cite{9912358,brancolep4z,fxwb1,fxwb2} that the inclusion
of an additional texture zero in these approaches, like texture 4
zero Fritzsch-like mass matrices, does have physical implications.
One may note that the above two possibilities of WB
transformations are equivalent, however the Fritzsch-Xing
approach\cite{fxwb1,fxwb2} not only exhibits parallel texture
structure but can also be diagonalized exactly, making the
construction of corresponding CKM matrix rather simple. As a
result, we would like to discuss the general texture specific
fermion mass matrices based on the Fritzsch-Xing approach for
investigating the implications of these\cite{lepnonfritex5zero}.

In the leptonic sector there is an extra motivation for
introducing texture zeros\cite{brancolep4z}, namely the fact that
without an appeal to theory it is not possible to fully
reconstruct the neutrino mass matrix from experimental inputs.
This is further reinforced by the fact that if Dirac neutrino and
the Majorana mass matrix $ M_{D} $ and $ M_{R} $ are assumed to
have zero texture structure then the neutrino mass matrix remains
seesaw invariant under these transformations. This not only
simplifies the analysis of texture specific mass matrices in the
leptonic sector, but also suggests the possibility of quarks and
leptons having texture universality. Further, this provides a
motivation for exploring the compatibility of texture specific
mass matrices with GUT models such as SO(10), etc..

After having observed that the most general mass matrices can be
reduced to texture 1 zero hermitian mass matrices given by
Eq.~(\ref{tex1z}), it becomes desirable to check the implications
of quark and lepton mixing data on such texture specific fermion
mass matrices. In particular, if the mass matrices are `natural',
it becomes interesting to check the parameter space available to
the additional non zero diagonal elements `$e_q$' appearing in the
(1,1) positions, as compared to the texture  4 zero Fritzsch-like
fermion mass matrices, while fitting the precisely measured CKM
elements and other CKM parameters. In the case of leptons, it is
easy to check that such mass matrices would be able to explain the
neutrino mixing data as well. To this end, in the sequel we shall
discuss the attempt by Gupta\cite{lepnonfritex5zero} wherein the
texture 2 zero fermion mass matrices with parallel texture
structures obtained using the Fritzsch-Xing
approach\cite{fxwb1,fxwb2} have been considered. Following
Ref.~\refcite{lepnonfritex5zero}, beginning with the diagonalizing
transformation of texture 1 zero hermitian mass matrices,
presented in Appendix C, the implications for texture 2 zero
natural mass matrices for both quarks and leptons have been
discussed below.

\subsection{Texture 2 zero natural quark mass matrices and CKM matrix}
Considering texture 2 zero fermion mass matrices with parallel
texture structures obtained using the Fritzsch-Xing
approach\cite{fxwb1,fxwb2}, for $q= U / D$, the texture 1 zero
hermitian mass matrix can be expressed as \be M_{q} = \left( \ba
{ccc} e_q & a_{q} & 0 \\ a^{*}_{q} & d_{q} & b_{q} \\ 0 &
b^{*}_{q} & c_{q}
\\ \ea \right), \ee where $a_{q} =|a_{q}|e^{i\alpha_{q}}$
 and $b_{q} = |b_{q}|e^{i\beta_{q}}$.
One may note that in comparison to Fritzsch-like texture 4 zero
mass matrices, now one has an additional element $e_q$ appearing
at the (1,1) positions both in $ M_{u}$, $ M_{d}$. To study the
implications of this element $e_q$ one needs to first find the
diagonalizing transformation of these matrices, presented in
Appendix C, so as to construct the corresponding mixing matrix.

Following Appendix C, in order to deduce the quark mixing matrix,
one can easily use the approximations for the elements of the
diagonalizing transformations, e.g., $m_1 \ll m_2 \ll m_3 $ and
$c_q \gg m_1$, resulting into \be O_q = \left(\ba{ccc}
   1 &
    {\sqrt \frac{ (m_1-e_q)(c_q+m_2)}{c_q~m_2}} &
  \frac{m_2}{m_3}{\sqrt \frac{(m_1-e_q)(m_3-c_q)}{c_q~m_2}}\\

  {\sqrt \frac{c_q(m_1-e_q)}{m_2~m_3}} &
-{\sqrt \frac{(c_q+m_2)}{m_3}} &
 {\sqrt \frac{(m_3-c_q) }{m_3}} \\

  -{\sqrt \frac{(m_1-e_q)(m_3-c_q)(c_q+m_2)}{m_2~m_3~c_q}} &
 {\sqrt \frac{(m_3-c_q)}{m_3}} &
   {\sqrt \frac{(c_q+m_2)}{m_3}} \ea \right). \label{appou} \ee
This can be further simplified to the following form \be O_q =
\left(\ba{ccc}
   1 &
    k_{1q}k_{4q} &
 k_{2q}k_{4q}m_2 / k_{3q}{m_3}\\

  k_{3q}k_{4q} &
-k_{1q}k_{3q} &
 k_{2q} \\

  -k_{1q}k_{2q}k_{4q} &
 k_{2q} &
   k_{1q}k_{3q} \ea \right), \label{simpou} \ee
where the parameters $k_{1q}$, $k_{2q}$, $k_{3q}$ and $k_{4q}$ are
defined as follows \be k_{1q}= {\sqrt \frac {c_q+m_2}{c_q}},
~~k_{2q}={\sqrt\frac{m_3-c_q}{m_3}},
~~k_{3q}={\sqrt\frac{c_q}{m_3}}= 1 - k_{2q}^2,
~~k_{4q}={\sqrt\frac{m_1-e_q}{m_2}}. \ee

One can now compute the quark mixing matrix or the CKM matrix
through
\be
 V_{CKM} = O_u^{\dagger} P_u P_ d^{\dagger} O_d .  \ee
In general these can be expressed as
\be
 V_{i\sigma} = O_{1i}^{u} O_{1\sigma}^{d}  e^{-i\phi_{1}}+ O_{2i}^{u} O_{2\sigma}^{d}
 + O_{3i}^{u}O_{3\sigma}^{d}  e^{i\phi_{2}}. \ee
Explicitly, the various CKM matrix elements may be re-expressed in
terms of the quark mass ratios $ k_{1q}$, $k_{2q}$, $ k_{3q}$, $
k_{4q}$ and the phases $\phi_{1}$ and $ \phi_{2}$, e.g.,
\be
V_{ud}=
e^{-i\phi_{1}}+k_{4u}k_{4d}[k_{3u}k_{3d}+k_{1u}k_{2u}k_{1d}k_{2d}e^{i\phi_{2}}],
\ee
\be
V_{us}=
k_{1d}k_{4d}e^{-i\phi_{1}}-k_{4u}[k_{3u}k_{1d}k_{3d}+k_{1u}k_{2u}k_{2d}e^{i\phi_{2}}],
\ee
\be
V_{ub}={\frac{m_s}{m_b}}{\frac{k_{2d}k_{4d}}{k_{3d}}}e^{-i\phi_{1}}+
k_{4u}[k_{3u}k_{2d}-k_{1u}k_{2u}k_{1d}k_{3d}e^{i\phi_{2}}], \ee
\be
V_{cd}= k_{1u}k_{4u}e^{-i\phi_{1}}-k_{4d}[k_{1u}k_{3u}k_{3d}+
k_{2u}k_{1d}k_{2d}e^{i\phi_{2}}], \ee
\be
V_{cs}=
k_{1u}k_{1d}k_{4u}k_{4d}e^{-i\phi_{1}}+[k_{1u}k_{3u}k_{1d}k_{3d}
+k_{2u}k_{2d}e^{i\phi_{2}}], \ee
\be
V_{cb}={\frac{m_s}{m_b}}{\frac{k_{1u}k_{2d}k_{4u}k_{4d}}{k_{3d}}}e^{-i\phi_{1}}-
[k_{1u}k_{3u}k_{2d}-k_{2u}k_{1d}k_{3d}e^{i\phi_{2}}], \ee
\be
V_{td}={\frac{m_c}{m_t}}{\frac{k_{2u}k_{4u}}{k_{3u}}}e^{-i\phi_{1}}+
k_{4d}[k_{2u}k_{3d}-k_{1u}k_{3u}k_{1d}k_{2d}e^{i\phi_{2}}], \ee
\be
V_{ts}={\frac{m_c}{m_t}}{\frac{k_{2u}k_{1d}k_{4u}k_{4d}}{k_{3u}}}e^{-i\phi_{1}}-
[k_{2u}k_{1d}k_{3d}-k_{1u}k_{3u}k_{2d}e^{i\phi_{2}}], \ee
\be
V_{tb}={\frac{m_c}{m_t}}{\frac{m_s}{m_b}}{\frac{k_{2u}k_{2d}k_{4u}k_{4d}}{k_{3u}k_{3d}}}e^{-i\phi_{1}}+
[k_{2u}k_{2d}+k_{1u}k_{3u}k_{1d}k_{3d}e^{i\phi_{2}}]. \ee

To facilitate the understanding of CKM elements in terms of the
quark mass ratios, phases $\phi_{1}$, $\phi_{2}$ and the hierarchy
among the elements of the mass matrices, we define
\be
\xi_q= \frac {e_q}{m_1}~~~{\rm and}~~~~  \zeta_q= \frac{d_q}{c_q}.
\ee Subsequently, the parameters $ k_{1q}$, $ k_{2q} $, $ k_{3q} $
and $ k_{4q} $ appearing in the diagonalizing transformation $O_q$
and the elements of the CKM matrix may be rewritten as
 \be
  k_{1q}={\sqrt{1-(\frac{m_2}{m_3}(1+\zeta_q))}},
  ~~k_{2q}={\sqrt{\frac{\zeta_q}{1+\zeta_q}}
  },
  \ee
  \be
  k_{3q}= 1-k^{2}_{2q}= {\sqrt \frac{1}{1+\zeta_q}},
  k_{4q}={\sqrt{{\frac{m_1}{m_2}}(1-\xi_q)}}.
  \ee
As a result, the various CKM matrix elements can be expressed in
terms of these, e.g.,
\be
  V_{ud}=e^{-i\phi_1}+{\sqrt\frac{m_u m_d}{m_c m_s}}
  {\sqrt\frac{(1-\xi_u)(1-\xi_d)}{(1+\zeta_u)
  (1+\zeta_d)}}[1+k_{1u} k_{1d} \sqrt{\zeta_u \zeta_d}
   e^{i\phi_2}],
  \ee
the others can be written in a similar manner.

It has already been shown in the context of texture 4 zero quark
mass matrices that the precisely known parameter sin$\,2\beta$
gives vital clues for the structural features of mass matrices. It
was also mentioned that Verma {\it et al.}\cite{s2b} had
incorporated certain vital corrections to the formula of
sin$\,2\beta$, ignored in the earlier attempts of texture 4 zero
quark mass matrices, showing unambiguously that both the phases of
the mass matrices are essential for compatibility with the quark
mixing data. In the context of texture 2 zero natural quark mass
matrices, we would like to present the generalization of the
formula of sin$\,2\beta$ given in Ref.~\refcite{s2b}. It is easy
to check that for the natural mass matrices considered here, the
parameter $k_{1q}$ $\approx$ 1. Using the above approximations,
the formula of angle $\beta$ now becomes
\be
  \beta = {\rm arg}\left(1-\sqrt\frac {m_u m_s}{m_c m_d} \sqrt
  \frac{(1-\xi_u)}{(1-\xi_d)} e^{-i(\phi_1 + \phi_2)}\right)+{\rm
  arg}\left(\frac{1-r_2e^{i\phi_2}}{1-r_1
  e^{i\phi_2}}\right),
  \ee
 where the parameters $ r_1 $ and $ r_2 $ are defined as
  \be
  r_1=\frac{1}{\sqrt{\zeta_u/\zeta_d}}\left(1+\frac {m_s}{m_b}
  \frac{(1+\zeta_d)}{2}\right),
  \ee
  and
  \be
 r_2=\frac{1}{\sqrt{\zeta_u/
 \zeta_d}}\left(1-\frac {m_s}{m_b}
  \frac{(1+\zeta_d)}{2}\right).
  \ee
It is interesting to note that the various CKM elements and
related parameters can now be expressed entirely in terms of the
quark masses, the hierarchy parameters $ \xi_q $, $\zeta_q $ and
the phases $ \phi_1$ and $\phi_2 $  . Furthermore the above
relations incorporate the `next to leading order' terms and are
found to hold remarkably well within an error of less than
fraction of a percent, consistent with the present experimental
bounds. Apart from clearly underlying the dependence on quark
masses and both the phases $ \phi_1$ and $\phi_2 $, the above
formulae clearly depict the corrections induced by the parameters
$ e_q$ through the terms $ \xi_u $ and $ \xi_d$ viz-a-viz  the
corresponding relations obtained from the texture 4 zero
Fritzsch-like quark mass matrices.  It is easy to check that the
above relations reduce to the corresponding texture 4 zero
relations in case $ e_q = 0$ implying vanishing $ \xi_u $ and $
\xi_d$. Also it is interesting to observe that the relation for $
V_{cb} $ is the same in the texture 2 zero case and the texture 4
zero case. This is justified since the elements $ e_q $ only lead
to a mixing among the first and second generations as well as
first and third generations and not among the second and the third
generations. This is further evident from the minor corrections
appearing in the expressions for $V_{us} $ and $ V_{ub} $ in
relations in comparison to their counterparts in texture 4 zero
case.

In case one uses the quark masses at $ M_Z $ scale and the values
of the quark mixing parameters $ V_{us}$, $ V_{cb}$, $ V_{ub}$ and
sin$\,2\beta$, one can make an attempt to study the implications
of the additional element $ e_q $ on the quark mixing parameters.
The calculations suggest that the parameters $\xi_u$ and $\xi_d$
can assume only very small values and also that their parameter
space is very limited, this being compatible with the condition of
`naturalness' of the mass matrices. One arrives at similar
conclusions in case one makes an attempt to study the implications
of $ \xi_u $ and $ \xi_d$ on some of the vital CKM parameters such
as $ V_{us}$, $ V_{cb}$ ,$ V_{ub}$ and sin$\,2\beta$. In
particular, it is observed that $\xi_u$ and $\xi_d$ do not have
any pronounced effect on $V_{us }$, $V_{cb}$ as well as
sin$\,2\beta$. However, for the case of $ V_{ub}$ one finds that
for its inclusive values, the allowed ranges of the parameters
$\xi_u$ and $\xi_d$ are somewhat limited whereas in case one
restrict to exclusive values of $ V_{ub}$ then almost the entire
ranges of $\xi_u$ and $\xi_d$ are able to reproduce the results.

\subsection{Texture 2 zero lepton mass matrices and PMNS matrix}
Following the case of quarks, it becomes interesting to check
whether we arrive at the same conclusions in the case of texture 2
zero lepton mass matrices. For this purpose, one need not look
into an extensive analysis, rather it is informative to discuss
the effects of additional parameter $e_{\nu}$ in the case of
leptons. To this end, it is interesting to discuss the attempt by
Ref.~\refcite{lepnonfritex5zero} wherein the compatibility of such
texture 2 zero lepton mass matrices for Dirac as well as Majorana
neutrinos has been explored. It may be mentioned that while
carrying out the diagonalization of these mass matrices, the
approximations used in the quarks case do not seem to be valid,
therefore the exact diagonalizing transformations given in
Eq.~(\ref{ou}) need to be considered. It may be noted that in case
neutrinos are considered to be Dirac-like, the procedure for the
calculation of the mixing matrix is essentially the same as that
for the case of quarks considered in the previous section.
However, if neutrinos are considered to be Majorana-like, one
first obtains the light Majorana neutrino mass matrix $M_\nu$
using the type-I seesaw mechanism, $ M_\nu = -M^{T}_{\nu
D}M^{-1}_{\nu R}M_{\nu D}$, where $ M_{\nu D} $ and $ M_{\nu R} $
are respectively the Dirac neutrino mass matrix and right handed
Majorana neutrino mass matrix. For the case of normal hierarchy of
neutrinos characterized by $ m_{\nu_1}< m_{\nu_2}\ll m_{\nu_3} $,
one of the element of the diagonalizing transformation for the
neutrino mass matrix is
\be
O_{\nu M}(1,1)= \sqrt \frac{(e_{\nu M} +{\sqrt {m_{\nu M_2})}}
{(\sqrt{ m_{\nu M_3}}} - e_{\nu M}) (e_{\nu M}-{\sqrt {m_\nu
M_1)}}} {(e_{\nu M}-e_{\nu M}) {({\sqrt {m_{\nu M_3}}}} -{ \sqrt
{m_\nu M_1)}} {(\sqrt{ m_{\nu M_2}}} - {\sqrt{ m_\nu M_1)}}} ,\ee
the other elements can be found in a similar manner.

As already mentioned, texture 4 zero Fritzsch-like lepton mass
matrices with parallel texture structure for neutrinos and charged
leptons are able to accommodate the lepton mixing data quite well,
it is therefore expected that the same can also be achieved using
texture 2 zero hermitian lepton mass matrices with parallel
texture structure. In this context, it becomes interesting to find
the viable range for the additional  elements $ e_q$, $q = e, \nu
$ in these mass matrices, both for the case of Dirac neutrinos as
well as for Majorana neutrinos, especially when the condition of
`naturalness' is invoked on these mass matrices.

To this end, for the purpose of numerical calculations, using the
recent values of the lepton mixing angles as well as neutrino mass
square differences as inputs, one can examine the implications of
the lepton mixing angle $(s^{l}_{13})^2 $ (the superscript $l$
denotes the `lepton' mixing angle in order to differentiate it
from the corresponding `quark' mixing angle) on the (1,1) elements
in these matrices for the Dirac neutrinos as well as for the
Majorana neutrinos.  In the case of Dirac neutrinos, it is
observed that the Dirac neutrino masses take the following values
\be
m_{\nu D_1}=2.3 \times 10^{-13} -   9.9 \times 10^{-12} {\rm GeV},
\ee

\be
 m_{\nu D_2}= 8.3 \times 10^{-12} - 1.3 \times 10^{-11} {\rm GeV},
 \ee

 \be
  m_{\nu D_3}= 4.4 \times 10^{-11} - 5.3 \times 10^{-11} {\rm
  GeV}.
  \ee
It is interesting to note that the parameter $e_e$ takes all
values between 0 to $m_e \sim 5 \times 10^{-4}$ GeV, while the
parameter $e_{\nu D}$ takes values between 0 to 7 $\times 10
^{-12}$ GeV which is about 70$\%$ of the value of the lightest
Dirac neutrino mass $ m_{\nu D_1} $, consistent with the condition
of `naturalness' imposed on these mass matrices. It can also be
seen that the additional free parameters $ e_e $ and $ e_{\nu D} $
do not show any pronounced effect on the lepton mixing angle
$(s^{l}_{13})^2 $. This clearly indicates that provided the
condition of `naturalness' is obeyed by these mass matrices, the
results obtained using texture 2 zero hermitian lepton mass
matrices and those obtained using texture 4 zero Fritzsch-like
lepton mass matrices are essentially the same since the additional
(1,1) elements $ e_e $ and $ e_{\nu D} $  in the texture 2 zero
hermitian lepton mass matrices do not have any pronounced effect
on the lepton mixing data.

In a similar manner, for the Majorana neutrinos it is observed
that the light Majorana neutrino masses, obtained through the
type-I seesaw mechanism, take the following values
 \be
     m_{\nu M_1}=2.8 \times 10^{-13} - ~~~ 9.9 \times 10^{-12} {\rm GeV},
     \ee
     \be
 m_{\nu M_2}=8.3 \times 10^{-12} - ~~~1.3 \times 10^{-11} {\rm GeV},
 \ee
 \be
  m_{\nu M_3}=4.4 \times 10^{-11} - ~~~5.3 \times 10^{-11} {\rm GeV}.
  \ee
  In this case, the parameter $ e_e $ again takes all values between 0 to
   $ m_e \ \sim 5 \times 10 ^{-4} $   GeV, while the parameter $
   e_{\nu D} $
takes a maximum value which is about 50$ \% $ of the value of the
square root of the lightest Majorana neutrino mass i.e. $ \sqrt
{m_{\nu M_1}} $ consistent with the condition of `naturalness'
imposed on these mass matrices. Additionally, in case one assumes
that the right handed Majorana neutrino mass matrix elements $ m_R
$ are of the order of 2 $ \times  10^{14}$ GeV, corresponding to
the intermediate energy scale\cite{pati} in GUTs, then the values
of the Dirac neutrino masses predicted in this case are as follows
\be
     m_{\nu D_1} = 7.6 - ~~~44.5~{\rm GeV},
     \ee
     \be
 m_{\nu D_2} = 40.9 -~~~ 51.6~{\rm GeV},
 \ee
 \be
  m_{\nu D_3} = 94.5 -~~~ 102.8~{\rm GeV}.
  \ee
It is interesting to note that the order of these masses is the
same as that of the up sector of quarks, a feature that emerges
naturally\cite{pati}\cdash\cite{joshipurapatel} in GUTs like
SO(10).

These calculations reveal that even in the case of Majorana
neutrinos, both the additional parameters $e_e$ and $ e_{\nu M} $
have almost no pronounced effect on the lepton mixing data as
compared to the texture 4 zero Fritzsch-like lepton mass matrices.
In conclusion, we can state that the condition of `naturalness'
re-ensures that the results obtained using texture 2 zero lepton
mass matrices and those obtained using texture 4 zero
Fritzsch-like lepton mass matrices are essentially the same.

\section{SO(10) and texture specific fermion mass matrices \label{texso10}}
After having discussed that the phenomenological texture specific
mass matrices are able to accommodate the fermion mixing data and
further noting that these are also compatible with the
`naturalness' condition and weak basis transformations, one may
now mention that recently a few
authors\cite{leptex5,joshipurapatel}\cdash\cite{sdevso10} have
also observed the importance of texture 4 zero Fritzsch-like mass
matrices in the context of SO(10). It may be noted that an
extensive and detailed review of some of these as well as some
other attempts to explain fermion masses and mixings within the
framework of SO(10) GUTs has already been carried out by Chen and
Mahanthappa\cite{chen}, however in the present case, our emphasis
will be on the issue of the compatibility of the textures of the
mass matrices within the constraints of the SO(10) formalism.

In particular, Fukuyama {\it et al.}\cite{fukuyamaso10} have
carried out an analysis wherein they have investigated the
compatibility of texture specific mass matrices, having similar
forms for quarks and leptons, with the SO(10) inspired mass
matrices. However, their analysis is not able to simultaneously
fit the quark mixing and the lepton mixing data, in particular
they are not able to reproduce the solar mixing angle with the
constraints on the neutrino masses coming from the oscillation
data. Further, Joshipura and Patel\cite{joshipurapatel} have also
recently carried out an extensive analysis of mass matrices based
on SO(10), however, they do not apply any `textures' as well as
the condition of `naturalness' on these mass matrices. Similarly,
the emphasis of analysis by Dev {\it et al.}\cite{sdevso10} is to
fit the lepton mixing data in the context of mass matrices based
on SO(10), however again the condition of `naturalness' on these
mass matrices has not been imposed as well as the quark textures
have not been taken into consideration. Very recently a detailed
and comprehensive analysis of the texture 4 zero mass matrices in
the context of SO(10), with the latest constraints of quark and
lepton mixing parameters has been carried out by Verma {\it et
al.}~\cite{throhitso10} wherein the condition of `naturalness' on
these mass matrices has been imposed as well. In the sequel, we
present the essentials of SO(10) and SO(10) based mass matrices
and broad conclusions of the analysis carried out by
Ref.~\refcite{throhitso10}.

\subsection{Introduction to SO(10)} It may be noted that the SO(10)
group incorporates several interesting features that makes it a
very promising and a leading GUT group\cite{pati1}. For example,
the SO(10) group achieves complete quark-lepton symmetry by
unifying all the 15 known fermions, along with the right handed
neutrino of each family, into one sixteen dimensional spinor
representation denoted by 16. Similarly, it is not only compatible
with supersymmetry, but also includes the seesaw
mechanism\cite{seesaw1}\cdash\cite{seesaw5} for explaining the
small neutrino masses.

The SO(10) group is a rank 5 orthogonal group with 10 dimensional
fundamental or vector representation. The fifteen fermions of each
generation that belong to the $\bar{5} +10$ of SU(5) and CP
conjugate of right handed neutrino can be accommodated into the 16
dimensional spinor representation of SO(10) as
\be
\left(
\begin{array}{cccc}
 u_1 & u_2 & u_3 & \nu_e\\
d_1 & d_2 & d_3 & e^-\\
\end{array}
\right) + \left(
\begin {array}{cccc}
  u_1^c & u_2^c & u_3^c & \nu_e^c\\
d_1^c & d_2^c & d_3^c & e^+\\
\end{array}
\right) \ee\\ where the symbols have their usual
meanings{\cite{sarkar}}.

Since SO(10) is a rank 5 gauge group and the SM has a rank 4,
there are many possible chains of symmetry breaking through which
SO(10) can descend to the gauge group for the Standard Model,
$G_{SM}=SU(3)_c \times SU(2)_L \times U(1)_Y$\cite{mohapatra}. The
two usually considered symmetry breaking chains of SO(10) are
\\
\\
Chain 1 : $ SO(10) \longrightarrow SU(5) \times U(1)
\longrightarrow SU(5) \longrightarrow G_{SM},$
\\
\\
Chain 2 : $ SO(10) \longrightarrow SU(4)_c \times SU(2)_L \times
SU(2)_R
 \longrightarrow \\~~~~~~~~~~~~~~SU(3)_c \times  SU(2)_L \times SU(2)_R \times U(1)_{B-L}
  \longrightarrow \\~~~~~~~~~~~~~~G_{SM} \longrightarrow G_{EW}. $
\\
\\
The first one is ruled out as the proton decays in this case much
faster than required, therefore the second possibility of symmetry
breaking in SO(10) is usually adopted.

\subsection{Yukawa sector in SO(10)} In renormalizable SO(10), only
three types of Higgs fields can couple to fermions, e.g.,
\be
16\times 16= 10_S + 120_A + {\overline{126}}_S, \ee where the
symbols $S$ and $A$ refer to the symmetry property under
interchange of two family indices in the Yukawa couplings
$Y_{AB}$. The gauge invariant Yukawa couplings are given
as\cite{chen}
\be
S_{AB}^{10} {(16)}_a {(16)}_b \phi_{10}, ~~A_{AB}^{120} {(16)}_a
{(16)}_b \phi_{120},
 ~~S_{AB}^{\overline{126}} {(16)}_a {(16)}_b \phi_{\overline{126}}. \ee
The relevant SO(10) representations have the following
decomposition in terms of the Pati Salam Group, $SU(4)_c \times
SU(2)_L \times SU(2)_R $,
 \be 10 = (6,1,1) +
(1,2,2), \ee \be 16 = (4,2,1) + (\bar{4},1,2), \ee \be 45 =
(1,3,1) + (1,1,3) + (15,1,1) + (6,2,2), \ee \be 54 = (1,1,1) +
(1,3,3) + (20 ^\prime,1,1) + (6,2,2), \ee \be 120 = (1,2,2)
+(10,1,1) + (\overline{10},1,1) + (6,1,3) + (6,3,1) + (15,2,2),
\ee \be 126 = (15,2,2) + (10,1,3) + (\overline{10},3,1) + (6,1,1).
\ee

We know that $ 16 = (4,2,1) + (\bar{4},1,2)$ and  $(4,2,1) \times
(\bar{4},1,2) = (15,2,2) + (1,2,2)$ so the Dirac masses for quarks
and leptons are generated when neutral components in $(1,2,2)$
multiplet in 10, $(15,2,2)$ and $(1,2,2)$ in 120 and $(15,2,2)$ in
126 dimensional representations acquire non vanishing expectation
values. On the other hand, the ($\overline{10}, 3, 1)$ and $(10,
1, 3)$ components of 126 dimensional Higgs break the $SU(2)_L$ and
$SU(2)_R$ symmetries and hence are responsible for the left and
right handed Majorana neutrino masses through the Higgs
lepton-lepton interactions ($\overline{10}, 3, 1)$ $(4,2,1)$
$(4,2,1)$ and $(10, 1, 3)$ ($\bar{4},1,2)$ $(\bar{4},1,2)$
respectively. Being able to achieve a complete quark-lepton
symmetry, SO(10) has the promise for explaining the pattern of
fermion masses and mixing in a renormalizable form wherein only
three Yukawa coupling matrices $ S^{10}, S^{\overline{126}},
A^{120} $ and relative strengths
 between them determine six physical mass matrices $M_f$ with f = u, d, e, $\nu_D, \nu_L, \nu_R$.
The fermion masses are generated when the Higgs fields of 10, 120
and  dimensional SO(10) representation (denoted by $\phi_{10}$,
$\phi_{120}$ and $\phi_{\overline{126}}$ respectively) develop non
vanishing expectation values and lead to the following mass
matrices\cite{botino,goran}
\be
M_u= S^{10}<\phi_{10}^{+}> + A^{120} ( < \phi_{120}^+> +
\frac{1}{3} <\phi_{120}^{\prime +}>) +
S^{\overline{126}}\frac{1}{3}<\phi_{\overline{126}^+}> ,
\label{so10mu} \ee

\be
M_d= S^{10}<\phi_{10}^{-}> + A^{120} ( -< \phi_{120}^-> +
\frac{1}{3} <\phi_{120}^{\prime -}>) -
S^{\overline{126}}\frac{1}{3}<\phi_{\overline{126}^-}> ,
\label{so10md} \ee

\be M_e= S^{10}<\phi_{10}^{-}> + A^{120} ( -< \phi_{120}^-> -
<\phi_{120}^{\prime -}>) +
S^{\overline{126}}<\phi_{\overline{126}^-}> , \label{so10me} \ee

\be
M_{\nu_D}= S^{10}<\phi_{10}^{+}> + A^{120} ( < \phi_{120}^+> -
<\phi_{120}^{\prime +}>) -
S^{\overline{126}}\frac{1}{3}<\phi_{\overline{126}^+}> .
\label{so10mnud} \ee

Note that a Clebsch-Gordon coefficient (-3) is generated in the
lepton sectors when the $SU(4)_c \times SU(2)_L \times SU(2)_R $
components (15,2,2) are involved. The Majorana neutrino mass
matrices are given by \be M_{\nu_L} = S^{\overline{126}} <
\phi_{\overline{126}^{\prime 0}}> ,\ee \be M_{\nu_R} =
S^{\overline{126}} < \phi_{\overline{126}^{\prime +}}> ,\ee where
the superscripts +/0/- refer to the sign of the hypercharge Y.
Furthermore $<\phi_{10}^{\pm}>$ are the vacuum expectation values
of the Higgs fields of  $\phi_{10}$, $\phi_{120}^{\prime \pm}$ are
those of $\phi_{120}$ and $<\phi_{\overline{126}}^{\pm}>$,
$<\phi_{\overline{126}} ^{\prime 0}>$,
$<\phi_{\overline{126}}^{\prime +}>$ are those of
$\phi_{\overline{126}}$. As already mentioned, the matrices
$S^{10}$ and $S^{\overline{126}}$ are complex symmetric while
$A^{120}$ is complex anti-symmetric in nature.

\subsection{SO(10) based mass matrices}
It is possible to define the fermion mass matrices by considering
the minimal and non minimal approaches, both in the context of non
Supersymmetric (non-SUSY) and SUSY SO(10) frameworks. The minimal
approaches involve fermion mass matrices with Yukawa contributions
coming from only two Higgs i.e. $ \phi_{10} $ and $\phi_{126}$ or
$\phi_{120}$ and $ \phi_{126}$. Keeping in mind the limitations
involved\cite{joshipurapatel,botino,goran} in these minimal
scenarios, we discuss here the non minimal case incorporating the
$S_{AB} ^{10}, A_{AB}^{120}$ and $S_{AB}^{126}$. Although some
investigations have been carried
out\cite{leptex5,fukuyamaso10,sdevso10} to check their
compatibility with texture specific mass matrices, a detailed and
comprehensive analysis is yet to be carried out. Further, the
compatibility of the mass matrices based on SO(10) has not been
carried out in the case of texture specific mass matrices
incorporating `weak' hierarchy.

To begin with, we consider the mass matrices formulated by several
authors\cite{joshipurapatel,fukuyamaso10,altarelli}\cdash\cite{grimus},
within the context of SO(10). The approach adopted by Fukuyama
{\it et al.}~\cite{fukuyamaso10} wherein the 6 physical mass
matrices in Eqs.~(\ref{so10mu})-(\ref{so10mnud}) can be reduced to
a simpler form as \be M_u = S + \sigma S^\prime + \xi^ \prime A
,\ee \be M_d = \eta S + S^{\prime} + A ,\ee \be M_e = \eta S - 3
S^\prime + \xi A ,\ee \be M_D = S - 3\sigma S^\prime + \xi^{\prime
\prime} A ,\ee \be M_{\nu_L} = \beta S^\prime , M_{\nu_R} = \gamma
S^ \prime ,\ee\\ where \be S = S^{10} < \phi_{10}^+ >, S^\prime =
-S^{\overline{126}} \frac{1}{3}
< \phi_{126} ^->, A= A^{120} (-<\phi_{120}^-> +
\frac{1}{3}<\phi_{120}^{\prime -}>), \ee \be \eta = (<\phi_{10}
^-> / <\phi_{10}^+>) ,\ee \be \sigma = - (<\phi_{126} ^+> /
<\phi_{126}^->) ,\ee \be \xi = (-<\phi_{120}^-> -
<\phi_{120}^{\prime -})/ (-<\phi_{120}^-> + 1/3
<\phi_{120}^{\prime -}> ) ,\ee \be \xi^{\prime} = (<\phi_{120}^+> +
1/3 <\phi_{120}^{\prime +}> ) / (-<\phi_{120}^-> + 1/3
<\phi_{120}^{\prime -}) ,\ee \be \xi^{\prime \prime } =
(<\phi_{120}^+> - <\phi_{120}^{\prime +}> ) / (-<\phi_{120}^-> +
1/3 <\phi_{120}^{\prime -}) ,\ee \be \beta = - (
<\phi_{\overline{126}}^{\prime 0}>)/( 1/3 <\phi_{\overline{126}}
^->), \ee \be \gamma = -(<\phi_{\overline{126}}^{\prime +})/( 1/3
<\phi_{\overline{126}} ^->) .\ee
The matrices $S$ and $S^\prime$ are complex symmetric while $A$ is
complex anti-symmetric in nature, and the parameters $\sigma,
\eta, \xi, \xi^\prime, \xi^{\prime \prime}, \beta, \gamma$ are
dimensionless complex parameters of which $\beta$ and $ \gamma$
can be chosen to be real without loss of
generality\cite{joshipurapatel}.

Similarly, using the approaches given by Altarelli and
Blankenburg\cite{altarelli} as well as by Dutta {\it et
al.}~\cite{dutta}, the above expressions take the form \be M_u =
r(H + sF + t_u G), \label{mualt} \ee \be M_d =  H + F + G, \ee \be
M_e = H -3F + t_e G, \ee \be M_{\nu_D} = r(H-3sF + t_D G), \ee \be
M_{\nu_L}= r_L F, \ee \be M_{\nu_R}= r_R^{-1} F, \label{mnualt}
\ee where H and F are symmetric coupling matrices while G is
anti-symmetric. It may also be mentioned that a few other
authors\cite{joshipurapatel,grimus} have also derived these
relations, however all these approaches can be shown to be
essentially equivalent. The parameters of the above mentioned two
approaches are related as \beqn S= rH = h,~~ S^\prime = F = r_1 f,
~~A = G = r_1 h^\prime, ~~\eta = 1/r = r_1, \\\sigma = sr = r_2/
r_1, ~~\xi = t_e = c_e, ~~\xi^\prime = r t_u =
r_3/r_1,~~~~~~~~~~~~~~~~\\ \xi^{\prime \prime} = rt_{D \nu} =
c/r_1, ~~\beta = r_L = v, ~~\nu_R = r_R ^{-1} =
v.~~~~~~~~~~~~~~~~~\eeqn It may be mentioned that the above mass
matrices can easily be related to hermitian mass matrices in
SO(10) incorporating L-R symmetry\cite{botino,goran,moorehouse}.
In such a case, the parameters $ \sigma, \eta, \xi, \xi^ \prime,
\xi^{\prime \prime}, \beta$ and $\gamma$ can be treated as real.
It is important to mention that the above mentioned approaches are
valid in both the non SUSY SO(10) as well as the SUSY SO(10)
scenarios.

\subsection{Compatibility of texture 4 zero hermitian mass matrices
with SO(10)} As a next step, it becomes an interesting exercise to
check the compatibility of texture specific mass matrices with
SO(10) based mass matrices. To this end, we present the essentials
of Refs.~\refcite{lepnonfritex5zero,throhitso10} wherein the
authors have imposed textures and hermiticity on the mass matrices
given in Eqs.~(\ref{mualt})-(\ref{mnualt}) and examined the
compatibility of these. It may be mentioned that in case the
matrices $ M_u, M_d, M_e$ and $M_{\nu D}$ are considered to be
texture 4 zero hermitian mass matrices, then the parameters  r, s,
$t_u, t_e, t_D, r_L, r_R$ have to be real and the matrices H and F
have to be real symmetric while G has to be purely imaginary and
anti-symmetric.

To understand the constraints imposed by SO(10) on texture 4 zero
fermion mass matrices, the authors have considered texture 4 zero
Fritzsch-like mass matrices as
\be
M_f=\left( \ba {ccc} 0 & a_{f}e^{i\alpha_f} & 0 \\
                     a_{f}e^{-i\alpha_{f}} & d_{f} & b_{f}e^{i\beta_f}\\
                     0 & b_{f}e^{-i \beta_f} & c_{f} \ea \right),\ee
where $f=u, d, l, \nu_D$. It may be noted that the notation used
in the above equation is somewhat different than the one used
earlier. This has been done keeping in mind the unified treatment
of quarks and leptons to be presented in the context of SO(10).
The corresponding left and right handed Majorana neutrino mass
matrices are real symmetric and defined as
\be
M_{k}=\left( \ba {ccc} 0 & a_{k} & 0 \\
                     a_{k} & d_{k} & b_{k}\\
                     0 & b_{k} & c_{k} \ea \right), \ee
where $k=\nu_L, \nu_R$. Keeping in mind the texture imposed on
$M_u$, $M_d$, etc., the matrices H, F and G can be defined as
\begin{equation}
H= \left( \begin{array}{ccc} 0 & a_H & 0\\ a_H & d_H & b_H\\ 0 &
b_H & c_H \end{array} \right), F = \left( \begin{array}{ccc} 0 &
a_F & 0\\ a_F & d_F & b_F\\ 0 & b_F & c_F \end{array} \right), G =
\left( \begin{array}{ccc} 0 & ia_G & 0\\ -ia_G & 0 & ib_G\\ 0 &
-ib_G & 0 \end{array} \right),
\end{equation}
resulting into Eqs.~(\ref{mualt})-(\ref{mnualt}) being
re-expressed as \be M_u = r(H + sF + t_u G) = S_u + A_u ,
\label{mualtreexp} \ee \be M_d = H + F + G = S_d + A_d ,\ee \be
M_e = H -3F + t_e G = S_e + A_e ,\ee \be M_{\nu_D} = r(H-3sF + t_D
G)= S_D + A_D ,\ee \be M_{\nu_L}= r_L F = S_L, M_{\nu_R}= r_R^{-1}
F = S_R, \label{mnualtreexp} \ee where the matrices $S_u$, $A_u$,
etc. respectively represent the real symmetric and the imaginary
anti-symmetric parts of $M_u$, etc.. These are defined as \beqn
S_u = r(H+ sF), ~~S_d = H + F, ~~S_e = H-3F, ~~S_D = r(H-3sF),\\
A_d = G, ~~A_u = r t_u G, ~~A_e = t_e G, ~~A_D = r t_D G.
 ~~~~~~~~~~~~~~~~~~~~~~~~~~~\eeqn
Using Eqs.~(\ref{mualtreexp})-(\ref{mnualtreexp}), one obtains
\begin{equation}
S_u = \left( \begin{array}{ccc} 0 & a_u cos \alpha_u & 0\\ a_u cos
\alpha_u & d_u & b_u cos \beta_u \\ 0 & b_u cos \beta_u & c_u
\end{array} \right) , \label{sumat} \end{equation}
\begin{equation}
S_d = \left( \begin{array}{ccc} 0 & a_d cos \alpha_d & 0\\ a_d cos
\alpha_d & d_d & b_d cos \beta_d \\ 0 & b_d cos \beta_d & c_d
\end{array} \right) ,\end{equation}
\begin{equation}
S_e = \left( \begin{array}{ccc} 0 & a_e cos \alpha_e & 0\\ a_e cos
\alpha_e & d_e & b_e cos \beta_e \\ 0 & b_e cos \beta_e & c_e
\end{array} \right) ,\end{equation}
\begin{equation}
S_D = \left( \begin{array}{ccc} 0 & a_D cos \alpha_D & 0\\ a_D cos
\alpha_D & d_D & b_D cos \beta_D \\ 0 & b_D cos \beta_D & c_D
\end{array} \right) ,\end{equation}
\begin{equation}
A_u = \left( \begin{array}{ccc} 0 & i a_u sin \alpha_u & 0\\ - i
a_u sin \alpha_u & d_u & i b_u sin \beta_u \\ 0 & -i b_u sin
\beta_u & c_u \end{array} \right) ,\end{equation}
\begin{equation}
A_d = \left( \begin{array}{ccc} 0 & i a_d sin \alpha_d & 0\\ - i
a_d sin \alpha_d & d_d & i b_d sin \beta_d \\ 0 & -i b_d sin
\beta_d & c_d \end{array} \right) ,\end{equation}
\begin{equation}
A_e = \left( \begin{array}{ccc} 0 & i a_e sin \alpha_e & 0\\ - i
a_e sin \alpha_e & d_e & i b_e sin \beta_e \\ 0 & -i b_e sin
\beta_e & c_e \end{array} \right) ,\end{equation}
\begin{equation}
A_D = \left( \begin{array}{ccc} 0 & i a_D sin \alpha_D & 0\\ - i
a_D sin \alpha_D & d_D & i b_D sin \beta_D \\ 0 & -i b_D sin
\beta_D & c_D \end{array} \right) ,\end{equation}
\begin{equation}
S_L= r_L\left( \begin{array}{ccc} 0 & a_F & 0\\ a_F & d_F & b_F\\
0 & b_F & c_F \end{array} \right) = \left( \begin{array}{ccc} 0 &
a_L & 0\\ a_L & d_L & b_L\\ 0 & b_L & c_L \end{array} \right),
\end{equation}
\begin{equation}
S_R= r_R ^{-1}\left( \begin{array}{ccc} 0 & a_F & 0\\ a_F & d_F &
b_F\\ 0 & b_F & c_F \end{array} \right) = \left(
\begin{array}{ccc} 0 & a_R & 0\\ a_R & d_R & b_R\\ 0 & b_R & c_R
\end{array} \right).  \label{srmat} \end{equation}
It may be noted that the 17 real free parameters of $M_u, M_d,
M_e, M_{\nu D}, M_{\nu L}$ and $M_{\nu R}$ correspond to 7 real
parameters $r, s, t_u, t_e, t_D, r_L, r_R$, 4 parameters $a_H,
b_H, c_H, d_H$ of H, 4 parameters  $a_F, b_F, c_F, d_F$ of F and 2
parameters $a_G, b_G$ of G. Using
Eqs.~(\ref{mualtreexp})-(\ref{mnualtreexp}) one obtains \beqn H =
\frac{1}{4}(S_d- S_e), F= \frac{1}{4}(3S_d + S_e), G = A_d, \\
r(1-s) S_e = 4 S_u - (3 + s) r S_d,\\ S_D = S_u - rs(S_d - S_e),\\
A_u = r t_u A_d, A_e = t_e A_d, A_D = r t_D A_d. \eeqn

Using the above equations, along with
Eqs.~(\ref{sumat})-(\ref{srmat}), one obtains the following
relations among the elements of the texture specific mass matrices
and the SO(10) based mass matrices, e.g.,
\be
r(1-s)a_e ~cos \alpha_e = 4 a_u ~cos\alpha_u -
(3+s)ra_d~cos\alpha_d,\ee \be r(1-s)b_e ~cos \beta_e = 4 b_u
~cos\beta_u - (3+s)rb_d ~cos\beta_d,\ee \be
r(1-s)d_e=4d_u-(3+s)rd_d,\ee \be r(1-s)c_e=4c_u-(3+s)rc_d,\ee \be
a_D~cos\alpha_D= a_u~cos\alpha_u-rs(a_d~
cos\alpha_d-a_e~cos\alpha_e), \ee \be b_D~cos\beta_D=
b_u~cos\beta_u-rs(b_d~ cos\beta_d-b_e~cos\beta_e), \ee \be
d_D=d_u-rs(d_d-d_e), \ee \be c_D=c_u-rs(c_d-d_e), \ee \be a_e~sin
\alpha_e=t_e a_d~sin\alpha_d,\ee \be b_e~sin \beta_e=t_e
b_d~sin\beta_d,\ee \be a_u~sin \alpha_u=rt_u a_d~sin\alpha_d,\ee
\be b_u~sin \beta_u=rt_u b_d~sin\beta_d,\ee \be a_D~sin
\alpha_D=rt_D a_d~sin\alpha_d,\ee \be b_D~sin \beta_D=rt_D
b_d~sin\beta_d.\ee The above large number of coupled equations
need to be solved for checking the compatibility of texture
specific mass matrices with the constraints imposed by SO(10). The
procedure being followed by
Refs.~\refcite{lepnonfritex5zero,throhitso10} is to construct the
texture 4 zero mass matrices and check their viability with the
available data. To this end, the authors consider the fermion
masses at the GUT scale ${\rm M_X} = 2 \times 10^{16}$ GeV as
provided in the Refs.~\refcite{xingmass} and
\refcite{joshipurapatel}. It may be mentioned that
Ref.~\refcite{joshipurapatel} considers the mixing angles to be
scale independent, however, it has been shown\cite{chen} that in
the case of quarks, there is a slight scale dependence of the
elements of the CKM matrix. Therefore, both the cases where the
CKM matrix elements $V_{ub}$, $V_{cb}$, $V_{td}$ and $V_{ts}$ are
scale independent as well as when these are scale dependent have
been considered. In the latter case, it can be shown\cite{chen}
that at the GUT scale these elements get re-scaled as \be V_{ij} =
V_{ij}^0 B_t^{-1}, \ee where $ij$= $ub$, $cb$, $td$, $ts$ and
$B_t$ is the running coupling constant induced by the top quark
Yukawa coupling and varies between 0.7 to 0.9. For the leptonic
sector, since the neutrino mass squared differences and the lepton
mixing parameters do not show much scale dependence, their $M_Z$
values as quoted by Fogli {\it et al.}~\cite{foglinew} have been
used for the purpose of calculations.

As a first step, the parameters of the mass matrices $M_u$ and
$M_d$, e.g., $a_u, b_u, c_u, d_u, a_d, b_d, c_d, d_d, \phi_1$ and
$\phi_2$ have been found by imposing the constraints of the quark
mixing data. It may be noted that the mixing data does not impose
any constraint on the absolute values of the phases $\alpha_u,
\alpha_d, \beta_u$ and $\beta_d$ appearing in the quark mass
matrices. As a next step, the element $d_e$ of the mass matrix
$M_e$ is considered as the free parameter to obtain the parameters
r and s in terms of the diagonal elements of the mass matrices
$M_{u,d,e}$, e.g., \beqn r= \frac{(d_u c_e - c_u d_e) +(d_d c_u -
d_u c_d)}{(c_e d_d - c_d d_e)},\\ s= 1- \frac{4(d_u c_d-d_d
c_u)}{r(d_e c_d- d_d c_e)}. \eeqn

It may further be noted that by varying $\alpha_u$, it is possible
to find $\alpha_d$ through the condition $\alpha_d = \alpha_u -
\phi_1$ . This allows the calculation of the parameter $t_u$,
e.g., \be t_u = \frac{a_u sin~\alpha_u}{r a_d sin \alpha_d}. \ee
Further, one can obtain the value of the phase $\beta_u$ as
\be
\beta_u = tan^{-1}[\frac{r t_u b_d sin \phi_2}{(r t_u b_d cos
\phi_2) - b_u}], \ee using which the phase $\beta_d$ can be found
through the relation \be \beta_d = \beta_u - \phi_2. \ee The
phases $\alpha_e$ and $\beta_e$, can also be found, e.g.,
\be
\alpha_e = tan^{-1} \left[ \frac{t_e a_d sin \alpha_d}{a_u cos
\alpha_u
          {(\frac{c_e d_d - d_e c_d}{c_u d_d- d_u c_d}})-{{a_d cos \alpha_d
           ( \frac{c_e d_u - d_e c_u}{c_u d_d - d_u c_d})}}} \right], \ee
\be
\beta_e = tan^{-1} \left[ \frac{t_e b_d sin \beta_d}{b_u cos
\beta_u
          {(\frac{c_e d_d - d_e c_d}{c_u d_d- d_u c_d}})-{{b_d cos \beta_d
           ( \frac{c_e d_u - d_e c_u}{c_u d_d - d_u c_d})}}}  \right], \ee
wherein the parameter $t_e$ can be considered to be free.
Subsequently, considering $\alpha_D$ as a free parameter, one gets
\be a_D= (a_u cos \alpha_u - r s(a_d cos \alpha_d - a_e cos
\alpha_e))/ cos \alpha_D. \ee This allows one to calculate the
parameter $t_D$ as \be t_D = \frac{a_D sin \alpha_D} {r a_d sin
\alpha_d}. \ee The remaining elements of the Dirac neutrino mass
matrix, in terms of the parameters already found, can be obtained
as
\be
b_D = \sqrt{(rt_D b_d sin \beta_d)^2 + {(b_u cos \beta_u -rs(b_d
cos \beta_d - b_e cos \beta_e))}^2},\ee \be d_D = d_u - rs(d_d -
d_e),\ee \be c_D = c_u - rs(c_d - c_e). \ee This allows one to
evaluate the elements of the mass matrices H and F as \beqn a_H =
\frac{1}{4}(3 a_d cos \alpha_d + a_e cos \alpha_e),\\ b_H =
\frac{1}{4}(3 b_d cos \beta_d + b_e cos \beta_e),\\ d_H =
\frac{1}{4}(3 d_d + d_e),\\ c_H = \frac{1}{4} (3 c_d + c_e),\\ a_F
= \frac{1}{4}( a_d cos \alpha_d - a_e cos \alpha_e),\\ b_F =
\frac{1}{4}( b_d cos \beta_d - b_e cos \beta_e),\\ d_F =
\frac{1}{4} (d_d - d_e),\\ c_F = \frac{1}{4} (c_d - c_e). \eeqn

As a next step, $r_R$ can be considered as a free parameter and
the elements of the right handed Majorana neutrino matrix can be
evaluated as \be a_R = r_R ^{-1} a_F, b_R = r_R ^{-1} b_F, d_R =
r_R ^{-1} d_F, c_R = r_R ^{-1} c_F. \ee The light Majorana
neutrino mass matrix may be obtained using the type-I seesaw
mechanism, \be M_\nu = - M_{\nu_D}^T M_{\nu_R}^{-1} M_{\nu_D} =
-r_R ^{-1} M_{\nu_D}^T F^{-1} M_{\nu_D} \ee or by using the
type-II seesaw mechanism defined as \be M_\nu = M_{\nu L} - M_{\nu
D}^T M_{\nu R} ^{-1} M_{\nu D}. \ee However, the results obtained
in the two cases are not very different, therefore
Refs.~\refcite{lepnonfritex5zero} and \refcite{throhitso10} have
discussed the results corresponding to the type-I seesaw mechanism
only. The eigenvalues $(m_{\nu1}, m_{\nu2}, m_{\nu3})$ and the
complex diagonalizing matrix $O_\nu$ of the matrix $M_\nu$ are
then numerically computed and are used along with the
diagonalizing transformations for the charged lepton matrix to
compute the PMNS mixing matrix.

It may be noted that there are only 16 free parameters in case one
uses type-I seesaw mechanism which further increases to 17 in the
case of type-II seesaw mechanism. For the purpose of calculations,
along with type-I seesaw mechanism, the parameters $a_u, b_u, c_u,
d_u, a_d, b_d, c_d, d_d, a_e, b_e, c_e, d_e, \alpha_u, t_e,
\alpha_D$ and $r_R^{-1}$ have been used. Interestingly, one finds
that the present fermion mixing data is very well accommodated by
the texture 4 zero SO(10) inspired mass matrices. The constraints
from the CKM matrix elements have implications only on the
hierarchy of the quark mass matrices and on the phase differences
$\phi_1 = \alpha_u -\alpha_d$ and $\phi_2 = \beta_u - \beta_d$,
without bearing any impressions on the absolute values of the
phases $\alpha_u, \alpha_d, \beta_u$ and $\beta_d$ involved in the
quark mass matrices. However, the fermion mixing data as well as
the observed neutrino mass square differences have implications
not only on the hierarchy of the charged lepton mass matrices and
the light neutrino mass matrix but also on the phases involved in
these matrices. In particular, one finds that the quark mass
matrices follow the `natural' hierarchy whereas the lepton mass
matrices do not exhibit the same, as is expected. Specifically the
neutrinos follow a `normal' hierarchy, whereas the hierarchy for
the charged lepton mass matrix is different.

The analysis also leads to several interesting points regarding
the phase structure of the elements of the mass matrices. In
particular, the lepton mixing angle ${(s_{12}^l)}^2$ supports the
large values of $d_e$ only wherein $d_e > c_e$. However, these
large values of $d_e$ are compatible with small values of $d_u$
and $d_d$ suggesting that the quark mass matrices continue to
possess a `weak' hierarchy as one goes from $M_Z$ to $M_X$ scale.
Furthermore, it is observed that the parameter $t_e$ along with
the phase $\alpha_D$ of the Dirac neutrino mass matrix has to be
non zero in order to reproduce the observed values of the ratio
$\Delta m_{13}^2 / \Delta m_{12}^2$ indicating that real $M_e$ and
$M_{\nu D}$ may not be able to accommodate the lepton mixing data
under the constraints of SO(10). Likewise it is also observed that
the mixing angle ${(s_{23}^l)}^2$ can be reproduced only by non
vanishing values of the phase $\alpha_u$ of the mass matrix $M_u$
suggesting that within the SO(10) framework, the real $M_u$ may
not be allowed.

For the sake of completion, the allowed ranges of the elements of
the various mass matrices in the SO(10) framework have also been
presented, e.g.,
 \be M_u = \left( \ba{ccc}
 0 & (0.0114~$-$~0.0125)e^{i(162)} & 0\\
 (0.0114~$-$~0.0125)e^{-i(162)} & 9~$-$~20 & (24.5~$-$~33.06)e^{i(-0.06)}\\
 0 & (24.5~$-$~33.06)e^{-i(-0.06)} & 53.77~$-$~64.77
 \ea \right), \ee
\be M_d = \left( \ba{ccc}
 0 & (0.0053~$-$~0.0057)e^{i(74)} & 0\\
 (0.0053~$-$~0.0057)e^{-i(74)} & 0.1~$-$~0.22 & (0.33~$-$~0.43)e^{i(-7)}\\
 0 & (0.33~$-$~0.43)e^{-i(-7)} & 0.76~$-$~0.88
 \ea \right), \ee
 \be M_e = \left( \ba{ccc}
 0 & (0.0117~$-$~0.0182)e^{i(-37)} & 0\\
 (0.0117~$-$~0.0182)e^{-i(-37)} & 1.01~$-$~1.35 & (0.697~$-$~0.865)e^{i(-39)}\\
 0 & (0.697~$-$~0.865)e^{-i(-39)} & 0.2369~$-$~0.5769
 \ea \right), \ee
\be M_{\nu d} = \left( \ba{ccc}
 0 & (3.19~$-$~6.65)e^{i(83)} & 0\\
 (3.19~$-$~6.65)e^{-i(83)} & 71.17~$-$~90.24 & (-84.78~$-$~-47.21)e^{i(45)}\\
 0 & (-84.78 ~$-$~ -47.21)e^{-i(45)} & 22.74~$-$~44.16
 \ea \right), \ee
\be M_{\nu} = \left( \ba{ccc}
 0 & (0.53~$-$~1.02)\times 10^{-11} & 0\\
 (0.53~$-$~1.02)\times 10^{-11} & (0.40~$-$~1.69)\times 10^{-11} & (0.270~$-$~2.19)\times 10^{-11}\\
 0 & (0.270~$-$~2.19)\times 10^{-11} & (3.38~$-$~5.19)\times 10^{-11}
 \ea \right), \ee
\be M_{R} = \left( \ba{ccc}
 0 & (-1.42~$-$~6.89)\times 10^{12} & 0\\
 (-1.42~$-$~6.89)\times 10^{12} & (-1.5~$-$~0.283)\times 10^{14} & (-16.11 ~$-$~ -4.03)\times 10^{12}\\
 0 & (-16.11 ~$-$~ -4.03)\times 10^{12} & (1.28~$-$~4.87)\times 10^{13}
 \ea \right). \ee

The corresponding CKM and PMNS matrices as well as the associated
parameters are as follows, \be V_{{\rm
CKM}}=\left(\begin{array}{ccc}
               0.9738-0.9747 & 0.2235-0.2274 & 0.00367-0.00577\\
               0.2233-0.2272 & 0.9720-0.9736 & 0.0454-0.0618\\
               0.0009-0.0139 & 0.0446-0.0608 & 0.9981-0.9990 \end{array} \right), \ee
\be {\rm sin\,}2\beta=0.6561-0.7058, J_q=(3.266-6.935)\times
10^{-5}, \delta_{13}= 55.73^\circ~to~81.16^\circ.\ee \be V_{{\rm
PMNS}}= \left(\begin{array}{ccc}
                 0.7791-0.8540 & 0.5025-0.6009 & 0.0709-0.2236\\
                 0.3050-0.4286 & 0.4785-0.7552 & 0.5724-0.7958\\
                 0.3712-0.5124 & 0.4004-0.7042 & 0.5865-0.8022 \end{array} \right), \ee
\be  (s_{12}^l)^2=0.265-0.364, (s_{13}^l)^2=0.005-0.05,
(s_{23}^l)^2=0.3403-0.6399 ,\ee \be \Delta m
_{12}^2=(6.99-8.18)\times 10^{-23}{\rm GeV}^2, \Delta m
_{13}^2=(2.06-2.67) \times 10^{-21} {\rm GeV}^2 \ee
 \be m_{\nu 1}=(2.92-7.07) \times 10^{-12} {\rm GeV}, m_{\nu2}=(0.89-1.14)\times 10^{-11} {\rm GeV},\ee
 \be m_{\nu3}=(4.55-5.20) \times 10^{-11} {\rm GeV} \ee
\be J_l=-0.0428-0.0353, \delta_l=-87.16^\circ-71.31^\circ. \ee It
is easy to check that the CKM matrix and the related parameters
have good overlap with the PDG 2010 values. Similarly, the PMNS
matrix and the related parameters are also in good agreement with
a recent analysis\cite{othersmm3}.

\section{Summary and conclusion \label{summ}}
The fermion masses and mixings not only provide a fertile ground
to hunt for physics beyond the SM but also pose a big challenge to
understand these from more fundamental considerations. In the
present work, attempts have been made to present a comprehensive
review of some of the aspects of fermion mixing phenomenon and
texture specific mass matrices. In the context of fermion mixings,
keeping in mind the role played by unitarity of the CKM matrix and
unitarity triangles in establishing the CKM paradigm, implications
of these on parameters like sin$\,2\beta$, $V_{ub}$ and phase
$\delta$ have been discussed. Interestingly, one finds that
unitarity along with precisely measured $V_{us}$, $V_{cb}$,
sin$\,2\beta$ and angle $\alpha$ provides important constraints on
the CKM matrix element $V_{ub}$ and the CP violating phase
$\delta$.

The recent precision measurements of CKM phenomenological
parameters along with several developments in the lattice QCD
calculations of hadronic factors in the case of $K - \bar{K}$ and
the $B_d - \bar{B_d}$ mixings and the contribution of the long
distance effects in the $K - \bar{K}$ system provide motivation to
investigate the implications of these for the CKM phenomenology.
In this context, some authors\cite{buras,soni} point towards the
possibility of New Physics (NP) in these systems to the tune of
20$\%$ or so. However, a recent analysis\cite{ourepsilon} has made
an attempt to re-look this issue and interestingly, one finds that
both the $K - \bar{K}$ and $B_d - \bar{B_d}$ systems do not seem
to provide any significant clues regarding the possibility of
existence of NP and therefore, the presence of NP effects, if any,
would be less than a few percent only.

The issues of unitarity of the PMNS matrix and unitarity triangles
in the leptonic sector have also been discussed in the present
work.  To this end, attempts\cite{ourlepuni} have been made to
explore the possibility of the construction of the leptonic
unitarity triangle in the modified tribimaximal scenario of
Bjorken {\it et al.} \cite{bjorken}. In particular, using the PMNS
matrix constructed in this scenario and considering values of
$U_{e3}$ suggested by different theoretical models, the Dirac-like
CP violating phase $\delta$ in the leptonic sector has been found.
Further, in light of recent T2K, MINOS, DAYA BAY and RENO
observations regarding the mixing angle $s_{13}$, the possibility
of existence of CP violation in the leptonic sector has been
explored\cite{lepcp}, suggesting a good possibility of having non
zero CP violation.

Coming to the texture specific hermitian fermion mass matrices, in
the present work, we have given an overview of possible cases of
Fritzsch-like as well as non Fritzsch-like texture 6 and 5 zero
fermion mass matrices. Further, for the case of texture 4 zero
Fritzsch-like quark mass matrices, the issue of the hierarchy of
the elements of the mass matrices and the role of their phases
have been discussed. Furthermore, the case of texture 4 zero
Fritzsch-like lepton mass matrices has also been discussed with an
emphasis on the hierarchy of neutrino masses for both Majorana and
Dirac neutrinos.

For the case of quarks\cite{neelu56zeroquarks}, all the texture 6
zero combinations are completely ruled out whereas in the case of
texture 5 zero mass matrices the only viable possibility looks to
be that of Fritzsch-like matrices which shows only limited
viability, depending upon the light quark masses used as input.
Further, for the case of texture 4 zero quark mass
matrices\cite{s2b,cps}, including the case of `weak hierarchy'
along with the usually considered `strong hierarchy' case, one
finds that the weakly hierarchical mass matrices are able to
reproduce the strongly hierarchical mixing angles. Also, both the
phases having their origin in the mass matrices have to be non
zero to achieve compatibility of these matrices with the quark
mixing data, in particular with the parameter sin$\,2\beta$.

Similar investigations have been presented for the neutrino mixing
data considering normal/ inverted hierarchy and degenerate
scenario of neutrino masses for Majorana as well as Dirac
neutrinos. For the texture 6 zero case\cite{neelu6zerolep}, all
the possibilities pertaining to normal/ inverted hierarchy and
degenerate scenario of neutrino masses for Dirac neutrinos and
inverted hierarchy as well as degenerate scenarios in the case of
Majorana neutrinos are ruled out. Normal hierarchy of neutrino
masses for Majorana neutrinos results into some combinations which
are in accordance with the neutrino oscillation data. Regarding
texture 5 zero lepton mass matrices\cite{lepnonfritex5zero},
interestingly, one finds that these can accommodate all
hierarchies of neutrino masses.

For the Fritzsch-like texture 4 zero neutrino mass
matrices\cite{ourneut4zero}, analysis pertaining to both Majorana
and Dirac neutrinos for different hierarchies of neutrino masses
reveals that for both types of neutrinos, all the cases pertaining
to inverted hierarchy and degenerate scenarios of neutrino masses
are ruled out at $3\sigma$ C.L. by the existing data. For the
normal hierarchy cases, one gets viable ranges of neutrino masses,
mixing angle $s_{13}$, Jarlskog's rephasing invariant parameter
$J_l$ and the CP violating Dirac-like phase $\delta_l$.
Interestingly, a measurement of $m_{\nu_1}$ and further
refinements regarding mixing angle $\theta_{13}$ could have
important implications for the nature of neutrinos.

The success of texture 4 zero weakly hierarchical mass matrices
warrants a closer look at the origin of these from more
fundamental considerations. To this end, general concepts like
naturalness\cite{nmm} and weak basis
transformations\cite{9912358,brancolep4z,brancowb1}\cdash\cite{fxwb2}
for reducing the general mass matrices to texture specific form
have been discussed. Using the condition of naturalness as well as
the facility of WB transformations, the most general mass matrices
$M_{u,d}$ or/and $M_{\nu,e}$ can be reduced to texture 1 zero
hermitian mass matrices, the implications of fermion mixing data
on these texture structures have been discussed. One finds that
the additional (1,1) elements in these matrices do not show any
significant effect on the CKM parameters. Similar observations
have been made in the case of lepton mass matrices with similar
texture structures wherein the entire range of the various lepton
mixing parameters can be reproduced for the case of Dirac as well
as Majorana neutrinos. From these observations one can conclude
that for the purpose of accommodating the quark mixing data as
well as the lepton mixing data, without loss of generality, the
texture 4 zero hermitian mass matrices can be considered to be
equivalent to texture 2 zero hermitian mass matrices. This also
motivates one to understand the significance of texture 4 zero
mass matrices from the `top-down' perspective.

As a next step, the issue of compatibility of the texture 4 zero
Fritzsch-like hermitian mass matrices with the SO(10) inspired
mass matrices has been discussed. One notes that the texture 4
zero hermitian mass matrices $M_u$, $M_d$, $M_e$ and $M_{\nu D}$
can be expressed in terms of SO(10) inspired symmetric and anti-
symmetric texture 4 zero mass matrices. Interestingly, one finds
that a simultaneous fit to the fermion masses and mixings within
the constraints of SO(10) and naturalness can be arrived at. The
analysis also shows that quarks and the neutrino mass matrices
follow normal hierarchy. Further, in the case of quarks, there are
constraints on the hierarchy of the elements of the quark mass
matrices and on the phase differences. In the case of leptons, the
fermion mixing data has implications not only on the hierarchy of
the charged lepton mass matrices and the light neutrino mass
matrix but also on the phases involved in these. The weakly
hierarchical quark mass matrices continue to be supported within
the SO(10) framework.

In conclusion, we would like to remark that on the one hand there
is a need to take the analysis of texture specific mass matrices
towards completion. For example, besides carrying out the analysis
of texture 4 zero non Fritzsch-like fermion mass matrices, one has
to consider texture 3 zero cases also, the latter corresponding to
general mass matrices after carrying out weak basis rotations. On
the other hand, one may also consider breaking the hermiticity
condition perturbatively as has been done
recently\cite{frixingzhou} and to go into its detailed
implications. Similarly, the compatibility of the texture 4 zero
Fritzsch-like hermitian mass matrices with the SO(10) inspired
mass matrices motivates one to find deeper understanding of the
texture 4 zero {\it ans\"{a}tze}, may be within SO(10),
incorporating Abelian or Horizontal symmetries.

\section*{Acknowledgments}
The authors would like to thank Prof. K. K. Phua and the
organizers of the `1st IAS-CERN School on Particle Physics and
Cosmology and Implications for Technology' held in NTU, Singapore,
9-31 January 2012, for giving an opportunity to present several
aspects of the present work. The authors would also like to thank
Prof. N. P. Chang, Rohit Verma, Priyanka Fakay and Samandeep for
stimulating discussions and help as well as the Chairman,
Department of Physics for providing facilities to work in the
department. G.A. would also like to acknowledge DST, Government of
India (Grant No: SR/FTP/PS-017/2012) for financial support.

\appendix

\section{Diagonalizing transformation of texture 2 zero lepton mass matrices}

The Fritzsch-like texture 2 zero lepton mass matrices can be
expressed as
 \be
 M_{l}=\left( \ba{ccc}
0 & A _{l} & 0      \\ A_{l}^{*} & D_{l} &  B_{l}     \\
 0 &     B_{l}^{*}  &  C_{l} \ea \right), \qquad
M_{\nu D}=\left( \ba{ccc} 0 &A _{\nu} & 0      \\ A_{\nu}^{*} &
D_{\nu} &  B_{\nu}     \\
 0 &     B_{\nu}^{*}  &  C_{\nu} \ea \right),
 \label{frzmm5}
 \ee
$M_{l}$ and $M_{\nu D}$ respectively corresponding to charged
lepton and Dirac neutrino mass matrices. It may be noted that each
of the above matrix is texture 2 zero type with $A_{l(\nu)}
=|A_{l(\nu)}|e^{i\alpha_{l(\nu)}}$
 and $B_{l(\nu)} = |B_{l(\nu)}|e^{i\beta_{l(\nu)}}$, in case these
 are symmetric then $A_{l(\nu)}^*$ and $B_{l(\nu)}^*$ should be
 replaced by $A_{l(\nu)}$ and $B_{l(\nu)}$, as well as
 $C_{l(\nu)}$ and $D_{l(\nu)}$ should respectively be defined as $C_{l(\nu)}
 =|C_{l(\nu)}|e^{i\gamma_{l(\nu)}}$ and $D_{l(\nu)}
 =|D_{l(\nu)}|e^{i\omega_{l(\nu)}}$.

Texture 6 zero mass matrices can be obtained from the above
mentioned matrices by taking both $D_l$ and $D_{\nu}$ to be zero,
which reduces the matrices $M_{l}$ and $M_{\nu D}$ each to texture
3 zero type. Texture 5 zero matrices can be obtained by taking
either $D_l=0$ and $D_{\nu}\neq 0$ or $D_{\nu}=0$ and $D_l \neq
0$, thereby, giving rise to two possible cases of texture 5 zero
matrices, referred to as texture 5 zero $D_l=0$ case pertaining to
$M_l$ texture 3 zero type and $M_{\nu D}$ texture 2 zero type and
texture 5 zero $D_{\nu}=0$ case pertaining to $M_l$ texture 2 zero
type and $M_{\nu D}$ texture 3 zero type.

To fix the notations and conventions, we detail the formalism
connecting the mass matrix to the neutrino mixing matrix. The mass
matrices $M_l$ and $M_{\nu D}$ given in Eq.~(\ref{frzmm5}), for
hermitian as well as symmetric case, can be exactly diagonalized.
To facilitate diagonalization, the mass matrix $M_k$, where $k=l,
\nu D$, can be expressed as
\be
M_k= Q_k M_k^r P_k \,  \label{mk1} \ee or  \be M_k^r=
Q_k^{\dagger} M_k P_k^{\dagger}\,, \label{mkr1} \ee where $M_k^r$
is a real symmetric matrix with real eigenvalues and $Q_k$ and
$P_k$ are diagonal phase matrices. For the hermitian case $Q_k=
P_k^{\dagger}$, whereas for the symmetric case under certain
conditions $Q_k= P_k$. In general, the real matrix $M_k^r$ is
diagonalized by the orthogonal transformation $O_k$, e.g., \be
M_k^{diag}= {O_k}^T M_k^r O_k \,, \label{mkdiag} \ee which on
using Eq.~(\ref{mkr1}) can be rewritten as \be M_k^{diag}= {O_k}^T
Q_k^{\dagger} M_k P_k^{\dagger} O_k \,. \label{mkdiag2} \ee To
facilitate the construction of diagonalization transformations for
different hierarchies, we introduce a diagonal phase matrix
$\xi_k$ defined as $ {\rm diag} (1,\,e^{i \pi},\,1)$ for the case
of normal hierarchy and as $ {\rm diag} (1,\,e^{i \pi},\,e^{i
\pi})$ for the case of inverted hierarchy. Eq.~(\ref{mkdiag2}) can
now be written as \be \xi_k M_k^{diag}= {O_k}^T Q_k^{\dagger} M_k
P_k^{\dagger} O_k \,, \label{mkdiag3} \ee which can also be
expressed as \be M_k^{diag}= \xi_k^{\dagger} {O_k}^T Q_k^{\dagger}
M_k P_k^{\dagger} O_k \,. \label{mkdiag4} \ee Making use of the
fact that $O_k^*=O_k$ it can be further expressed as
\be
M_k^{diag}=(Q_k O_k \xi_k)^{\dagger} M_k (P_k^{\dagger}
O_k),\label{mkeq} \ee from which one gets \be M_k=Q_k O_k \xi_k
M_k^{diag} O_k^T P_k.\label{mkeq2} \ee

The case of leptons is fairly straight forward, for the neutrinos
the diagonalizing transformation is hierarchy specific as well as
requires some fine tuning of the phases of the right handed
neutrino mass matrix $M_R$. To clarify this point further, in
analogy with Eq.~(\ref{mkeq2}), we can express $M_{\nu D}$ as \be
M_{\nu D}=Q_{\nu D} O_{\nu D} \xi_{\nu D} M_{\nu D}^{diag} O_{\nu
D}^T P_{\nu D}.\label{mnud} \ee Substituting the above value of
$M_{\nu D}$ in Eq.~(\ref{seesaweq2}) one obtains
\be
M_{\nu}=-(Q_{\nu D} O_{\nu D} \xi_{\nu D} M_{\nu D}^{diag} O_{\nu
D}^T P_{\nu D})^T (M_R)^{-1} (Q_{\nu D} O_{\nu D} \xi_{\nu D}
M_{\nu D}^{diag} O_{\nu D}^T P_{\nu D}). \ee On using $P_{\nu D}^T
= P_{\nu D}$, the above equation can further be written as
\be
M_{\nu}=-P_{\nu D} O_{\nu D} M_{\nu D}^{diag} \xi_{\nu D} O_{\nu
D}^T Q_{\nu D}^T (M_R)^{-1} Q_{\nu D} O_{\nu D} \xi_{\nu D} M_{\nu
D}^{diag} O_{\nu D}^T P_{\nu D}. \ee Assuming fine tuning, the
phase matrices $Q_{\nu D}^T$ and $Q_{\nu D}$ along with $-M_R$ can
be taken as $m_R ~{\rm diag} (1,1,1)$ as well as using the
unitarity of $\xi_{\nu D}$ and orthogonality of $O_{\nu D}$, the
above equation can be expressed as
\be
M_{\nu}= P_{\nu D} O_{\nu D} \frac{(M_{\nu
D}^{diag})^2}{(m_R)^{-1}} O_{\nu D}^T P_{\nu D}. \label{mnu} \ee

The lepton mixing matrix, obtained from the matrices used for
diagonalizing the mass matrices $M_l$ and $M_{\nu}$, is expressed
as
 \be
U =(Q_l O_l \xi_l)^{\dagger} (P_{\nu D} O_{\nu D}). \label{mix5}
\ee Eliminating the phase matrix $\xi_l$ by redefinition of the
charged lepton phases, the above equation becomes
\be
 U = O_l^{\dagger} Q_l P_{\nu D} O_{\nu D} \,, \label{mixreal} \ee
where $Q_l P_{\nu D}$, without loss of generality, can be taken as
$(e^{i\phi_1},\,1,\,e^{i\phi_2})$, $\phi_1$ and $\phi_2$ being
related to the phases of mass matrices and can be treated as free
parameters.

To understand the relationship between diagonalizing
transformations for different hierarchies of neutrino masses as
well as their relationship with the charged lepton case, we
reproduce the general diagonalizing transformation $O_k$, e.g.,
\be O_k= \left( \ba{ccc} \pm O_k(11)& \pm O_k(12)& \pm O_k(13) \\
 \pm O_k(21)& \mp O_k(22)& \pm O_k(23)\\
 \mp O_k(31) & \pm O_k(32) & \pm O_k(33) \ea \right), \ee
where \beqn O_k(11) & = & {\sqrt \frac{m_{2} m_{3}
(m_{3}-m_{2}-D_k)}
     {(m_{1}-m_{2}+m_{3}-D_k)
(m_{3}-m_{1})(m_{1}+m_{2})} } \nonum  \\ O_k(12) & = & {\sqrt
\frac{m_{1} m_{3}
 (m_{1}+m_{3}-D_k)}
   {(m_{1}-m_{2}+m_{3}-D_k)
 (m_{2}+m_{3})(m_{2}+m_{1})} }
\nonum   \\O_k(13) & = & {\sqrt \frac{m_{1} m_{2}
 (m_{2}-m_{1}+D_k)}
    {(m_{1}-m_{2}+m_{3}-D_k)
(m_{3}+m_{2})(m_{3}-m_{1})} } \nonum   \\ O_k(21) & = & {\sqrt
\frac{m_{1}
 (m_{3}-m_{2}-D_k)}
  {(m_{3}-m_{1})(m_{1}+m_{2})} }
\nonum  \\O_k(22) & = & {\sqrt \frac{m_{2} (m_{3}+m_{1}-D_k)}
  {(m_{2}+m_{3})(m_{2}+m_{1})} }
 \nonum    \\
O_k(23) & = & \sqrt{\frac{m_3(m_{2}-m_{1}+D_k)}
 {(m_{2}+m_{3})(m_{3}-m_{1})} }
\nonum   \\O_k(31) & = &
 \sqrt{\frac{m_{1} (m_{2}-m_{1}+D_k)
    (m_{1}+m_{3}-D_k)}
{(m_{1}-m_{2}+m_{3}-D_k)(m_{1}+m_{2})(m_{3}-m_{1})}} \nonum
\\O_k(32) & = & {\sqrt \frac{m_{2}(D_k-m_{1}+m_{2})
(m_{3}-m_{2}-D_k)}{(m_{1}-m_{2}+m_{3}-D_k)
 (m_{2}+m_{3})(m_{2}+m_{1})} }
 \nonum  \\
O_k(33) & = & {\sqrt \frac{m_{3}(m_{3}-m_{2}-D_k)
(m_{1}+m_{3}-D_k)}{(m_{1}-m_{2}+m_{3}-D_k)
 (m_{3}-m_{1})(m_{3}+m_{2})}} \label{diageq} \,,
 \eeqn  $m_1$, $-m_2$,
$m_3$ being the eigenvalues of $M_k$. In the case of charged
leptons, because of the hierarchy $m_e \ll m_{\mu} \ll m_{\tau}$,
the mass eigenstates can be approximated respectively to the
flavor eigenstates. Using the approximation, $m_{l1} \simeq m_e$,
$m_{l2} \simeq m_{\mu}$ and $m_{l3} \simeq m_{\tau}$, the first
element of the matrix $O_l$ can be obtained from the corresponding
element of Eq.~(\ref{diageq}) by replacing $m_1$, $-m_2$, $m_3$ by
$m_e$, $-m_{\mu}$, $m_{\tau}$, e.g.,
 \be  O_l(11) = {\sqrt
\frac{m_{\mu} m_{\tau} (m_{\tau}-m_{\mu}-D_l)}
     {(m_{e}-m_{\mu}+m_{\tau}-D_l)
(m_{\tau}-m_{e})(m_{e}+m_{\mu})} } ~. \ee

For normal hierarchy defined as $m_{\nu_1}<m_{\nu_2}\ll
m_{\nu_3}$, as well as for the corresponding degenerate case given
by $m_{\nu_1} \lesssim m_{\nu_2} \sim m_{\nu_3}$,
Eq.~(\ref{diageq}) can also be used to obtain the first element of
diagonalizing transformation for Dirac neutrinos as well as
Majorana neutrinos. The first element of the diagonalizing
transformation for Dirac neutrinos can be obtained from the
corresponding element of Eq.~(\ref{diageq}) by replacing $m_1$,
$-m_2$, $m_3$ by $m_{\nu 1}$, $-m_{\nu 2}$, $m_{\nu 3}$ and is
given by
 \be O_{\nu D}(11)  =  {\sqrt \frac{m_{\nu_2} m_{\nu 3} (m_{\nu 3}-m_{\nu 2}-D_{\nu})}
     {(m_{\nu 1}-m_{\nu 2}+m_{\nu 3}-D_{\nu})
(m_{\nu 3}-m_{\nu 1})(m_{\nu 1}+m_{\nu 2})} }, \ee where
$m_{\nu_1}$, $m_{\nu_2}$ and $m_{\nu_3}$ are neutrino masses.
Similarly, for Majorana neutrinos, replacing $m_1$, $-m_2$, $m_3$
by $\sqrt{m_{\nu 1} m_R}$, $-\sqrt{m_{\nu 2} m_R}$, $\sqrt{m_{\nu
3} m_R}$ in the equation, we get \be O_{\nu}(11) = {\sqrt
\frac{\sqrt{m_{\nu_2}}
    \sqrt{m_{\nu_3}}
( \sqrt{m_{\nu_3}}-\sqrt{ m_{\nu_2}}-D_{\nu})}
{(\sqrt{m_{\nu_1}}-\sqrt{m_{\nu_2}} + \sqrt{m_{\nu_3}}- D_{\nu})
(\sqrt{m_{\nu_3}}-\sqrt{m_{\nu_1}}) (\sqrt{m_{\nu_1}} +
\sqrt{m_{\nu_2}} )} } \label{omajnh}. \ee  The parameter $D_{\nu}$
is to be divided by $\sqrt{m_R}$, however as $D_{\nu}$ is
arbitrary therefore we retain it as it is.

In the same manner, for Dirac and Majorana neutrinos one can
obtain the elements of diagonalizing transformation for the
inverted hierarchy case defined as $m_{\nu_3} \ll m_{\nu_1} <
m_{\nu_2}$ as well as for the corresponding degenerate case given
by $m_{\nu_3} \sim m_{\nu_1} \lesssim m_{\nu_2}$. For the case of
Dirac neutrinos, the first element, obtained by replacing $m_1$,
$-m_2$, $m_3$ with $m_{\nu 1}$, $-m_{\nu 2}$, $-m_{\nu 3}$ in
Eq.~(\ref{diageq}), is given by
 \be O_{\nu D}(11)  =  {\sqrt \frac{m_{\nu_2} m_{\nu 3} (m_{\nu 3}+m_{\nu 2}+D_{\nu})}
     {(-m_{\nu 1}+m_{\nu 2}+m_{\nu 3}+D_{\nu})
(m_{\nu 3}+m_{\nu 1})(m_{\nu 1}+m_{\nu 2})} }. \ee For Majorana
neutrinos, by replacing $m_1$, $-m_2$, $m_3$ in Eq.~(\ref{diageq})
with $\sqrt{m_{\nu_1} m_R}$, $-\sqrt{m_{\nu_2} m_R}$,
$-\sqrt{m_{\nu_3} m_R}$, we obtain \be O_{\nu}(11) = {\sqrt
\frac{\sqrt{m_{\nu_2}}
    \sqrt{m_{\nu_3}}
(D_{\nu}+\sqrt{ m_{\nu_2}} + \sqrt{m_{\nu_3}} )}
{(-\sqrt{m_{\nu_1}}+\sqrt{m_{\nu_2}} + \sqrt{m_{\nu_3}}+ D_{\nu})
(\sqrt{m_{\nu_1}}+\sqrt{m_{\nu_3}}) (\sqrt{m_{\nu_1}} +
\sqrt{m_{\nu_2}} )} } \label{omajih}. \ee The other elements of
diagonalizing transformations in the case of neutrinos as well as
charged leptons can similarly be found.

\section{Elements of the PMNS mixing matrix}
 In this Appendix, we present the elements of the PMNS mixing
matrix in the case of Dirac neutrinos corresponding to texture 4
zero mass matrices. For Majorana neutrinos, the elements of the
PMNS mixing matrix can be derived from those presented below by
replacing $m_1$, $-m_2$, $m_3$ by $\sqrt{m_{\nu 1} m_R}$,
$-\sqrt{m_{\nu 2} m_R}$, $\sqrt{m_{\nu 3} m_R}$.

Further, the corresponding relations for the texture 5 and 6 zero
mass matrices can be easily derived from these. For example,
considering both $D_l$ and $D_{\nu}$ to be zero the relations for
texture 6 zero mass matrices are obtained, whereas for the texture
5 zero mass matrices either $D_l$ or $D_{\nu}$ is considered to be
zero. The expressions for the elements of the PMNS mixing matrix
are given by
 \beqn
 U_{e1}=\sqrt{\frac{m_1 (-D_{\nu} + m_2 +
m_3)}{(m_1-m_2)(-m_1+m_3)}} \sqrt{\frac{m_e (-D_{l} + m_{\mu} +
m_{\tau})}{(m_e-m_{\mu})(-m_e+m_{\tau})}}+~~~~~~~~~~~ \nonum \\
\sqrt{\frac{-m_2 m_3 (-D_{\nu} + m_2 +
m_3)}{C_{\nu}(m_1-m_2)(-m_1+m_3)}} \sqrt{\frac{-m_{\mu}
m_{\tau}(-D_{l} + m_{\mu} +
m_{\tau})}{C_l(m_e-m_{\mu})(-m_e+m_{\tau})}}~\phi_1 + \nonum
\\ \sqrt{\frac{m_1 (D_{\nu} - m_1 - m_2)(-D_{\nu} + m_1 +
m_3)}{C_{\nu}(m_1-m_2)(-m_1+m_3)}} \times
~~~~~~~~~~~~~~~~~~~~~~~~~~\nonum \\ \sqrt{\frac{m_e (D_{l} - m_e -
m_{\mu})(-D_{l} + m_e +
m_{\tau})}{C_{l}(m_e-m_{\mu})(-m_e+m_{\tau})}}~\phi_2~~~~~~~~~~~~~~~~~~~~~~~~~~~
 \eeqn
 \beqn
U_{e2}=\sqrt{\frac{-m_2 (-D_{\nu} + m_1 +
m_3)}{(m_1-m_2)(-m_2+m_3)}} \sqrt{\frac{m_e (-D_{l} + m_{\mu} +
m_{\tau})}{(m_e-m_{\mu})(-m_e+m_{\tau})}}-~~~~~~~~~~~ \nonum \\
\sqrt{\frac{m_1 m_3 (-D_{\nu} + m_1 +
m_3)}{C_{\nu}(m_1-m_2)(-m_2+m_3)}} \sqrt{\frac{-m_{\mu}
m_{\tau}(-D_{l} + m_{\mu} +
m_{\tau})}{C_l(m_e-m_{\mu})(-m_e+m_{\tau})}}~\phi_1 + \nonum
\\ \sqrt{\frac{-m_2 (D_{\nu} - m_1 - m_2)(-D_{\nu} + m_2 +
m_3)}{C_{\nu}(m_1-m_2)(-m_2+m_3)}} \times
~~~~~~~~~~~~~~~~~~~~~~~~~~\nonum \\ \sqrt{\frac{m_e (D_{l} - m_e -
m_{\mu})(-D_{l} + m_e +
m_{\tau})}{C_{l}(m_e-m_{\mu})(-m_e+m_{\tau})}}~\phi_2~~~~~~~~~~~~~~~~~~~~~~~~~~~
 \eeqn
 \beqn
  U_{e3}=\sqrt{\frac{m_3 (D_{\nu} - m_1 -
m_2)}{(-m_1+m_3)(-m_2+m_3)}} \sqrt{\frac{m_e (-D_{l} + m_{\mu} +
m_{\tau})}{(m_e-m_{\mu})(-m_e+m_{\tau})}}+~~~~~~~~~~~ \nonum \\
\sqrt{\frac{-m_1 m_2 (D_{\nu} - m_1 -
m_2)}{C_{\nu}(-m_1+m_3)(-m_2+m_3)}} \sqrt{\frac{-m_{\mu}
m_{\tau}(-D_{l} + m_{\mu} +
m_{\tau})}{C_l(m_e-m_{\mu})(-m_e+m_{\tau})}}~\phi_1 - \nonum
\\ \sqrt{\frac{m_3 (-D_{\nu} + m_1 + m_3)(-D_{\nu} + m_2 +
m_3)}{C_{\nu}(-m_1+m_3)(-m_2+m_3)}} \times
~~~~~~~~~~~~~~~~~~~~~~~~~~\nonum \\ \sqrt{\frac{m_e (D_{l} - m_e -
m_{\mu})(-D_{l} + m_e +
m_{\tau})}{C_{l}(m_e-m_{\mu})(-m_e+m_{\tau})}}~\phi_2~~~~~~~~~~~~~~~~~~~~~~~~~~~
 \eeqn
 \beqn
 U_{\mu 1}=\sqrt{\frac{m_1 (-D_{\nu} + m_2 +
m_3)}{(m_1-m_2)(-m_1+m_3)}} \sqrt{\frac{-m_{\mu} (-D_{l} + m_{e} +
m_{\tau})}{(m_e-m_{\mu})(-m_{\mu}+m_{\tau})}}-~~~~~~~~~~~ \nonum
\\ \sqrt{\frac{-m_2 m_3 (-D_{\nu} + m_2 +
m_3)}{C_{\nu}(m_1-m_2)(-m_1+m_3)}} \sqrt{\frac{m_{e}
m_{\tau}(-D_{l} + m_{e} +
m_{\tau})}{C_l(m_e-m_{\mu})(-m_{\mu}+m_{\tau})}}~\phi_1 + \nonum
\\ \sqrt{\frac{m_1 (D_{\nu} - m_1 - m_2)(-D_{\nu} + m_1 +
m_3)}{C_{\nu}(m_1-m_2)(-m_1+m_3)}} \times
~~~~~~~~~~~~~~~~~~~~~~~~~~\nonum \\ \sqrt{\frac{-m_{\mu} (D_{l} -
m_e - m_{\mu})(-D_{l} + m_{\mu} +
m_{\tau})}{C_{l}(m_e-m_{\mu})(-m_{\mu}+m_{\tau})}}~\phi_2~~~~~~~~~~~~~~~~~~~~~~~~~~~
 \eeqn
 \beqn
U_{\mu 2}=\sqrt{\frac{-m_2 (-D_{\nu} + m_1 +
m_3)}{(m_1-m_2)(-m_2+m_3)}} \sqrt{\frac{-m_{\mu} (-D_{l} + m_{e} +
m_{\tau})}{(m_e-m_{\mu})(-m_{\mu}+m_{\tau})}}+~~~~~~~~~~~ \nonum
\\ \sqrt{\frac{m_1 m_3 (-D_{\nu} + m_1 +
m_3)}{C_{\nu}(m_1-m_2)(-m_2+m_3)}} \sqrt{\frac{m_{e}
m_{\tau}(-D_{l} + m_{e} +
m_{\tau})}{C_l(m_e-m_{\mu})(-m_{\mu}+m_{\tau})}}~\phi_1 + \nonum
\\ \sqrt{\frac{-m_2 (D_{\nu} - m_1 - m_2)(-D_{\nu} + m_2 +
m_3)}{C_{\nu}(m_1-m_2)(-m_2+m_3)}} \times
~~~~~~~~~~~~~~~~~~~~~~~~~~\nonum \\ \sqrt{\frac{-m_{\mu} (D_{l} -
m_e - m_{\mu})(-D_{l} + m_{\mu} +
m_{\tau})}{C_{l}(m_e-m_{\mu})(-m_{\mu}+m_{\tau})}}~\phi_2~~~~~~~~~~~~~~~~~~~~~~~~~~~
 \eeqn
\beqn
 U_{\mu 3}=\sqrt{\frac{m_3 (D_{\nu} - m_1 -
m_2)}{(-m_1+m_3)(-m_2+m_3)}} \sqrt{\frac{-m_{\mu} (-D_{l} + m_{e}
+ m_{\tau})}{(m_e-m_{\mu})(-m_{\mu}+m_{\tau})}}-~~~~~~~~~~~ \nonum
\\ \sqrt{\frac{-m_1 m_2 (D_{\nu} - m_1 -
m_2)}{C_{\nu}(-m_1+m_3)(-m_2+m_3)}} \sqrt{\frac{m_{e}
m_{\tau}(-D_{l} + m_{e} +
m_{\tau})}{C_l(m_e-m_{\mu})(-m_{\mu}+m_{\tau})}}~\phi_1 - \nonum
\\ \sqrt{\frac{m_3 (-D_{\nu} + m_1 + m_3)(-D_{\nu} + m_2 +
m_3)}{C_{\nu}(-m_1+m_3)(-m_2+m_3)}} \times
~~~~~~~~~~~~~~~~~~~~~~~~~~\nonum \\ \sqrt{\frac{-m_{\mu} (D_{l} -
m_e - m_{\mu})(-D_{l} + m_{\mu} +
m_{\tau})}{C_{l}(m_e-m_{\mu})(-m_{\mu}+m_{\tau})}}~\phi_2~~~~~~~~~~~~~~~~~~~~~~~~~~~
 \eeqn
 \beqn
  U_{\tau 1}=\sqrt{\frac{m_1 (-D_{\nu} + m_2 +
m_3)}{(m_1-m_2)(-m_1+m_3)}} \sqrt{\frac{m_{\tau} (D_{l} - m_{e} -
m_{\mu})}{(-m_e+m_{\tau})(-m_{\mu}+m_{\tau})}}+~~~~~~~~~~~ \nonum
\\ \sqrt{\frac{-m_2 m_3 (-D_{\nu} + m_2 +
m_3)}{C_{\nu}(m_1-m_2)(-m_1+m_3)}} \sqrt{\frac{-m_{e}
m_{\mu}(D_{l} - m_{e} -
m_{\mu})}{C_l(-m_e+m_{\tau})(-m_{\mu}+m_{\tau})}}~\phi_1 - \nonum
\\ \sqrt{\frac{m_1 (D_{\nu} - m_1 - m_2)(-D_{\nu} + m_1 +
m_3)}{C_{\nu}(m_1-m_2)(-m_1+m_3)}} \times
~~~~~~~~~~~~~~~~~~~~~~~~~~\nonum \\ \sqrt{\frac{m_{\tau} (-D_{l} +
m_e + m_{\tau})(-D_{l} + m_{\mu} +
m_{\tau})}{C_{l}(-m_e+m_{\tau})(-m_{\mu}+m_{\tau})}}~\phi_2~~~~~~~~~~~~~~~~~~~~~~~~~~~
 \eeqn
 \beqn
  U_{\tau 2}=\sqrt{\frac{-m_2 (-D_{\nu} + m_1 +
m_3)}{(m_1-m_2)(-m_2+m_3)}} \sqrt{\frac{m_{\tau} (D_{l} - m_{e} -
m_{\mu})}{(-m_e+m_{\tau})(-m_{\mu}+m_{\tau})}}-~~~~~~~~~~~ \nonum
\\ \sqrt{\frac{m_1 m_3 (-D_{\nu} + m_1 +
m_3)}{C_{\nu}(m_1-m_2)(-m_2+m_3)}} \sqrt{\frac{-m_{e}
m_{\mu}(D_{l} - m_{e} -
m_{\mu})}{C_l(-m_e+m_{\tau})(-m_{\mu}+m_{\tau})}}~\phi_1 - \nonum
\\ \sqrt{\frac{-m_2 (D_{\nu} - m_1 - m_2)(-D_{\nu} + m_2 +
m_3)}{C_{\nu}(m_1-m_2)(-m_2+m_3)}} \times
~~~~~~~~~~~~~~~~~~~~~~~~~~\nonum \\ \sqrt{\frac{m_{\tau} (-D_{l} +
m_e + m_{\tau})(-D_{l} + m_{\mu} +
m_{\tau})}{C_{l}(-m_e+m_{\tau})(-m_{\mu}+m_{\tau})}}~\phi_2~~~~~~~~~~~~~~~~~~~~~~~~~~~
 \eeqn
 \beqn
  U_{\tau 3}=\sqrt{\frac{m_3 (D_{\nu} - m_1 -
m_2)}{(-m_1+m_3)(-m_2+m_3)}} \sqrt{\frac{m_{\tau} (D_{l} - m_{e} -
m_{\mu})}{(-m_e+m_{\tau})(-m_{\mu}+m_{\tau})}}+~~~~~~~~~~~ \nonum
\\ \sqrt{\frac{-m_1 m_2 (D_{\nu} - m_1 -
m_2)}{C_{\nu}(-m_1+m_3)(-m_2+m_3)}} \sqrt{\frac{-m_{e}
m_{\mu}(D_{l} - m_{e} -
m_{\mu})}{C_l(-m_e+m_{\tau})(-m_{\mu}+m_{\tau})}}~\phi_1 - \nonum
\\ \sqrt{\frac{m_3 (-D_{\nu} + m_1 + m_3)(-D_{\nu} + m_2 +
m_3)}{C_{\nu}(-m_1+m_3)(-m_2+m_3)}} \times
~~~~~~~~~~~~~~~~~~~~~~~~~~\nonum \\ \sqrt{\frac{m_{\tau} (-D_{l} +
m_e + m_{\tau})(-D_{l} + m_{\mu} +
m_{\tau})}{C_{l}(-m_e+m_{\tau})(-m_{\mu}+m_{\tau})}}~\phi_2~~~~~~~~~~~~~~~~~~~~~~~~~~~
\eeqn

\section{Diagonalizing transformation of texture 1 zero hermitian mass
matrix} To facilitate diagonalization, the mass matrices $M_{q}$
may be expressed as $ M_q =  P_q^{\dagger} M_q^r P_q $ or $
M^{r}_q =  P_q M_q P_q^{\dagger}$ where $ M^{r}_q$ are real
symmetric matrices with real eigenvalues and $ P_q $ are the
diagonal phase matrices, e.g., \be M^{r}_{q} = \left(
 \ba {lll}
e_q &| a_{q}| & 0 \\ |a_{q}| & d_{q} &| b_{q}| \\ 0 & |b_{q}| &
c_{q}\ea \right), \qquad P_{q}= \left( \ba  {ccc} e^{i\alpha_{q}}
& 0 & 0 \\ 0 & 1 & 0 \\
            0 & 0 & e^{i\beta_{q}} \ea \right). \label{1fritzsch} \ee
The matrix $ M_{q}^r$ can be diagonalized using the following
transformations
\be
M^{diag}_{q}= O^{T}_{q}M^{r}_{q}O_{q}= O^{T}_{q}P_{q}
M_{q}P^{\dagger}_q O_{q}= {\rm Diag}(m_{1}, -m_{2}, m_{3}), \ee
where the subscripts 1, 2 and 3 refer respectively to $u$, $c$,
$t$ for the up sector, $d$, $s$, $b$ for the down sector, $e$,
$\mu$, $\tau $ for the charged lepton sector and $\nu_{1}$, $
\nu_{2} $, $\nu_{3}$ for the neutrino sector. The exact
diagonalization of the mass matrix $M^{r}_{q}$ can be carried out
using the three invariants, Trace $ M^{r}_{q}$, Trace $
(M^{r}_{q})^2 $ and Determinant $ M^{r}_{q}$. Using these, the
elements of the mass matrix $ |a_{q}|$, $ |b_{q}|$  and $ c_{q}$
can be expressed in terms of the free parameters $ e_{q}$, $
d_{q}$ and the respective fermion mass eigenvalues as
\be
c_q = (m_1-m_2+m_3-d_k-e_q)\,,\nonum \ee
\be
|a_k| = \sqrt\frac {(m_1-e_q)( m_2+e_q)(m_3-e_q)} {c_k-e_q}\,
   \nonum\ee
\be
|b_k| = \sqrt\frac {(c_q-m_1)( m_3-c_q)(c_q+m_2)} {c_q+e_q)}.
\label{elements1} \ee In order that the diagonalizing
transformations remain real, the free parameters $e_{q} $ and
$d_q$ get constrained within the limits \be
 (m_3-m_2-e_q)>d_q>(m_1-m_2-e_q),
 \ee
 and
 \be
 m_1>e_q>-m_2. \ee
The exact diagonalizing transformation $O_q$ for the matrix $
M^{r}_{q}$ is given by
 \be O_q = \left(\ba{lll}
   {\sqrt \frac{(e_q+m_2)(m_3-e_q)(c_q-m_1)}{c_q-e_q)(m_3-m_1)(m_2+m_1)}} &
    {\sqrt \frac{ (m_1-e_q)(m_3-e_q)(c_q+m_2)}{(c_q-e_q)(m_3+m_2)(m_2+m_1)}} &
  {\sqrt \frac{(m_1-e_q)(e_q+m-2)(m_3-c_q)}{(c_q-e_q)(m_3+m_2)(m_3-m_1)}}\\

  {\sqrt \frac{(c_q-m_1)(m_1-e_q)}{((m_3-m_1)(m_2+m_1)}} &
-{\sqrt \frac{(e_q+m_2)(c_q+m_2)}{(m_3+m_2)(m_2+m_1)}} &
 {\sqrt \frac{(m_3-e_q)(m_3-c_q) }{(m_3+m_2)(m_3-m_1)}} \\

  -{\sqrt \frac{(m_1-e_q)(m_3-c_q)(c_q+m_2)}{(c_q-e_q)((m_3-m_1)(m_2+m_1)}} &
 {\sqrt \frac{(e_q+m_2)(c_q-m_1)(m_3-c_q)}{(c_q-e_q)((m_3+m_2)(m_2+m_1)}} &
   {\sqrt \frac{(m_3-e_q)(c_q-m_1)(c_q+m_2)}{(c_q-e_q)
   (m_3+m_2)(m_3-m_1)}}  \ea \right). \label{ou} \ee

\end{document}